\documentclass{article}

\usepackage{arxiv}

\usepackage[utf8]{inputenc} 
\usepackage[T1]{fontenc}    
\usepackage{hyperref}       
\usepackage{url}            
\usepackage{booktabs}       
\usepackage{amsfonts}       
\usepackage{nicefrac}       
\usepackage{microtype}      
\usepackage{graphicx}
\usepackage{natbib}
\usepackage{doi}
\usepackage{multirow}
\usepackage{color,array,dcolumn}
\usepackage{amsmath}
\usepackage{amssymb}
\usepackage[export]{adjustbox}
\usepackage{bbm}
\usepackage{caption}

\newcolumntype{L}[1]{>{\raggedright\let\newline\\\arraybackslash\hspace{0pt}}m{#1}}
\newcolumntype{C}[1]{>{\centering\let\newline\\\arraybackslash\hspace{0pt}}m{#1}}
\newcolumntype{R}[1]{>{\raggedleft\let\newline\\\arraybackslash\hspace{0pt}}m{#1}}

\newcommand{\boldSigma}{\mbox{\boldmath$\mathsf{\Sigma}$}}

\title{Inversion of sea surface currents from satellite-derived SST-SSH synergies with 4DVarNets}

\author{ Ronan Fablet \\
	IMT Atlantique, Lab-STICC, INRIA team Odyssey\\
	Brest, France \\
	\texttt{ronan.fablet@imt-atlantique.fr} \\
	\And
	Bertrand Chapron \\
	Ifremer, LOPS, INRIA team Odyssey\\
	Brest, France \\
	\texttt{bertrand.chapron@ifremer.fr} \\
	\And
	Julien Le Sommer \\
	CNRS, IGE\\
	Grenoble, France \\
	\texttt{julien.lesommer@univ-grenoble-alpes.fr} \\
	\And
	Florian S\'{e}vellec \\
	CNRS, LOPS\\
	Brest, France \\
	\texttt{florian.sevellec@univ-brest.fr} \\
}

\hypersetup{
pdftitle={Inversion of sea surface currents from satellite-derived SST-SSH synergies with 4DVarNets},
pdfsubject={4DVarNet and Sea surface Dynamics},
pdfauthor={Fablet et al.},
pdfkeywords={sea surface velocities, satellite ocean remote sensing, deep learning, 4DVarNet, geostrophy, ageostrophy},
}

\begin{document}
\maketitle

\begin{abstract}
	Satellite altimetry is a unique way for direct observations of sea surface dynamics. This is however limited to the surface-constrained geostrophic component of sea surface velocities. Ageostrophic dynamics are however expected to be significant for horizontal scales below 100~km and time scale below 10~days. The assimilation of ocean general circulation models likely reveals only a fraction of this ageostrophic component. Here, we explore a learning-based scheme to better exploit the synergies between the observed sea surface tracers, especially sea surface height (SSH) and sea surface temperature (SST), to better inform sea surface currents. More specifically, we develop a 4DVarNet scheme which exploits a variational data assimilation formulation with trainable observations and {\em a priori} terms. An Observing System Simulation Experiment (OSSE) in a region of the Gulf Stream suggests that SST-SSH synergies could reveal sea surface velocities for time scales of 2.5-3.0 days and horizontal scales of 0.5$^\circ$-0.7$^\circ$, including a significant fraction of the ageostrophic dynamics ($\approx$ 47\%). The analysis of the contribution of different observation data, namely nadir along-track altimetry, wide-swath SWOT altimetry and SST data, emphasizes the role of SST features for the reconstruction at horizontal spatial scales ranging from \nicefrac{1}{20}$^\circ$ to \nicefrac{1}{4}$^\circ$.   
\end{abstract}


\section{Introduction}
\label{sec:intro}

Satellite altimeters provide the main source of observations to inform sea surface dynamics on a regional and global scale \citep{chelton_satellite_2001}. Their scarce space-time sampling of the sea surface prevents to recover spatial scales below 100~km and time scales below 10~days in general \citep{ballarotta_resolutions_2019}. Also altimetry can only reconstruct geostrophic velocities. As stressed by simulation and observational studies, the ageostrophic components are however critical features of upper ocean dynamics regarding for instance vertical mixing properties \citep{mahadevan_analysis_2006}, Lagrangian dynamics at sea surface \citep{baaklini_blending_2021,sun_impacts_2022}.

Retrieving sea surface currents at finer scales with both their geostrophic and ageostrophic components naturally arises as a key challenge. This has motivated a large research effort both in terms of simulation studies \citep{uchida_cloud-based_2022}, observational effort \citep{villas_boas_integrated_2019,ardhuin_skim_2019}, and data assimilation methods \citep{moore_synthesis_2019,storto_ocean_2019}. Regarding the latter aspect, state-of-the-art approaches mostly rely on the one hand on optimal interpolation approaches \citep{taburet_duacs_2019} and on the other hand on data assimilation schemes combined with ocean general circulation models \citep{baaklini_blending_2021,benkiran_assessing_2021,fujii_observing_2019}. As mentionned above, both approaches 
still show limitations in the ability to retrieve fine-scale patterns, whereas both observation-driven and theoretical studies evidence the interplay between fine-scale sea surface dynamics and some observed processes such as sea surface tracers \citep{ciani_ocean_2021,isern-fontanet_potential_2006} and drifters' trajectories \citep{sun_impacts_2022}.

From a methodological point of view, data-driven and learning-based schemes have also received a growing attention to solve inverse problems in geoscience \citep{alvera-azcarate_multivariate_2007,lguensat_analog_2017,barth_dincae_2020}. Especially, deep learning schemes appears as appealing schemes for the reconstruction of sea surface dynamics from irregularly-sampled satellite-derived observations \citep{fablet_end--end_2021,fablet_multimodal_2022,george_deep_2021,manucharyan_deep_2021}. Interestingly, these studies open 
new research avenues to make the most of available simulation and observation datasets. They also suggest potential breakthrough through the ability to exploit the synergies between different sea surface observational fields with no explicitly-known relationship \citep{fablet_multimodal_2022}.  

In this study, we exploit these recent methodological advances to explore satellite-derived SST-SSH synergies to inform sea surface currents, including their ageostrophic component. We exploit and adapt multi-modal 4DVarNet schemes introduced in \citep{fablet_multimodal_2022}. Through an observing system simulation experiment for a region of the Gulf Stream, our key contributions are four-fold:
\begin{enumerate}
\item We stress the potential of physics-informed deep learning schemes to enhance the reconstruction of sea surface currents (SSC) with a relative improvement greater than 50\% in terms of resolved space-time scales and mean-square-error metrics compared with the state-of-the-art products;

\item Our results also support wide-swath SWOT altimetry data to improve the reconstruction of SSC fields with a potential gain of $\approx 30\%$ for most performance metrics, though the reported improvement is marginal for the divergent component of the SSC.

\item We emphasize the contribution of SST-SSH synergies to retrieve a significant fraction of the ageostrophic component of the SSC, typically $\approx$ 47\% in terms of divergence of the SSC fields, with a major contribution of SST features in the spatial scale range of \nicefrac{1}{20}$^\circ$-\nicefrac{1}{4}$^\circ$.


\item We point out that, as hypothesized from theoretical considerations \citep{Sevellec:22}, the strain of sea surface dynamics partially explains ($\approx$ 60\%) the time-averaged  mean square error of the SSC.
\end{enumerate}
We further discuss the implications of these results in the context of ongoing research efforts towards the monitoring of upper ocean dynamics. 

\section{Problem statement}
\label{sec:statement}

{\bf Geostrophic and ageostrophic sea surface dynamics:} 
The horizontal momentum equations together with the hydrostatic equilibrium for the vertical momentum and the non-divergence leads to the following set of equations to describe sea surface dynamics:
\begin{subequations}
  \begin{equation}
    D_tu-fv=-\frac{1}{\rho_0}\partial_xP+\mathcal{F}_x,
  \end{equation}
  \begin{equation}
    D_tv+fu=-\frac{1}{\rho_0}\partial_yP+\mathcal{F}_y,
  \end{equation}
  \begin{equation}
    0=-\frac{1}{\rho_0}\partial_zP-\frac{g}{\rho_0}\rho,
  \end{equation}
   \begin{equation}
    \partial_xu+\partial_yv+\partial_zw=0,
    \label{eq:nondiv}
  \end{equation}
  \label{eq:momentum}
\end{subequations}
where $u$, $v$ and $w$ are the zonal, meridional, and vertical velocities, respectively, $t$ is time, $x$, $y$ and $z$ are the longitude, latitude, and depth, respectively, $\rho_{\left(0\right)}$ is the (reference) density for seawater, $P$ is the pressure, $f$ is the Coriolis  parameter, $g$ is the acceleration due to gravity, $\mathcal{F}_x$ and $\mathcal{F}_x$ are the action of the zonal and meridional viscous forces, respectively, and $D_t$ (=$\partial_t$+$u\partial_x$+$v\partial_y$+$w\partial_z$) is the material derivative.

These equations are often simplified to reflect the geostrophic balance which occurs under the small Rossby number (Ro$\ll$1 i.e., inertial terms are negligible), slow dynamics ($\partial_t \rightarrow$0 for the horizontal momemtum equation), large Péclet number (Pe$\ll$1 i.e., viscous terms are negligible), and away from direct forcing. This reads:
\begin{subequations}
  \begin{equation}
    -fv_g=-\frac{1}{\rho_0}\partial_xP,
  \end{equation}
  \begin{equation}
    +fu_g=-\frac{1}{\rho_0}\partial_yP,
  \end{equation}
  \label{eq:geo}
\end{subequations}
where $u_g$ and $v_g$ are the zonal and meridional geostrophic velocities, respectively. This formulation still implies an horizontal divergence, which, using (\ref{eq:geo}) with (\ref{eq:nondiv}), reads:
\begin{equation}
    \partial_xu_g+\partial_yv_g=-\frac{\beta}{f}v_g,
    \label{eq:geohordiv}
\end{equation}
where $\beta$ (=$\partial_y\,f$) accounts for the meridional variation of Earth equivalent rotation rate. Hence, the horizontal momentum equations for the ageostrophic components become:
\begin{subequations}
  \begin{equation}
    D_tu-fv_a=\mathcal{F}_x,
  \end{equation}
  \begin{equation}
    D_tv+fu_a=\mathcal{F}_y,
  \end{equation}
  \label{eq:ageo}
\end{subequations}
where $u_a$ and $v_a$ are the zonal and meridional ageostrophic velocities, respectively. This last set of equations show the complexity of the ageostrophic components through the action of viscous terms and of the full inertial terms in (\ref{eq:ageo}), compared to the rather simple linear relationships between pressure gradients and geostrophic velocities in (\ref{eq:geo}). This is worth noting that boundary conditions, such as observed at the ocean surface, occured though the actions of the viscous terms ($\mathcal{F}_x$ and $\mathcal{F}_y$). This further suggests the key role of the ageostrophic terms when studying the surface velocities.

In the subsequent, for the sake of simplicity, we will refer to geostrophically-derived sea surface velocities from SSH fields from (\ref{eq:geo}) using $\mathrm{SSH}=\nicefrac{(P|_{z=0}-P_\mathrm{atm})}{g\rho_s}$ [where $P|_{z=0}$ is the pressure at the geoide ($z=0$), $P_\mathrm{atm}$ is the atmospheric pressure, and $\rho_s$ is the assumed-vertically-constant density of the surface layer] as SSH-derived sea surface currents.


{\bf Data assimilation for sea surface dynamics:} Classically, inverse problems in geoscience \citep{evensen_data_2009} are stated as data assimilation problems through some underlying state-space formulation
\begin{equation}
\label{eq: state space}
\left \{\begin{array}{ccl}
    \displaystyle \frac{\partial \mathbf{x}(t)}{\partial t} &=& {\cal{M}}\left (\mathbf{x}(t) \right )+ \eta(t)\\~\\
    \mathbf{y}_m(t) &=& {\cal{H}}_m\left ( \mathbf{x}(t) \right ) + \epsilon_m(t), \forall t ,m\\
\end{array}\right.
\end{equation}
where $\mathbf{x}$ is the space-time process to be reconstructed and $\mathbf{y}_m$ is an observation process which relates to state $\mathbf{x}$ through observation operator ${\cal{H}}_m$ for observation modality $k$. When dealing with irregularly-sampled observations, operator ${\cal{H}}_m$ accounts for sampling masks. Processes $\eta$ and $\epsilon$ refer to random processes to account for modeling uncertainties and observation noise, respectively. ${\cal{M}}$ refers to the dynamical prior on state $x$. Given this state-space formulation, the data assimilation problem for the reconstruction of  state $x$ given observation data  $\{y_k\}_k$ comes to the resolution of a minimization problem. Within a variational data assimilation framework, it generally writes as:
\begin{equation}
\label{eq: 4dVar}
\widehat{\mathbf{x}} = \arg \min_{\mathbf{x}} \sum_m \lambda_m \left \| \mathbf{y}_m - {\cal{H}}_m(\mathbf{x})  \right \|^2 + \gamma \left \| \mathbf{x} - \Phi_{\cal{M}}(\mathbf{x})  \right \|^2 
\end{equation}
with $\{\lambda_m\}$ and $\gamma$ Lagrangian multipliers, $\Phi_{\cal{M}}(\mathbf{x})$ the time-stepping operator to propagate one-step-ahead state $\mathbf{x}$ at time $t$ to time $t+\Delta t$ based on dynamical prior ${\cal{M}}$. In the above formulation, we consider a matrix form and drop the time variable such that the norms are evaluated as a sum over a given time interval $[0,T]$ according to time step $\Delta t$. Given dynamical prior ${\cal{M}}$ and observation operators $\{{\cal{H}}_m\}_m$, data assimilation methods provide different algorithms \citep{carrassi_data_2018,evensen_data_2009} to solve this minimization problem especially using adjoint-based gradient descent schemes and Kalman methods. Formulation (\ref{eq: 4dVar}) also relates to Optimal interpolation \citep{cressie_statistics_2015} when considering a single observation term with a masking operator operator and a  prior given by a Gaussian process. Under these hypotheses, one can derive the analytical solution of the resulting linear-quadratic variational cost.        

The state-of-the-art methods for the reconstruction of sea surface dynamics from satellite-derived observations rely on such data assimilation schemes. While the reconstruction of geostrophic sea surface dynamics may be stated as a space-time interpolation of satellite altimetry data \citep{taburet_duacs_2019}, the inference of the total sea surface currents from satellite-derived observations using the set of equations introduced above implies to reconstruct not only these velocities, but the whole ocean state. The assimilation of satellite altimetry and satellite-derived SST observations, possibly complemented by other data sources, in ocean general circulation models \citep{benkiran_assessing_2021,fujii_observing_2019} follows this strategy. An alternative approach relies on assimilation schemes with a simplified prior on sea surface dynamics, which only depend on sea surface velocities. Quasi-geostrophic (QG) dynamics are examples of such priors \citep{le_guillou_mapping_2020,ubelmann_dynamic_2014}. Similarly to the above-mentionned OI schemes, the latter only applies to geostrophic velocities and cannot recover ageostrophic components.  
  
Recently, a rich literature has emerged to bridge data assimilation and deep learning \citep{abdalla_altimetry_2021,barthelemy_super-resolution_2021,boudier_dan_2020,bocquet_bayesian_2020,fablet_learning_2021,nonnenmacher_deep_2021}. It provides new minimization schemes as well as new means to explore data assimilation problems when the observation operators and/or the dynamical priors are not explicitly known. As such, it opens new avenues to balance the complexity of the inversion problem and the genericity of the underlying variational formulation in (\ref{eq: 4dVar}). Here, as detailed in the subsequent, we benefit from the generic 4DVarNet framework introduced in \citep{fablet_learning_2021} and explore a learning-based data assimilation schemes for the reconstruction of sea surface currents from multimodal satellite-derived observations.  

\section{Data}
\label{sec:data}

\subsection{Case-study region and NATL60 data}

\begin{figure}[htb]
    \centering
    \begin{tabular}{C{3.cm}C{3.cm}C{3.cm}C{3.cm}}
    \includegraphics[trim={210 65 210 100},clip,width=3.cm]{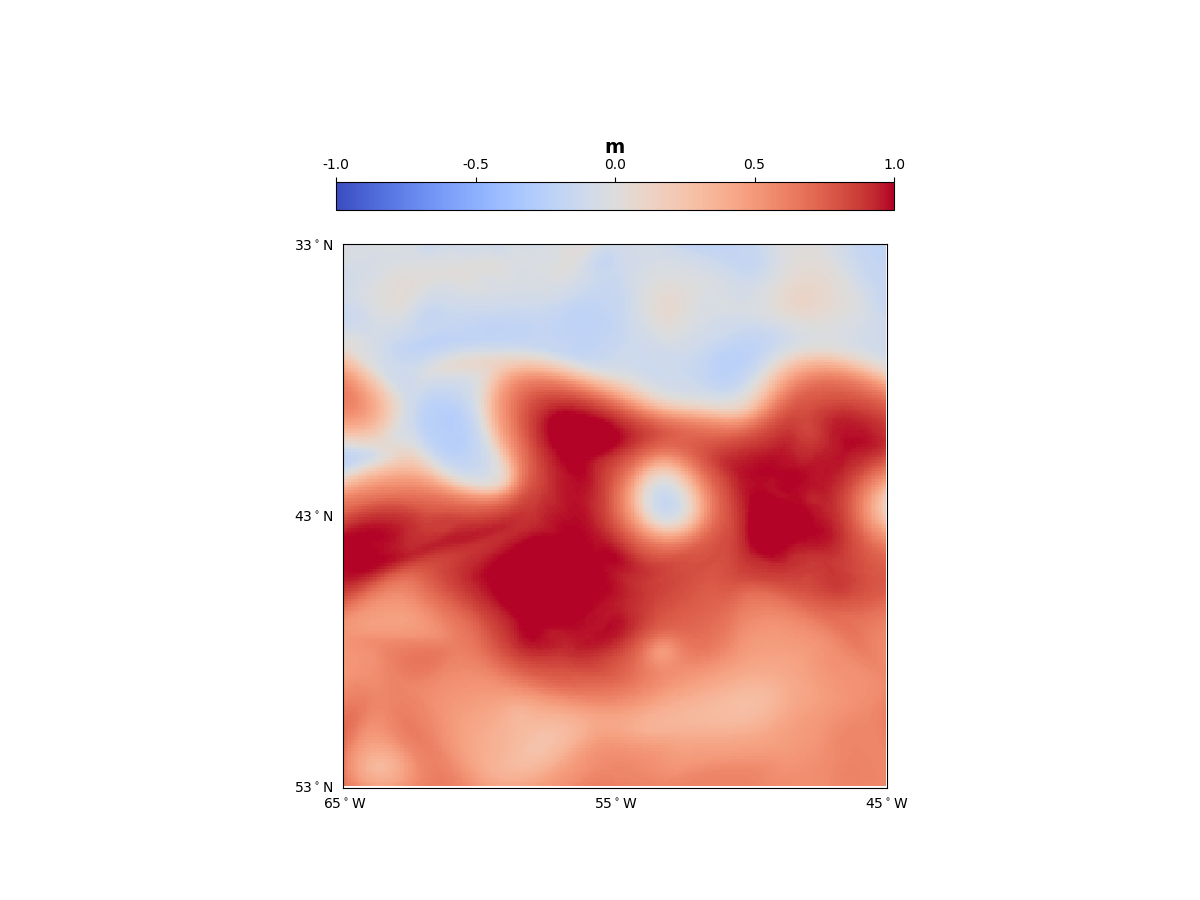}&
    \includegraphics[trim={210 65 210 100},clip,width=3.cm]{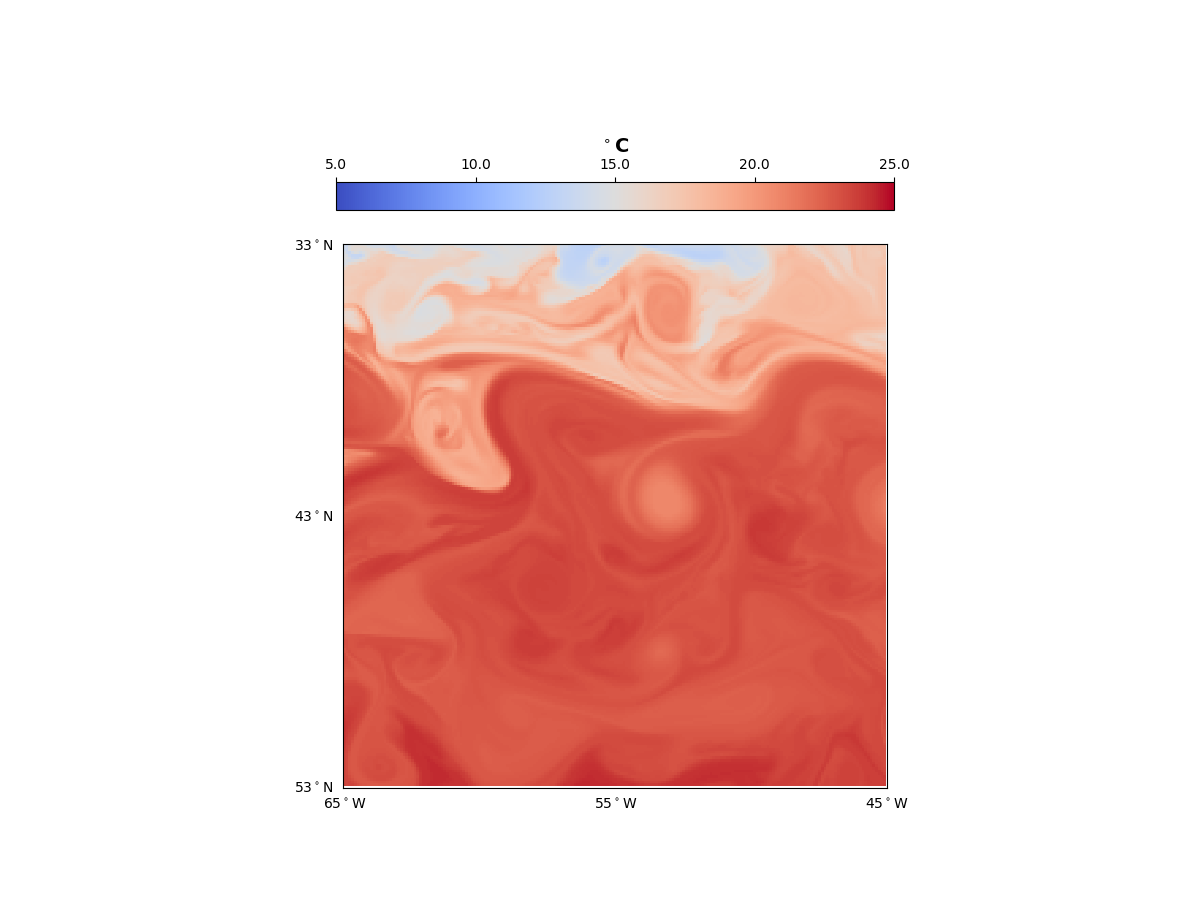}&
    \includegraphics[trim={210 65 210 100},clip,width=3.cm]{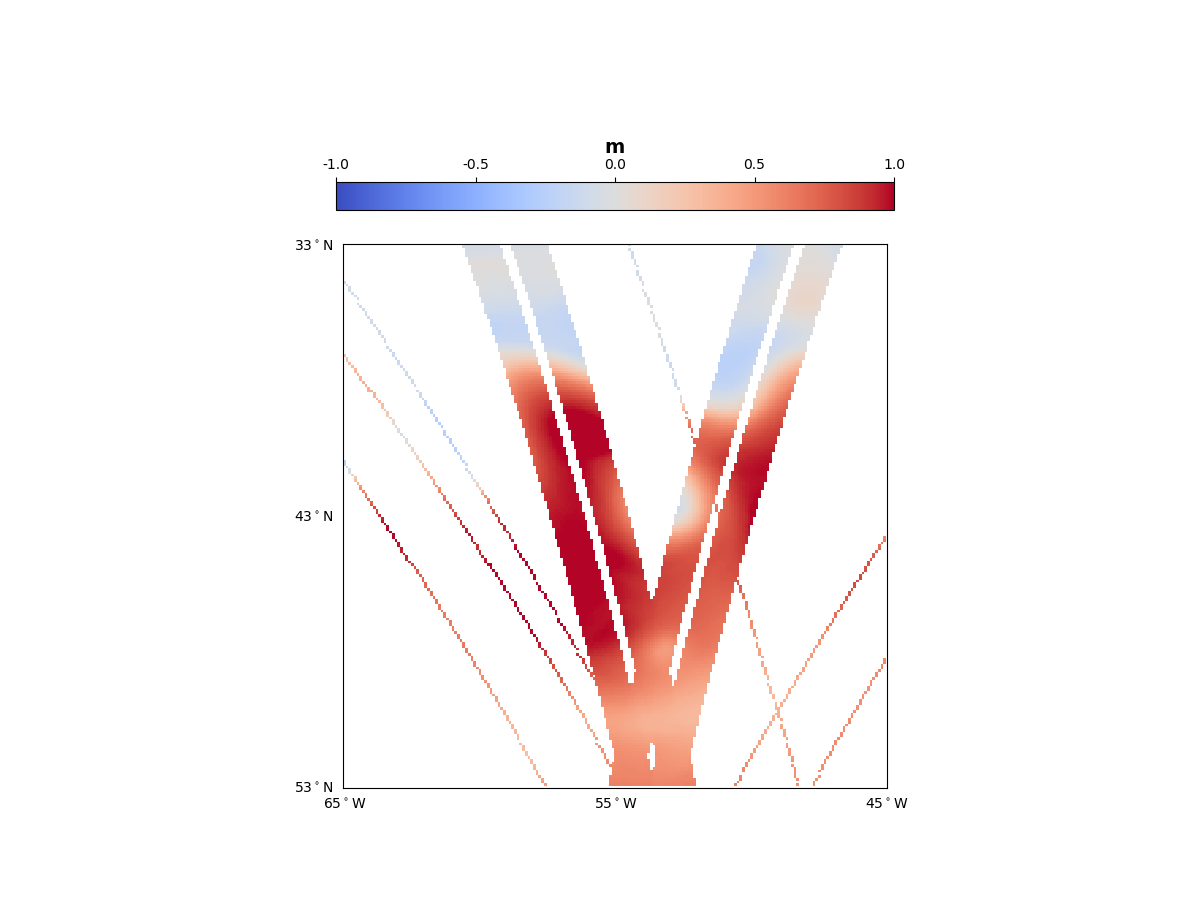}&
    \includegraphics[trim={210 65 210 100},clip,width=3.cm]{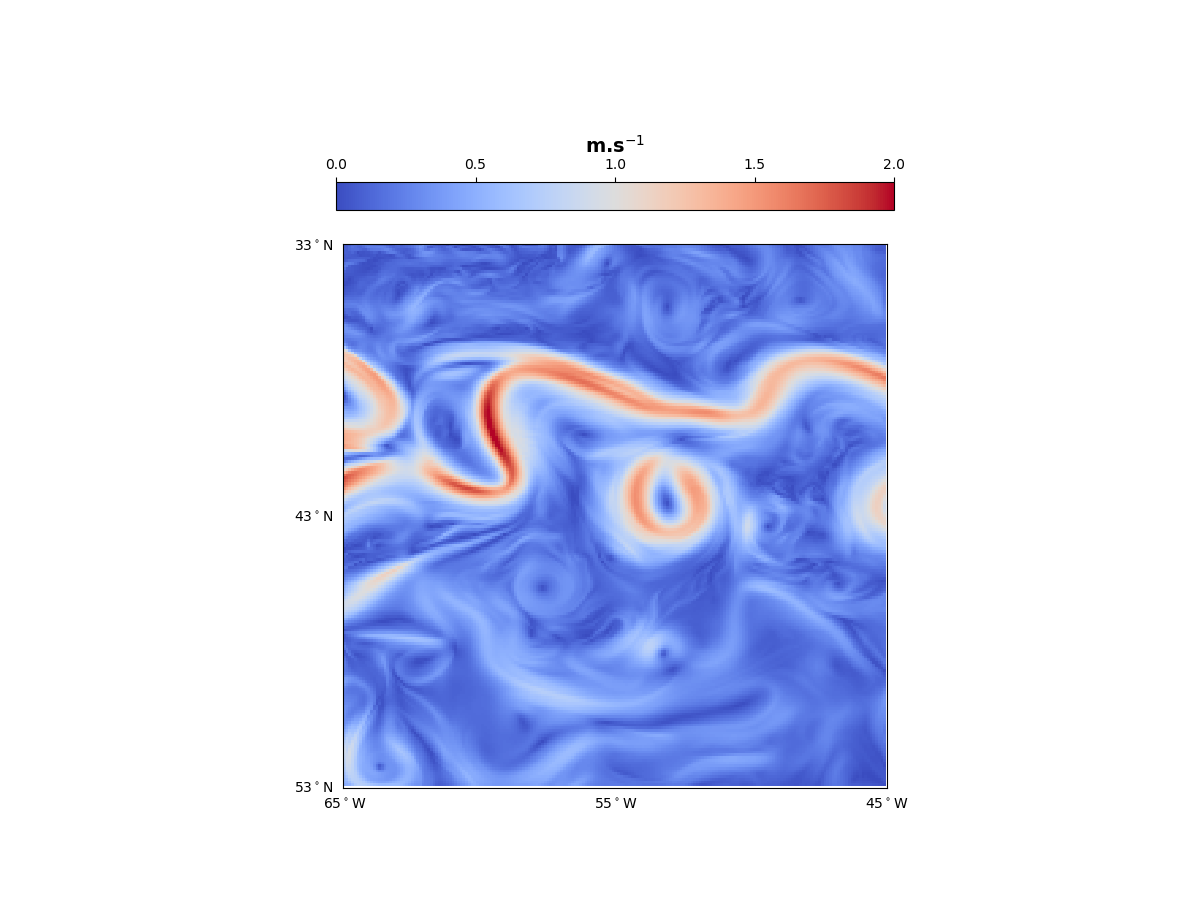}\\
    {\small \bf SSH}&{\small \bf SST}&{\small \bf Altimetry data}&{\small \bf SSC}\\~\\
    \includegraphics[trim={210 65 210 100},clip,width=3.cm]{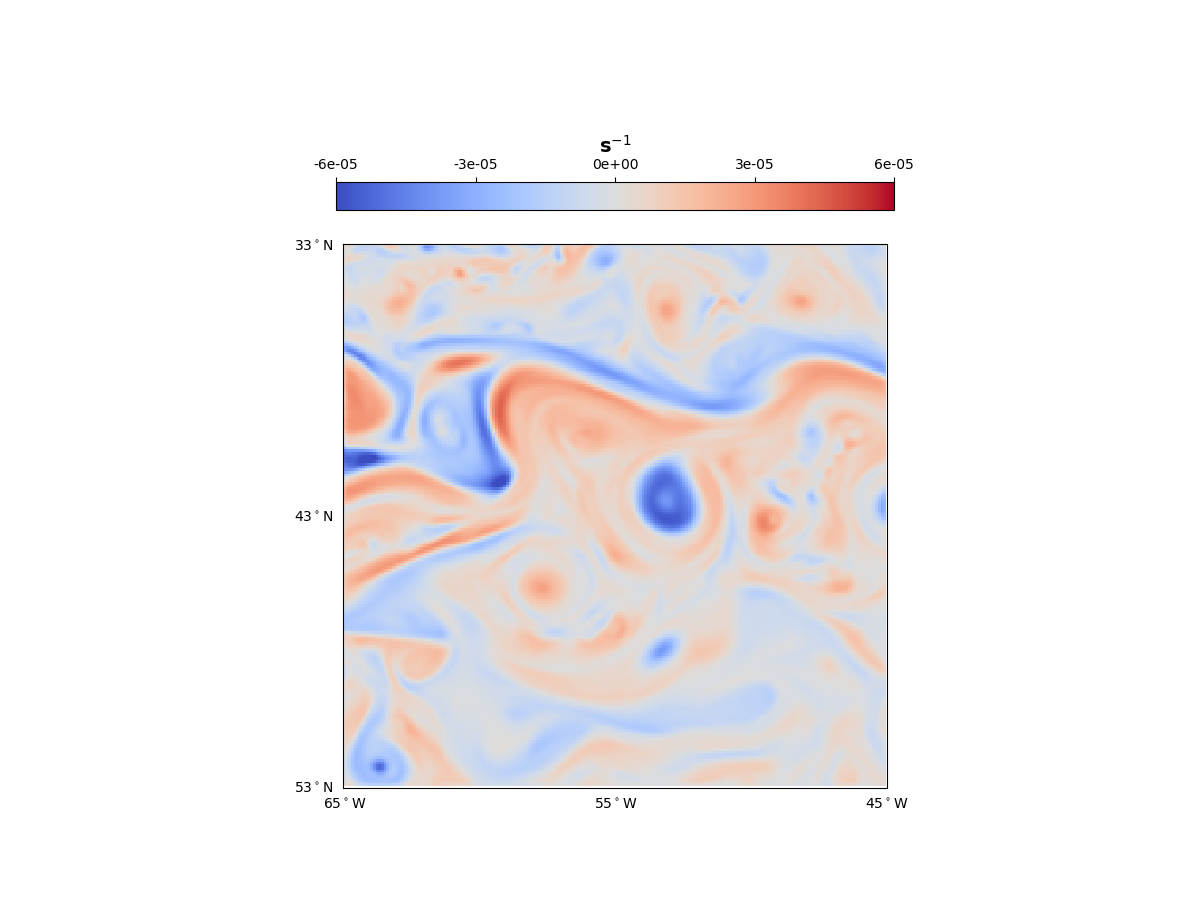}&    
    \includegraphics[trim={210 65 210 100},clip,width=3.cm]{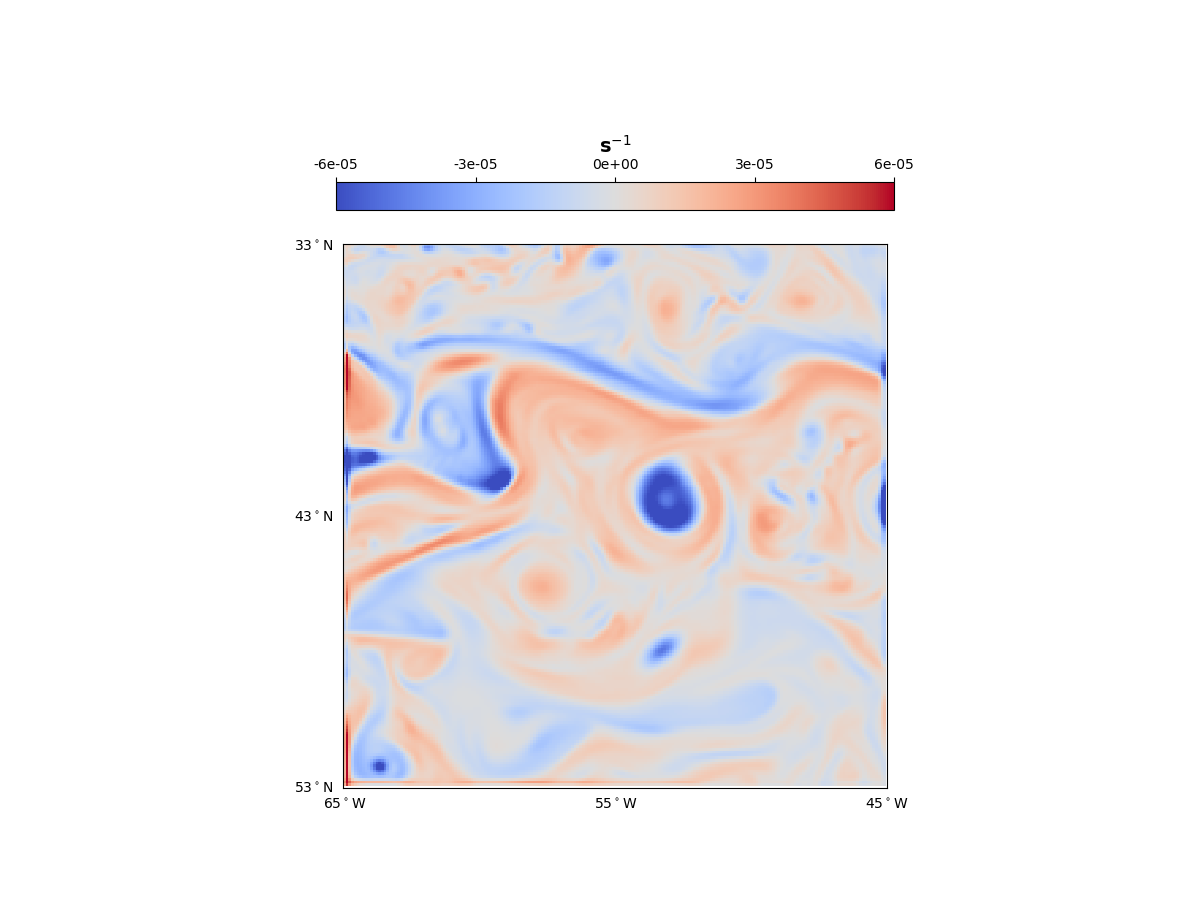}&
    \includegraphics[trim={210 65 210 100},clip,width=3.cm]{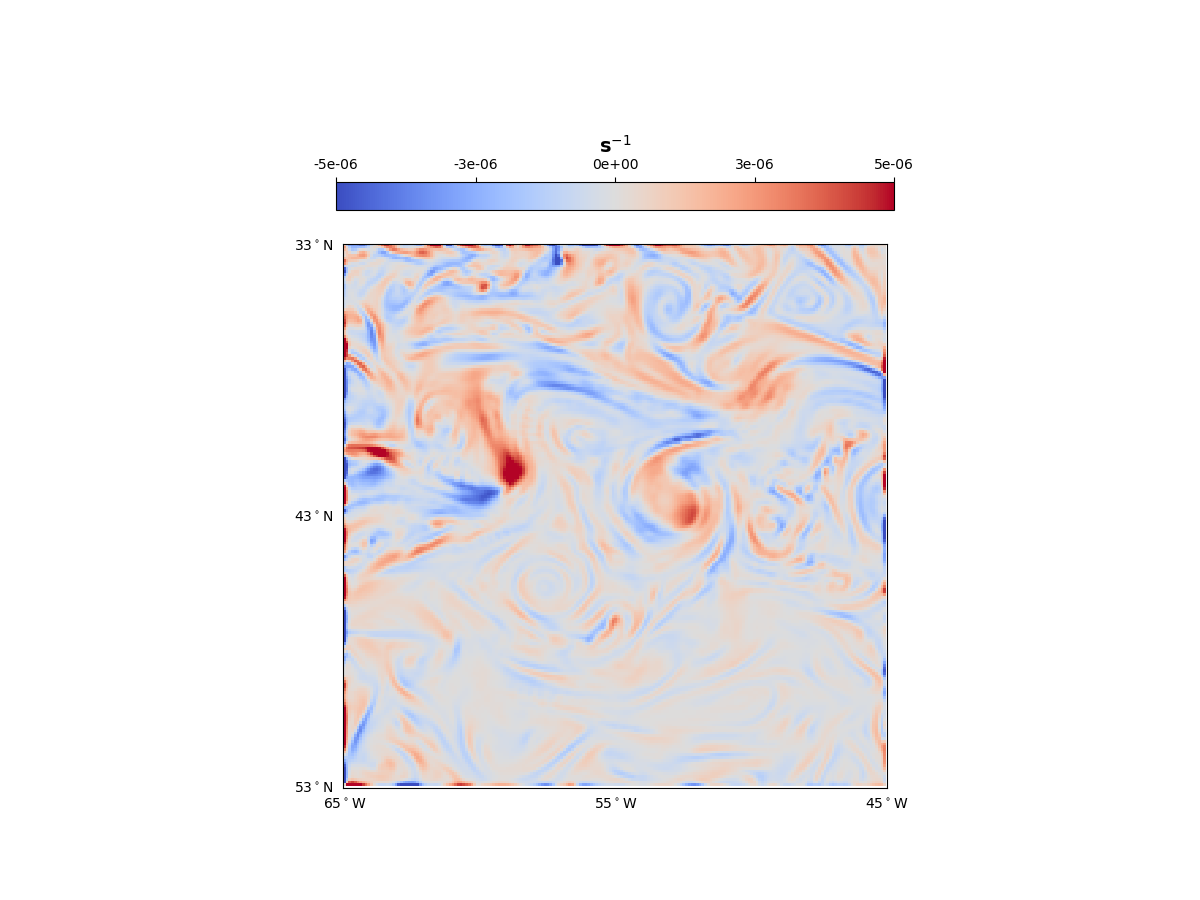}&
    \includegraphics[trim={210 65 210 100},clip,width=3.cm]{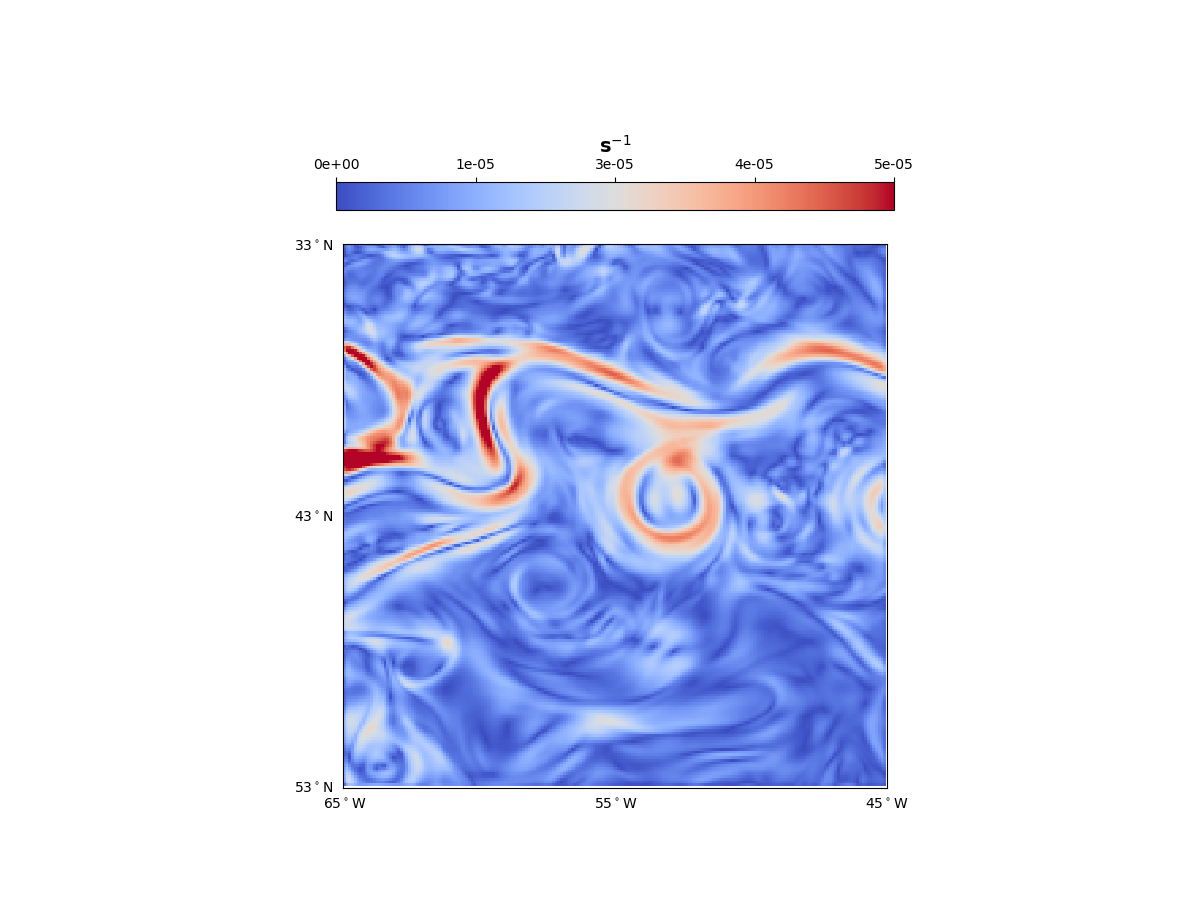}\\
    {\small \bf SSC vorticity}&{\small \bf SSH-derived vorticity}&{\small \bf SSC divergence}&{\small \bf SSC strain}\\
    \end{tabular}
    \caption{{\bf Illustration of the considered case-study using NATL60 simulation dataset:} first row, from left to right, NATL60 SSH field on November 12$^{th}$, associated SST field and SSH observation data from nadir and wide-swath SWOT altimeters, SSC field. The second row depicts the vorticity fields for the total currents and SSH-geostrophically-derived ones along with the divergence and strain of the total currents.}
    \label{fig:obs data}
\end{figure}

The considered case study focuses on a $10^\circ$$\times$$10^\circ$ region between (33$^\circ$N, 65$^\circ$W) and  (43$^\circ$N, 55$^\circ$W). As illustrated in Fig.~\ref{fig:obs data}, this region involves the main meander of the Gulf Stream as well as a variety of mesoscale eddies and finer-scale sub-mesoscale filaments. It also comprises clear divergent features associated with the ageostrophic flow component, which makes it suitable for the current study.

The considered simulation dataset relies on a nature run of the NATL60 configuration \citep{ajayi_spatial_2020} of the NEMO (Nucleus for European Modeling of the Ocean) model \citep{madec_nemo_2022}. This simulation delivers a realistic hindcast simulation of ocean dynamics, including mesoscale-to-submesoscale ocean dynamics, \citep{ajayi_spatial_2020} over one year from October 2012 to September 2013. It has been used in numerous studies regarding the reconstruction and observability of sea surface dynamics (see below for the related OSSE data challenge stated in \citep{le_guillou_mapping_2020}).   
 
\subsection{OSSE setting}
 
Our numerical experiments exploit an Observing System Simulation Experiment (OSSE). We rely on the OSSE setting \footnote{\url{https://github.com/ocean-data-challenges/2020a_SSH_mapping_NATL60}} proposed in \citep{le_guillou_mapping_2020} for the benchmarking of SSH mapping methods in the context of upcoming wide-swath altimetry mission SWOT \citep{gaultier_challenge_2015}. For a \nicefrac{1}{20}$^\circ$ spatial resolution, this OSSE dataset comprises daily-averaged SSH fields and simulated altimetry data. The latter combine simulated nadir along-track data according to real 4-altimeter configuration and SWOT altimetry data using SWOT simulator \citep{gaultier_challenge_2015}. We may point out that simulated altimetry data are created from hourly SSH fields.

In this study, we complement this initial OSSE dataset with two additional data sources with the same \nicefrac{1}{20}$^\circ$ spatial resolution: the series of daily-averaged sea surface currents (SSC) and the series of daily-averaged sea surface temperature (SST). We also generate SST observations with resolutions of \nicefrac{1}{10}$^\circ$, \nicefrac{1}{5}$^\circ$, \nicefrac{1}{4}$^\circ$ and \nicefrac{1}{2}$^\circ$ using coarsening and subsampling operations to provide a more realistic simulations of the diversity of operational L4 SST products \citep{donlon_operational_2012,ocarroll_observational_2019}. We report in Fig.~\ref{fig:obs data} an illustration of the considered dataset.

For the training configuration, we split the one-year time series into training, validation, and test datasets according to the following time periods: from  February 4$^\mathrm{th}$ 2013 to  September 30$^\mathrm{th}$ 2013, from January 1$^\mathrm{st}$  2013 to  February 4$^\mathrm{th}$ 2013, and from October 20$^\mathrm{th}$ 2012 to December 4$^\mathrm{th}$ 2012, respectively.

\section{Methods}
\label{sec:methods}

This Section presents the proposed 4DVarNet scheme for the reconstruction of sea surface currents from satellite-derived SST-SSH synergies. We first detail the considered 4DVarNet architecture before describing our learning scheme.

\subsection{Proposed 4DVarNet scheme}

\begin{figure}[htb]
    \centering
    \includegraphics[width=14.5cm]{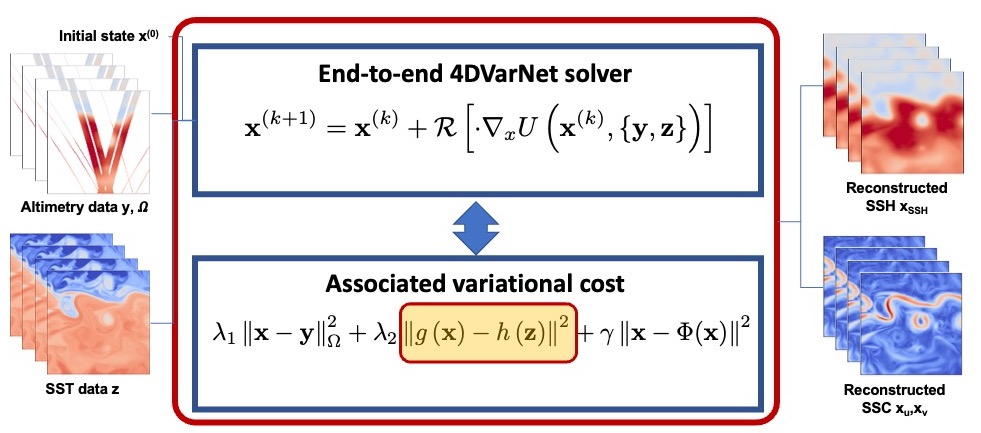}
    \caption{{\bf Sketch of 4DVarNet scheme for the reconstruction of sea surface current fields}: 4DVarNet provides an end-to-end neural architecture to map multimodal observation data, here satellite-derived SSH and SST observations, to SSH and SSC fields. It relies on a variational data assimilation and on the resolution of the associated minimization issue with a trainable gradient-based descent. We refer the reader to \citep{fablet_learning_2021} for a detailed presentation of 4DVarNet scheme.}
    \label{fig:4dvarnet scheme}
\end{figure}

4DVarNet schemes refer to end-to-end neural architectures introduced in \citep{fablet_learning_2021} to solve data assimilation problems and further extended to multimodal inversion schemes in \citep{fablet_multimodal_2022}. As sketched in Fig.~\ref{fig:4dvarnet scheme}, the proposed 4DVarNet scheme exploits as inputs time series of satellite altimetry data and SST fields and outputs gap-free SSH fields and SSC fields. Importantly, it combines two main components: the definition of the underlying variational formulation (\ref{eq:4dvar model}), the iterative update rule of the associated trainable gradient-based solver. The latter relies at each iteration on the gradient of the variational cost w.r.t. the state using automatic differentiation tools embedded in deep learning frameworks such as PyTorch. Overall, the 4DVarNet scheme implements a predefined number, typically from 5 to 15, of the iterative update rule to map the input partial observations to the reconstructed sea surface dynamics. We refer the reader to \citep{fablet_learning_2021} for a detailed presentation of the 4DVarNet framework along with its link to variational data assimilation formulations.  

Let us formally introduce the considered variational formulation. Let $\mathbf{x}$ denote hereafter the space-time state to be reconstructed, $\mathbf{y}$ the observed altimetry data and $\mathbf{z}$ the SST observations. The considered variational cost, denoted by $U_\Phi\left ( \mathbf{x} , \mathbf{y} , \mathbf{z} \right )$, is given by:
\begin{equation}
\label{eq:4dvar model}
\displaystyle U_\Phi\left ( \mathbf{x} , \mathbf{y} ,\mathbf{z} \right ) = \displaystyle \lambda_{1} \left \|  \mathbf{x} - \mathbf{y} \right \|^2_\Omega + \displaystyle \lambda_{2} \left  \|g\left ( \mathbf{x} \right) - h\left ( \mathbf{z} \right) \right \|^2 + \displaystyle\gamma \left \|\mathbf{x} - \Phi(\mathbf{x}) \right \|^2,
\end{equation}
where $\lambda_1$, $\lambda_2$, and $\gamma$ are Lagrangian multipliers. State $\mathbf{x}$ combines SSH denoted by $\mathbf{x}_\mathrm{SSH}$ and the meridional and zonal SSC components denoted by $\mathbf{x}_\mathrm{u}$ and $\mathbf{x}_\mathrm{v}$, such that 

and SSC components referred to as $\mathbf{x}_\mathrm{SSH}$ and $\mathbf{x}_\mathrm{SSC}$ such that $\mathbf{x}=(\mathbf{x}_\mathrm{SSH},\mathbf{x}_\mathrm{u},\mathbf{x}_\mathrm{v})$. Besides, as in \citep{fablet_multimodal_2022}, we adopt a two-scale decomposition of the SSH fields to explicitly use both raw altimetry data and DUACS coarse-scale interpolation as observation data\footnote{We refer the reader to \citep{fablet_multimodal_2022}  for a detailed description of this two-scale decomposition of the SSH component of the state $\mathbf{x}$.}.
$\Omega$ refers to a masking operator to account for observation gaps in the altimetry data as well as for the fact that the SSC component is never directly observed. 

Operators $g()$ and $h()$ are trainable operators which aim at extracting relevant features from SST observations $z$ to match features extracted from state $x$. Following \citep{fablet_multimodal_2022}, we parameterize $g()$ and $h()$ as non-linear convolutional networks using four convolutional layers and tanh activations. In our experiments, we exploit a 20-dimensional feature space for multimodal term $\left \|g\left ( \mathbf{x} \right) - h\left ( \mathbf{z} \right) \right \|^2$.

Operator $\Phi$ states the considered prior onto the space-time dynamics of state $\mathbf{x}$. Within a classic model-driven approach, $\Phi$ would refer to the time-stepping operator to forecast future states from an initial condition. Following \citep{fablet_learning_2021,fablet_multimodal_2022}, we rather consider a purely data-driven parameterization with a two-scale U-Net architectures \citep{cicek_3d_2016}. The latter proved to be more relevant in previous numerical experiments for both Lorenz's systems \citep{fablet_learning_2021} and SSH mapping \citep{beauchamp_end--end_2022,fablet_multimodal_2022}.

As sketched in Fig.~\ref{fig:4dvarnet scheme}, the considered 4DVarNet schemes implement the following iterative gradient-based solver to minimize variational cost (\ref{eq:4dvar model}):
\begin{equation}
\label{eq: lstm update}
\left \{\begin{array}{ccl}
     \mathbf{h}(k+1) , \mathbf{c}(k+1) &=&   {\cal{R}} \left[ \cdot \nabla_x U \left ( \mathbf{x}^{(k)},\{\mathbf{y},\mathbf{z}\} \right),  \mathbf{h}(k) , \mathbf{c}(k) \right ]  \\~\\
     \mathbf{x}^{(k+1)} &=& \mathbf{x}^{(k)} - {\cal{L}}  \left( \mathbf{h}^{(k+1)} \right )  \\
\end{array} \right.\hspace*{-0.2cm}
\end{equation}
with ${\cal{R}}$ a LSTM cell, ${\cal{L}}$ a linear operator, $(\mathbf{h}(k) , \mathbf{c}(k))$ the internal state of the LSTM cell at iteration $k$ and $\mathbf{x}^{(k)}$ the reconstructed state at iteration $k$. As mentioned above, we consider 4DVarNet schemes with 5 to 15 iterations. The LSTM cell is a 2D-convolutional LSTM cell with 150-dimensional internal states. The considered iterative rule can be regarded as a momentum-based gradient-based descent which has been widely explored for optimizer learning problems \citep{hospedales_meta-learning_2020}.

Overall, the trainable components of the 4DVarNet scheme comprise: operators $g$ and $h$, prior $\Phi$, LSTM cell ${\cal{R}}$, and linear mapping ${\cal{L}}$. This amounts to a total number of parameters to be trained from data of about 1.4~M parameters for the considered case-study. 

\subsection{Learning setting}

We benefit from the end-to-end feature of the considered 4DVarNet scheme to run a supervised strategy. It learns all trainable components with a view to optimizing the reconstruction performance. The training loss then combines reconstruction losses for SSH and SSC fields accounting respectively for the SSH and its gradient, the zonal and meridional components of current state $\mathbf{x}_\mathrm{SSC}=(\mathbf{x}_u,\mathbf{x}_v)$, and the divergence of the SSC fields:
\begin{equation}
{\cal{L}}_{\nabla \mathrm{SSH}} = \sum_i \alpha_\mathrm{SSH} \left \| \mathbf{x}^\mathrm{true}_{\mathrm{SSH},i} - \widehat{\mathbf{x}}_{\mathrm{SSH},i} \right \|^2 + \alpha_{\nabla\mathrm{SSH}} \left \| \nabla \mathbf{x}^\mathrm{true}_{\mathrm{SSH},i} - \nabla\widehat{\mathbf{x}}_{\mathrm{SSH},i} \right \|^2
\end{equation}
\begin{equation}
{\cal{L}}_{u,v} = \sum_i \left \| \mathbf{x}^\mathrm{true}_{u,i} - \widehat{\mathbf{x}}_{u,i} \right \|^2 + \left \| \mathbf{\mathbf{x}}^\mathrm{true}_{v,i} - \widehat{\mathbf{x}}_{v,i} \right \|^2
\end{equation}
\begin{equation}
{\cal{L}}_\mathrm{div} = \sum_i \left \| \partial_x \mathbf{x}^\mathrm{true}_{u,i} + \partial_y \mathbf{x}^\mathrm{true}_{v,i} - \partial_x \widehat{\mathbf{\mathbf{x}}}_{u,i} -  \partial_y\widehat{\mathbf{x}}_{v,i} \right \|^2
\end{equation}
where superscript $^\mathrm{true}$ (resp. $\widehat{\ }$) refers to the true (resp. reconstructed) SSH or SSC fields. As proposed in \citep{fablet_learning_2021}, we complement these training losses with additional regularisation terms
\begin{equation}
{\cal{L}}_{\Phi} =  \sum_i \left \| \mathbf{x}^\mathrm{true}_{i} - \Phi \left ( \mathbf{x}^\mathrm{true}_i  \right ) \right \|^2+  \sum_i \left \| \widehat{\mathbf{x}}_{i}- \Phi \left ( \widehat{\mathbf{x}}_{i} \right ) \right \|^2
\end{equation}
These regularisation terms better constrain the training of prior $\Phi$.

Overall, we implement the 4DVarNet scheme and the associated learning strategy within Pytorch framework. We use Adam as optimizer with a fixed learning rate of $10^{-3}$. After a first training procedure over 200 epochs, we fine-tune the best model over the validation dataset over 200 more epochs. The Pytorch code of our implementation is open source \citep{febvre_pytorch_2022}.

\subsection{Evaluation framework and benchmarked approaches}

Our numerical experiments involve a quantitative evaluation of the reconstruction performance using metrics evaluated over the test dataset. We adapt the  metrics introduced in \citep{le_guillou_mapping_2020} for mapping SSH fields  and geostrophic SSC fields \footnote{We refer the reader to the following link for the detailed presentation of the evaluation experiment and benchmarked approaches \url{https://github.com/ocean-data-challenges/2020a_SSH_mapping_NATL60}}, especially: 
\begin{itemize}
    \item $\lambda_{t,u}$ and $\lambda_{t,v}$, the minimum time scale resolved in days for respectively the zonal and meridional velocities; 
    \item $\lambda_{x,u}$ and $\lambda_{x,v}$, the minimum spatial scale resolved in degrees for respectively the zonal and meridional velocities.
\end{itemize}
As described in \citep{le_guillou_mapping_2020}, these metrics rely on a  spectral analysis. We also evaluate the reconstruction performance in terms of explained variance, respectively: 
\begin{itemize}
    \item $\tau_{u,v}$, the explained variance for the reconstructed SSC;
    \item $\tau_\mathrm{vort}$, the explained variance for the vorticity of the reconstructed SSC;
    \item $\tau_\mathrm{div}$, the explained variance for the divergence of the reconstructed SSC;
     \item $\tau_\mathrm{strain}$, the explained variance for the strain of the reconstructed SSC \citep{balwada_vertical_2021,okubo_horizontal_1970,weiss_dynamics_1991}.
\end{itemize}
The last three metrics characterize the extent to which the reconstructed SSC capture the local deformation tensor of the true velocity fields. Numerically, the computation of the vorticity, divergence, and strain combines a Gaussian filtering and finite difference approximation of the first-order derivatives.   

Our numerical experiments evaluate two configurations of the proposed 4DVarNet framework: one using only SSH observation data and the other one exploiting SST-SSH synergies. For benchmarking purposes, we perform a quantitative comparison with respect to the SSH-derived sea surface currents from the true SSH fields and optimally-interpolated DUACS ones \citep{taburet_duacs_2019} using the geostrophic approximation introduced in Section \ref{sec:statement}. We also evaluate direct learning-based inversion schemes based on U-Net architectures \citep{cicek_3d_2016} to directly map observation data to SSC fields either considering only SSH observation data  or jointly SSH and SST observation data.  

\section{Results}
\label{sec:res}

In this Section, we report the considered numerical experiments for the evaluation of the proposed 4DVarNet schemes for the reconstruction of sea surface currents.   

\begin{table*}[tb]
    \footnotesize
    \centering
    \begin{tabular}{|C{1.7cm}|C{1.5cm}|C{1.cm}|C{1.cm}|C{1.cm}|C{1.cm}|C{1.cm}|C{1.cm}|C{1.cm}|C{1.cm}|}
    \toprule
    \toprule
     \bf Approach & \bf Data used&\bf $\lambda_{x,u}$ ($^\circ$)&\bf $\lambda_{x,v}$ ($^\circ$)& $\lambda_{t,u}$ (d)& $\lambda_{t,v}$ (d)&$\tau_{u,v}$&$\tau_\mathrm{vort}$&$\tau_\mathrm{div}$&$\tau_\mathrm{strain}$\\
    \toprule
    \toprule
     True SSH  
     & SSH only & \it 0.36 & \it 0.17 &  \it 19.6 &  \it 11.2 & \it 97.0\% &\it  96.3\%& \it $-$1.0\% & \it 92.1\% \\    
    \bottomrule
    \toprule

     DUACS 
     & SSH only & 1.72 &  1.24 &  12.4 &  11.6 & 83.7\% & 53.5\%& $-$0.5\% & 24.8\% \\    
    \toprule
      U-Net & SSH only  & 1.39 & 1.22 & 9.1 & 10.3 & 89.1\% & 72.3\% & $-$3.0\%& 65.0\%  \\ 
      & SSH-SST & 1.33 & 0.90 & 4.0 & 4.2 & 92.6\% & 79.4\% & 19.5\% & 72.0\%  \\ 
    \toprule
     4DVarNet & SSH only & 0.9 & 0.7 & 4.3 & 5.6 & 94.0\% & 86.1\% & 12.1\% & 81.3\%\\ 
     (ours) & SSH-SST  &  \bf 0.76& \bf  0.61& \bf 2.7 & \bf 2.5 & \bf  97.4\% & \bf  92.1\% & \bf 46.9\% & \bf 87.2\%\\ 
    \bottomrule
    \bottomrule
    \end{tabular}
    \caption{{\bf Synthesis of the reconstruction performance of the benchmarked approaches:} we report the performance metrics of the benchmarked approaches for the reconstruction of image time series of sea surface currents from satellite data. We refer the reader to the main text for the description of the different metrics. We highlight in bold the best score.} 
    \label{tab:res all}
\end{table*}

{\bf Synthesis of the benchmarking experiments:} Tab.~\ref{tab:res all} compares the  performance of the benchmarked schemes. DUACS SSH-derived geostrophic currents provide a baseline performance with reconstruction score in line with those reported for the metrics considered for SSH mapping \citep{fablet_multimodal_2022,le_guillou_mapping_2020} with resolved space-time scales above 1$^\circ$ and 10~days. This baseline accounts for more than 80\% of the variance of the total current and about 50\% of its vorticity and 25\% of its strain. The two 4DVarNet schemes clearly outperform this reconstruction performance with resolved space-time scales below 1$^\circ$ and 7~days. The relative improvement is particularly significant when exploiting SSH-SST synergies, especially for resolved time scales below 3~days. Interestingly, with this 4DVarNet configuration, the metrics for the explained variance of the different fields are better than those of the SSH-derived currents using the true SSH. In this respect, the poorly-resolved time scales of the latter are indicators of the relative contribution of the ageostrophic component of the total currents. Both 4DVarNet configurations learn to correct geostrophic velocities to better recover these temporal features. Though the configuration using only altimetry data only recover on average 12\% of the divergent component of the SSC, the SSH-SST synergies captured by the 4DVarNet through the multimodal observation term in variational formulation (\ref{eq:4dvar model}) greatly improve the ability to retrieve a significant fraction of this divergent component (here, $\approx 47\%$ on average). The reported relative improvement greater than 30\% for most of the performance metrics further highlights their contribution. 
This benchmarking experiment also clearly supports the relevance of 4DVarNet schemes compared with direct learning-based inversion schemes using off-the-shelf deep learning architecture. For instance, we report for all metrics a better reconstruction performance of the 4DVarNet scheme using only altimetry data than that of a direct UNet-based inversion with altimetry and SST inputs.  

\begin{figure}[htb]
    \centering
    {\footnotesize
    \begin{tabular}{C{3.25cm}C{3.25cm}C{3.25cm}C{3.25cm}}
    \multicolumn{4}{c}{\bf True SSC}\\
    \includegraphics[trim={210 65 210 100},clip,width=3.25cm]{./Figure/Fig_4dVarNet-UVwSST/norm_uv_gt_d25.png}&
    \includegraphics[trim={210 65 210 100},clip,width=3.25cm]{./Figure/Fig_4dVarNet-UVwSST/curl_uv_gt_d25.png}&
    \includegraphics[trim={210 65 210 100},clip,width=3.25cm]{./Figure/Fig_4dVarNet-UVwSST/strain_uv_gt_d25.png}&
    \includegraphics[trim={210 65 210 100},clip,width=3.25cm]{./Figure/Fig_4dVarNet-UVwSST/div_uv_gt_d25.png}\\
    \multicolumn{4}{c}{\bf True SSH-derived SSC}\\
    \includegraphics[trim={210 65 210 155},clip,width=3.25cm]{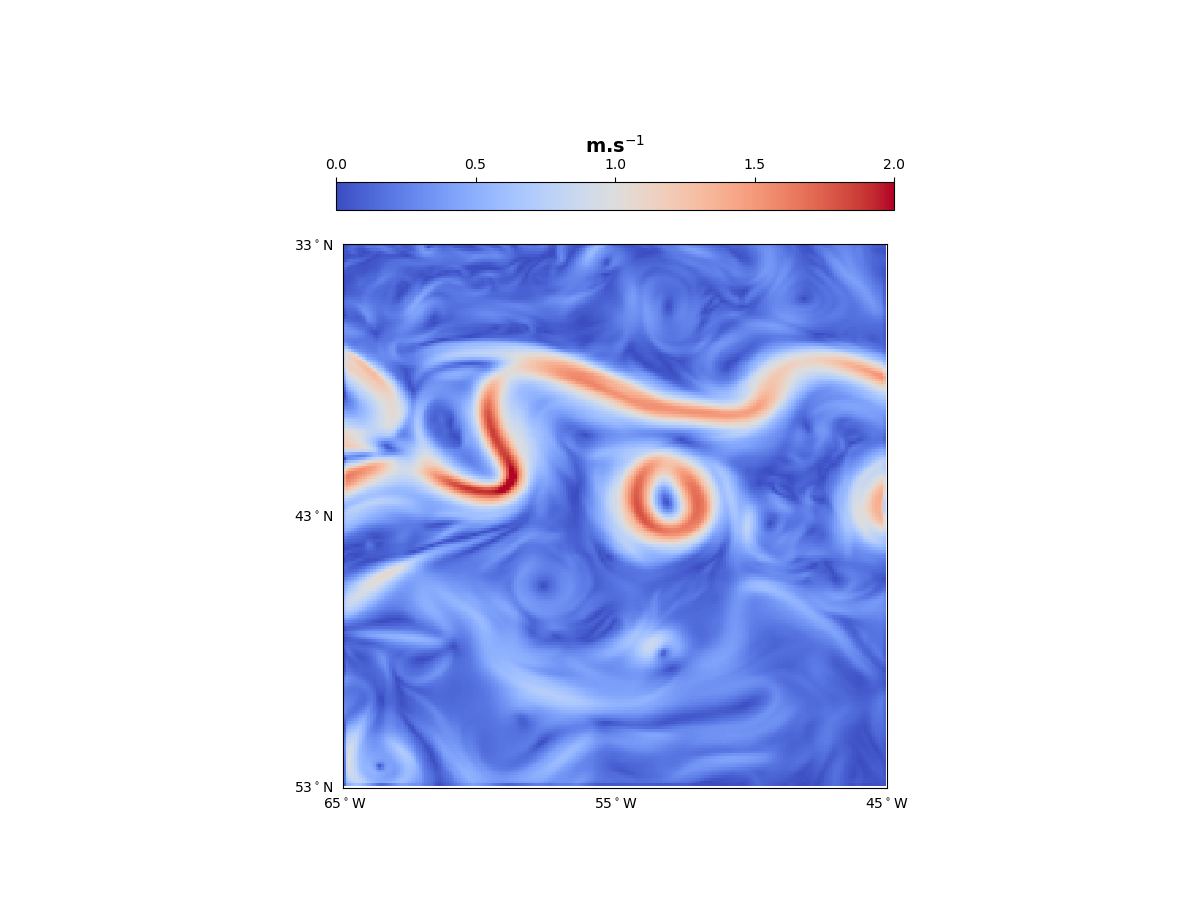}&
    \includegraphics[trim={210 65 210 155},clip,width=3.25cm]{./Figure/Fig_4dVarNet-UVwSST/curl_uv_ssh-gt_d25.png}&
    \includegraphics[trim={210 65 210 155},clip,width=3.25cm]{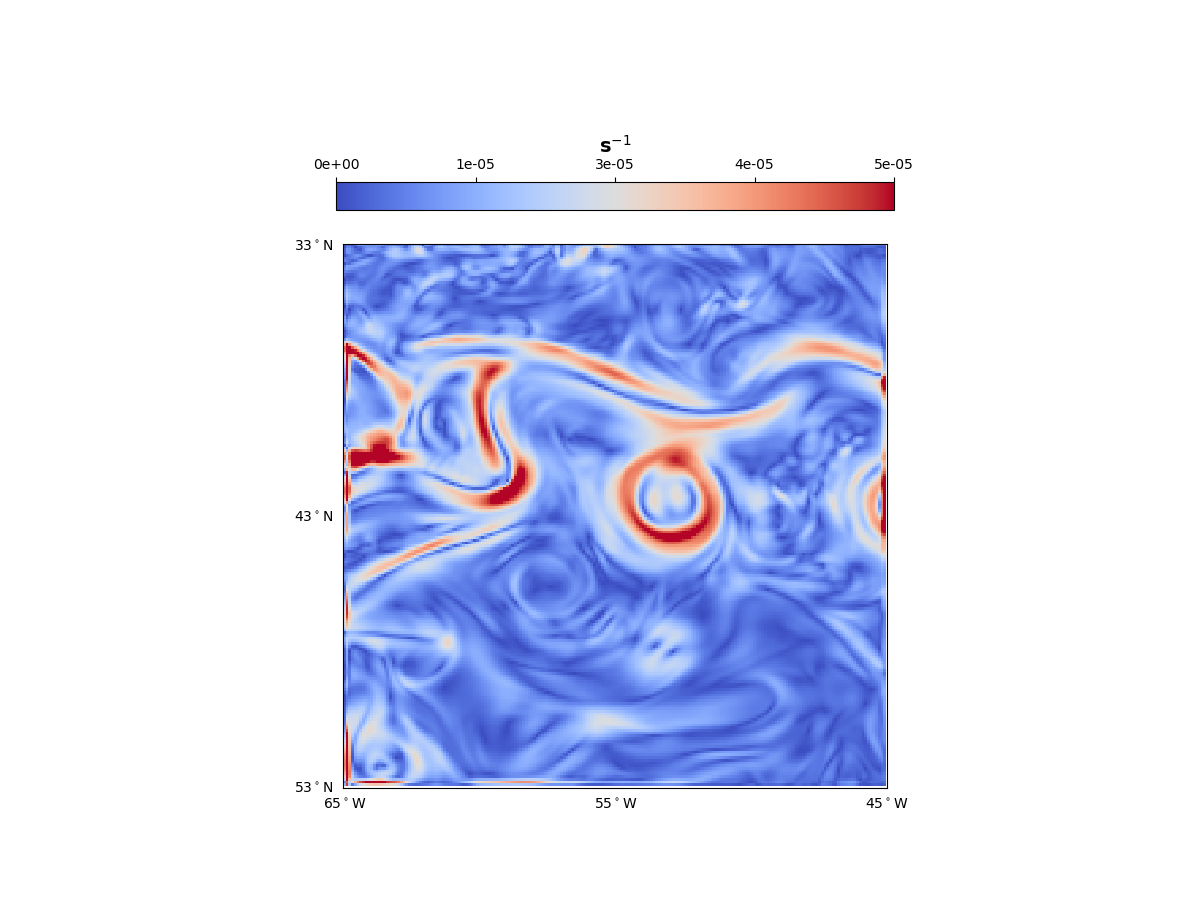}&
    \includegraphics[trim={210 65 210 155},clip,width=3.25cm]{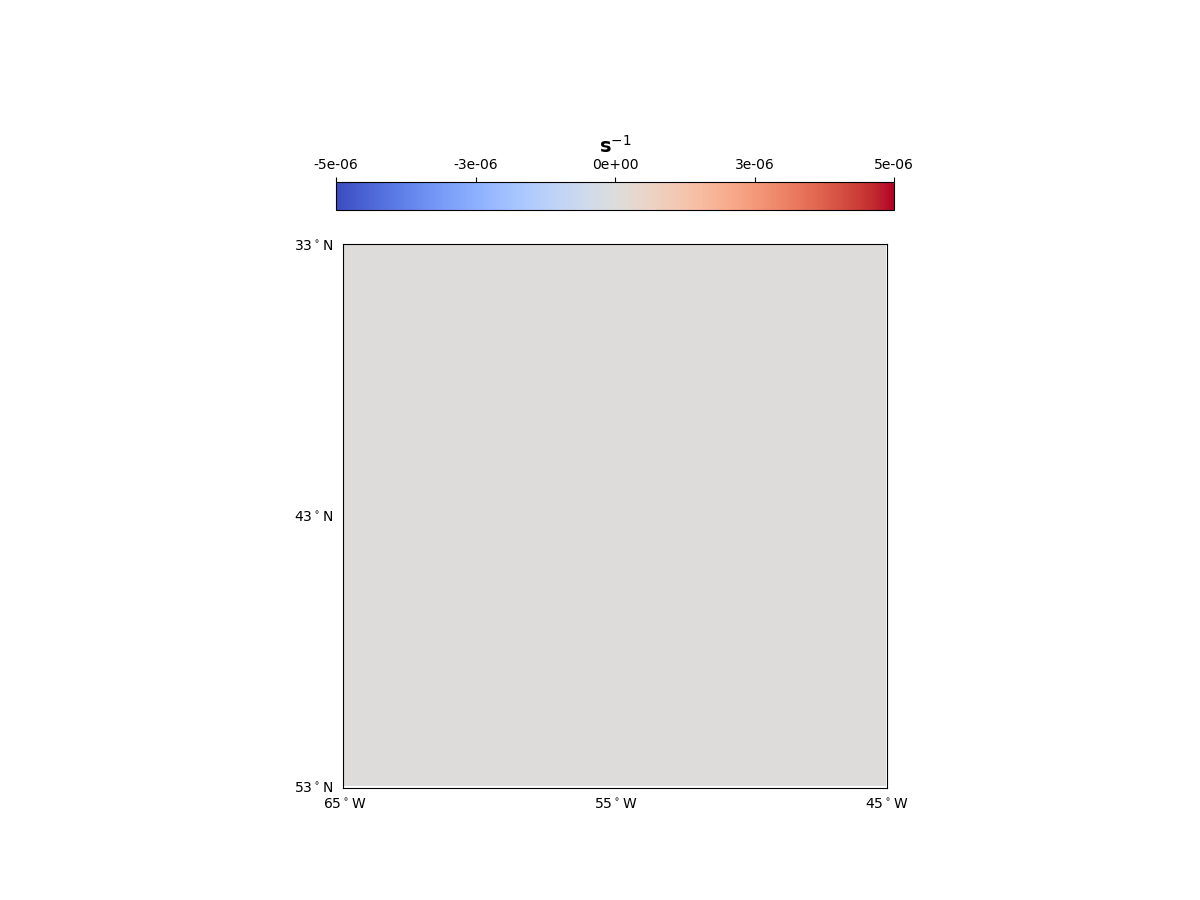}\\
    \multicolumn{4}{c}{\bf DUACS-based SSH-derived SSC}\\
    \includegraphics[trim={210 65 210 155},clip,width=3.25cm]{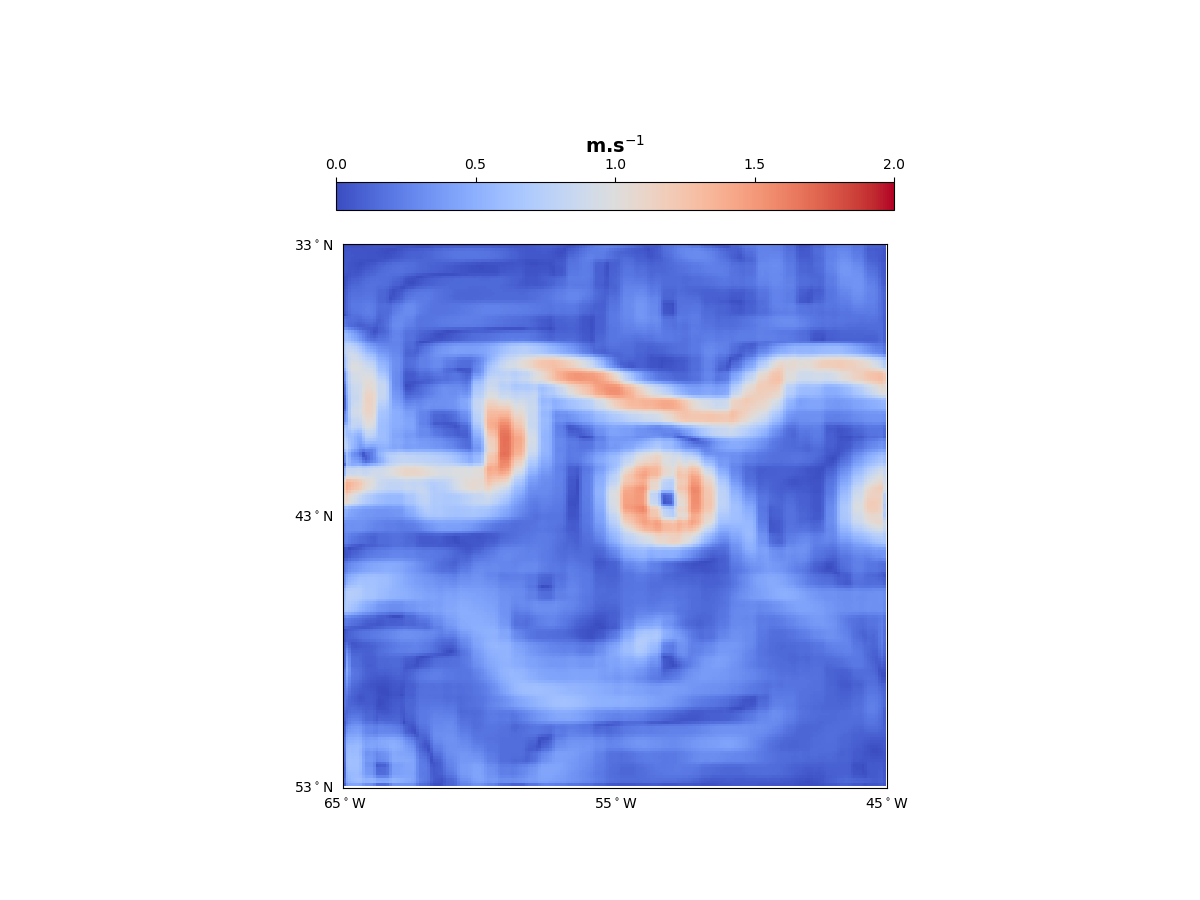}&
    \includegraphics[trim={210 65 210 155},clip,width=3.25cm]{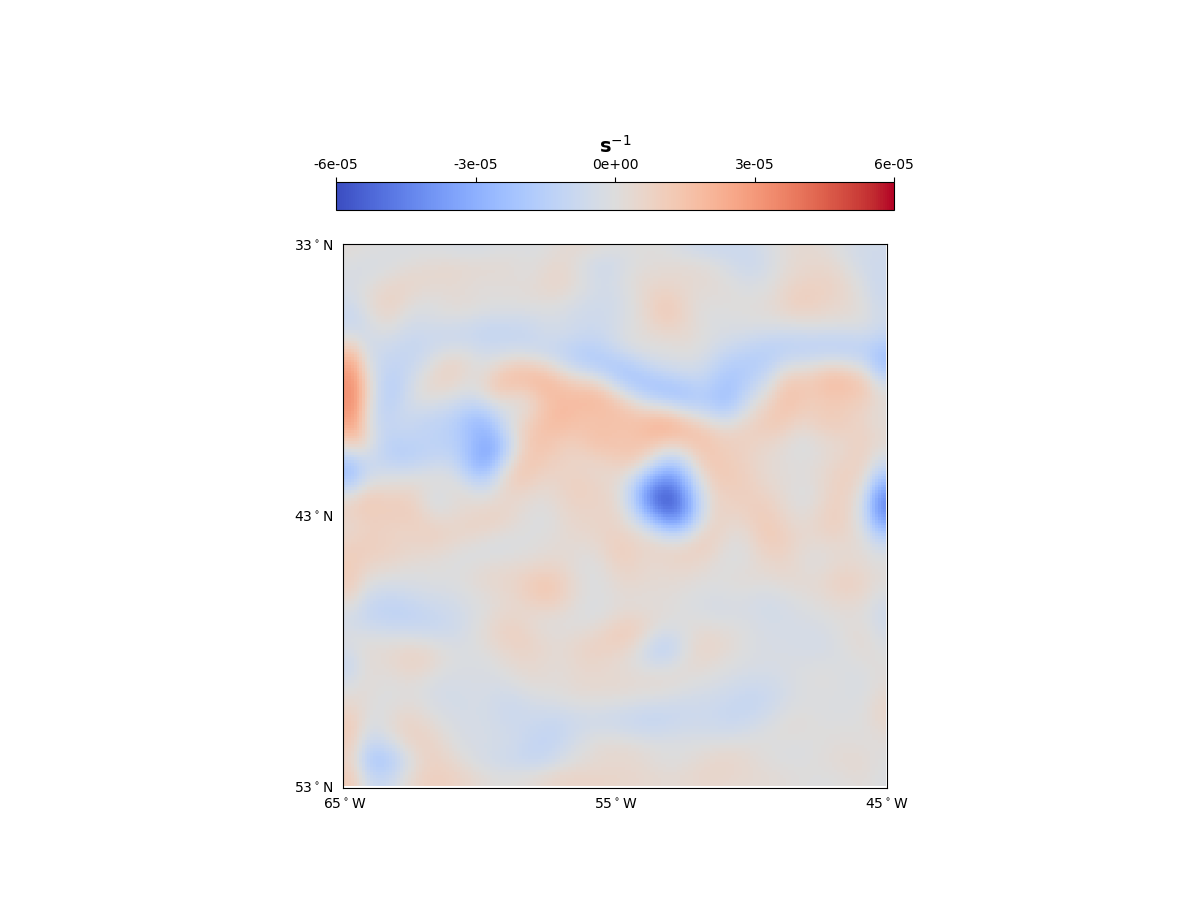}&
    \includegraphics[trim={210 65 210 155},clip,width=3.25cm]{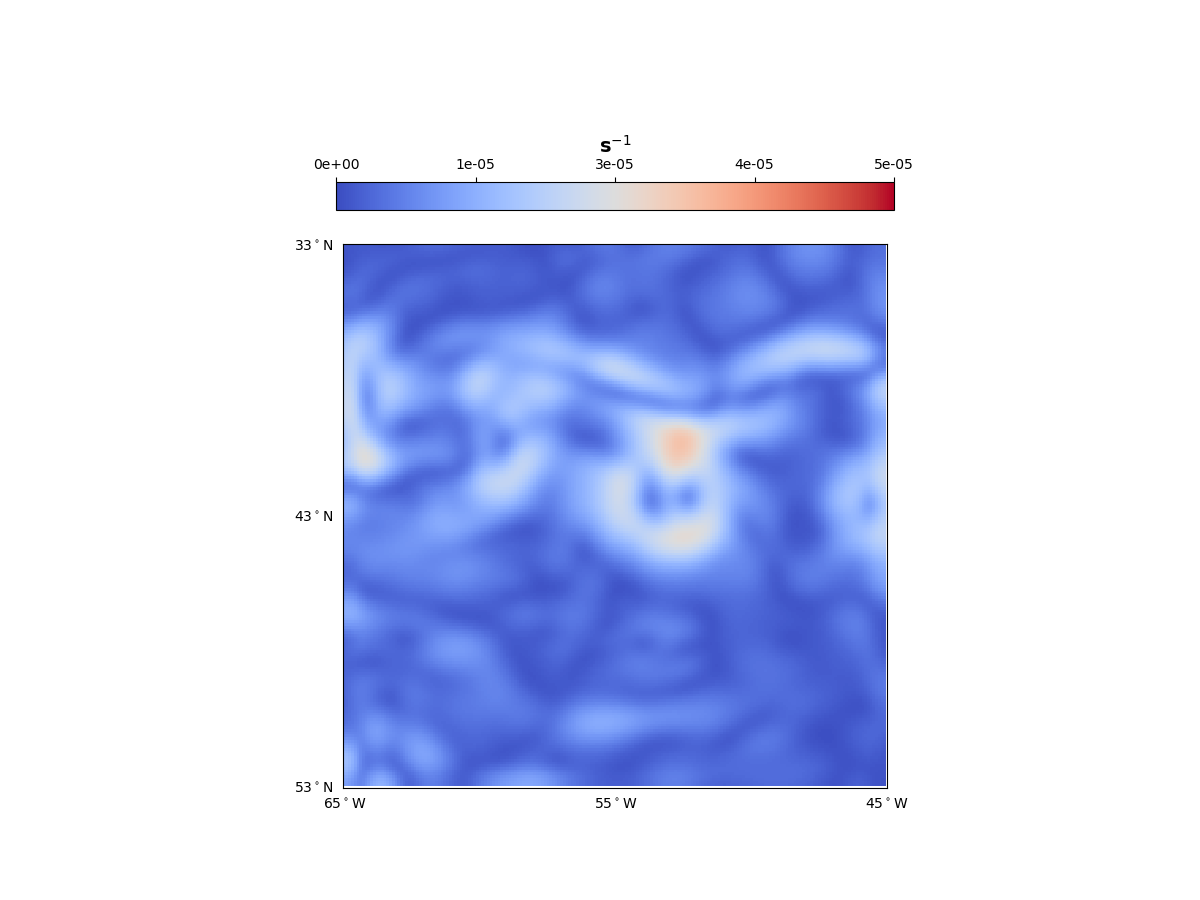}&
    \includegraphics[trim={210 65 210 155},clip,width=3.25cm]{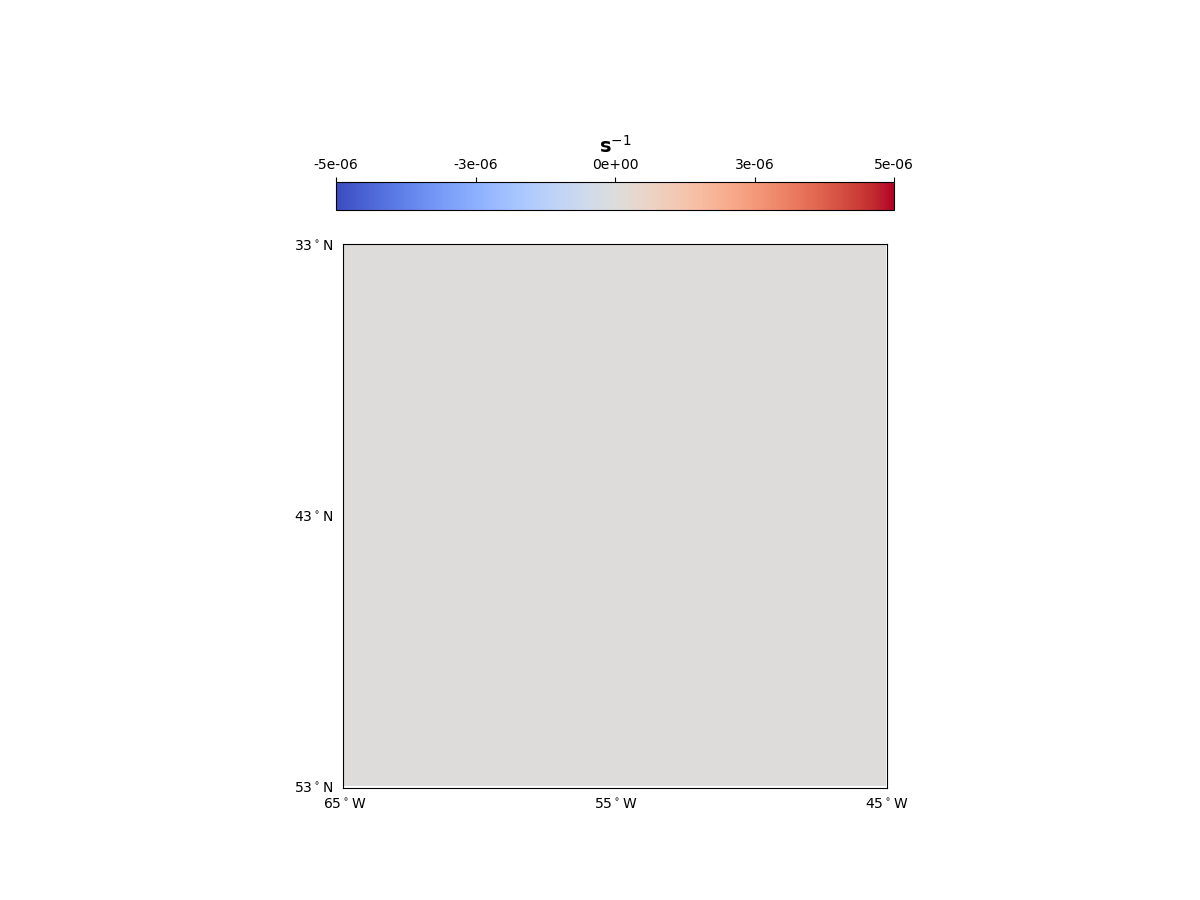}\\
    \multicolumn{4}{c}{\bf 4DVarNet SSC using SST}\\
    \includegraphics[trim={210 65 210 155},clip,width=3.25cm]{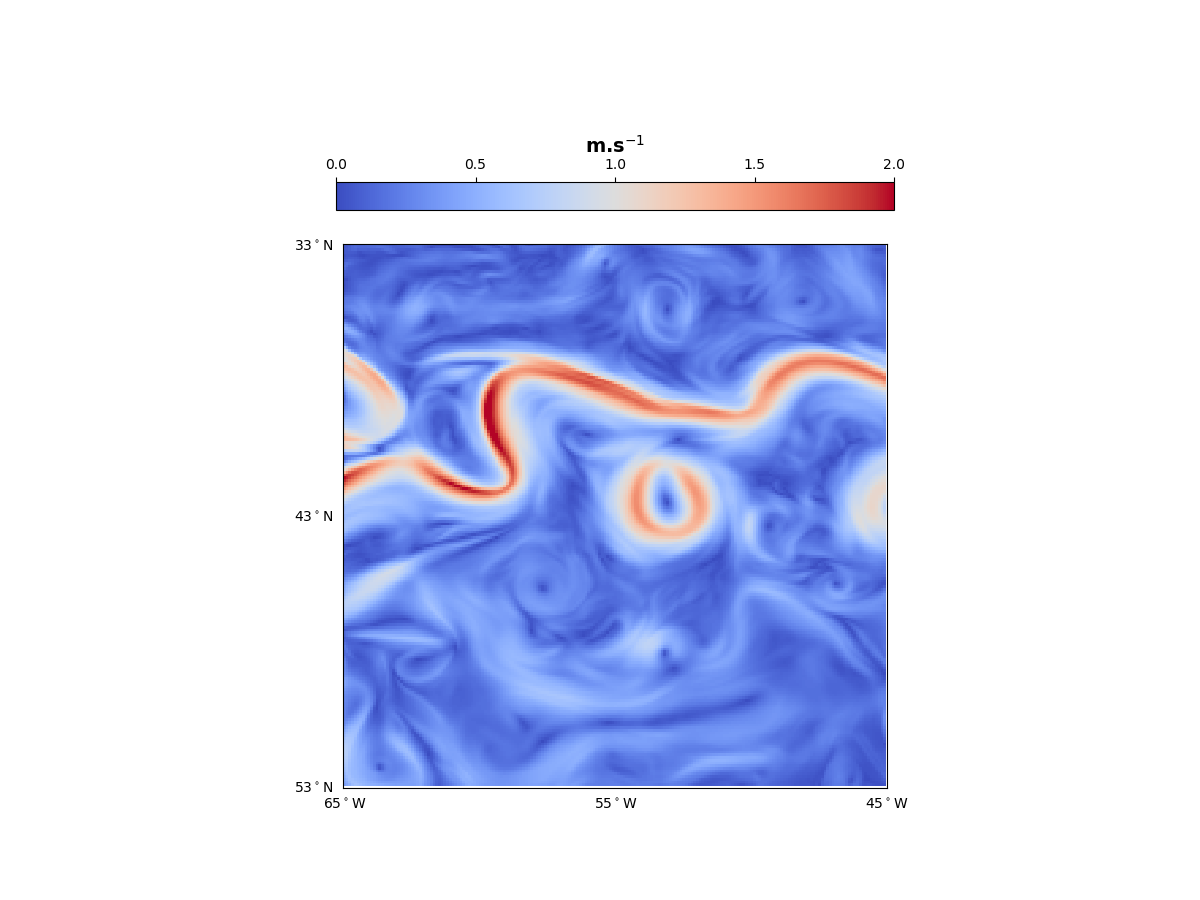}&
    \includegraphics[trim={210 65 210 155},clip,width=3.25cm]{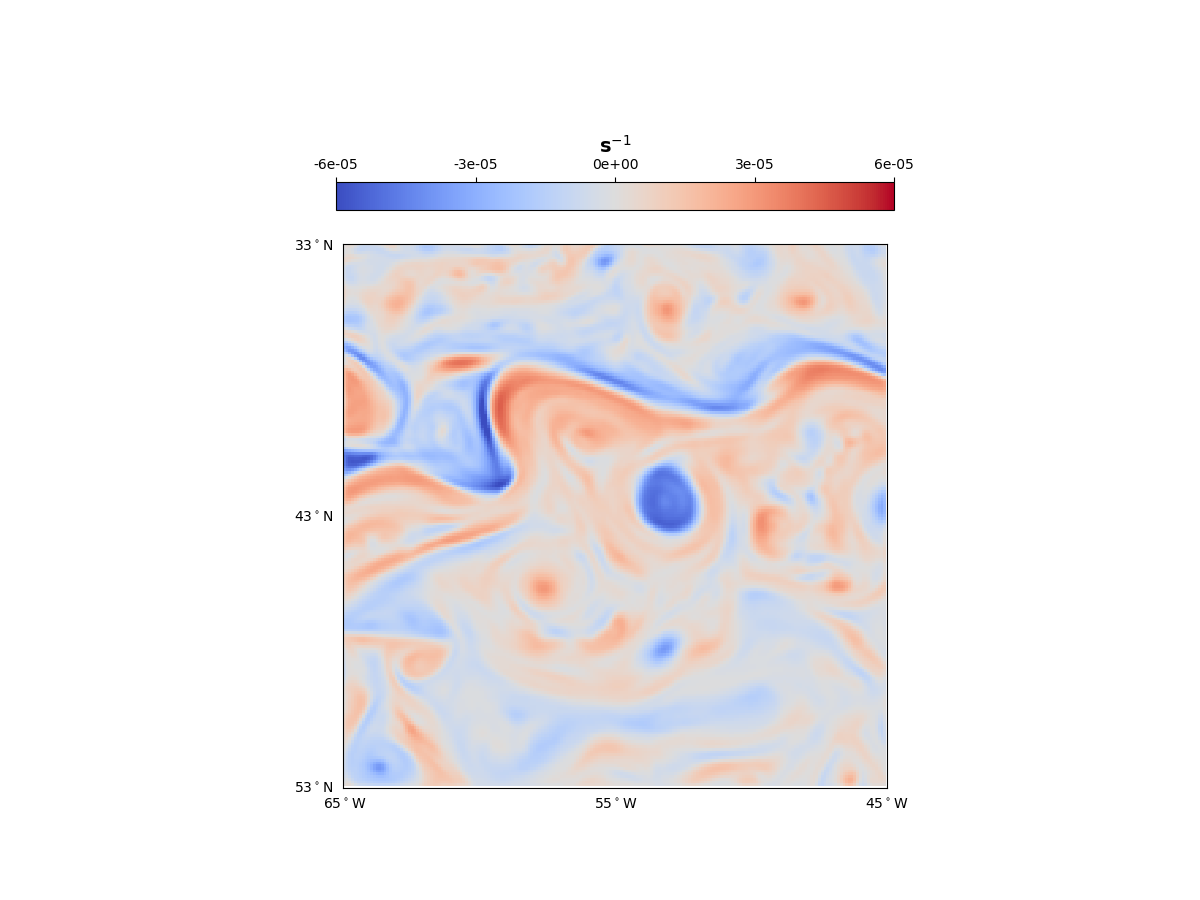}&
    \includegraphics[trim={210 65 210 155},clip,width=3.25cm]{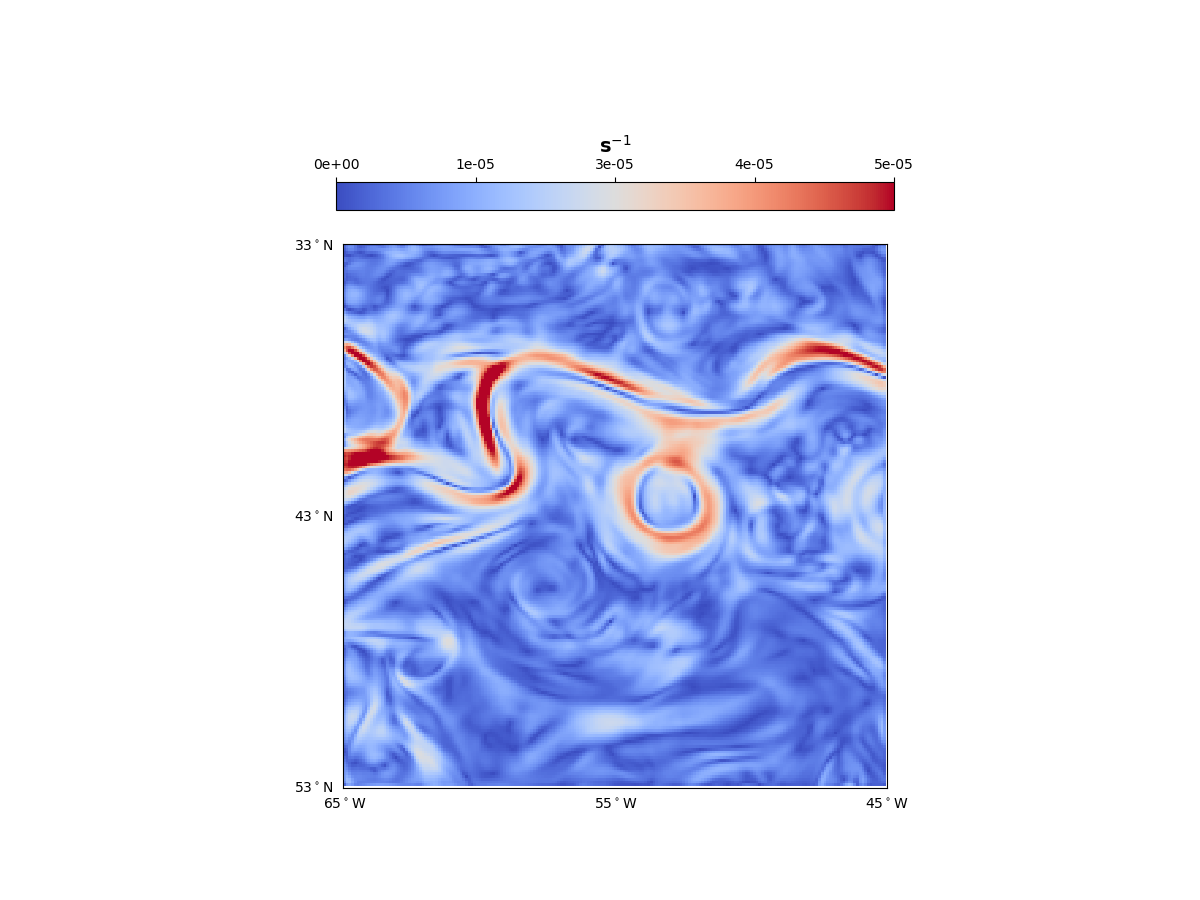}&
    \includegraphics[trim={210 65 210 155},clip,width=3.25cm]{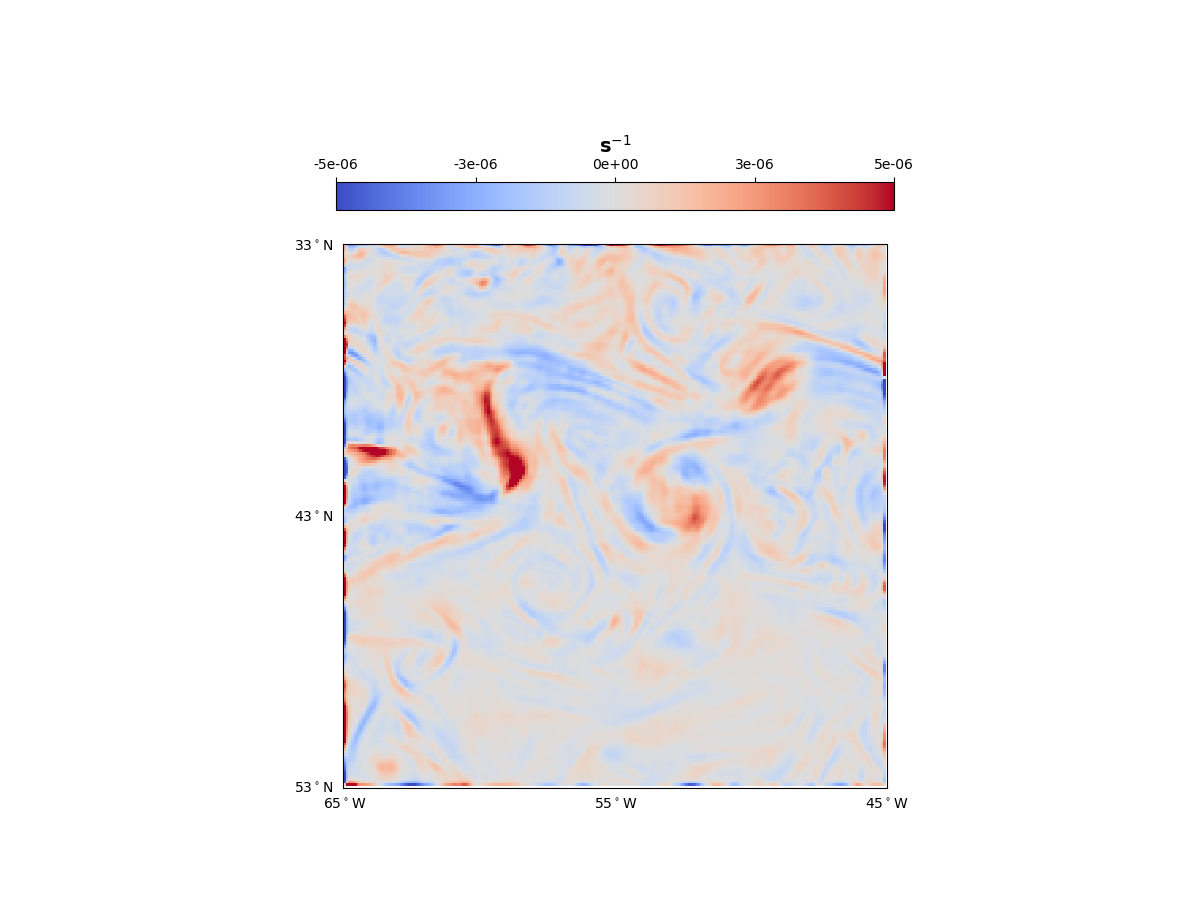}\\
    \multicolumn{4}{c}{\bf 4DVarNet SSH without SST}\\
    \includegraphics[trim={210 65 210 155},clip,width=3.25cm]{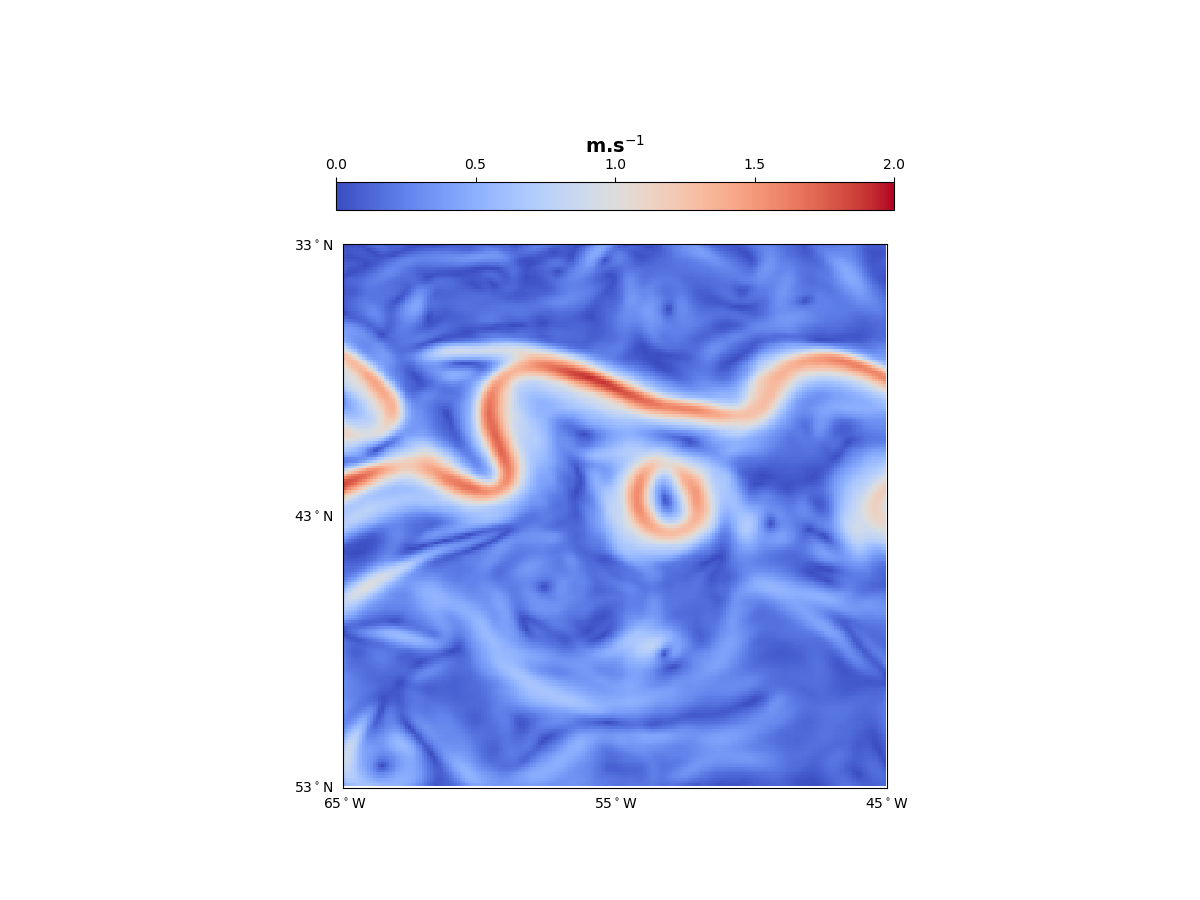}&
    \includegraphics[trim={210 65 210 155},clip,width=3.25cm]{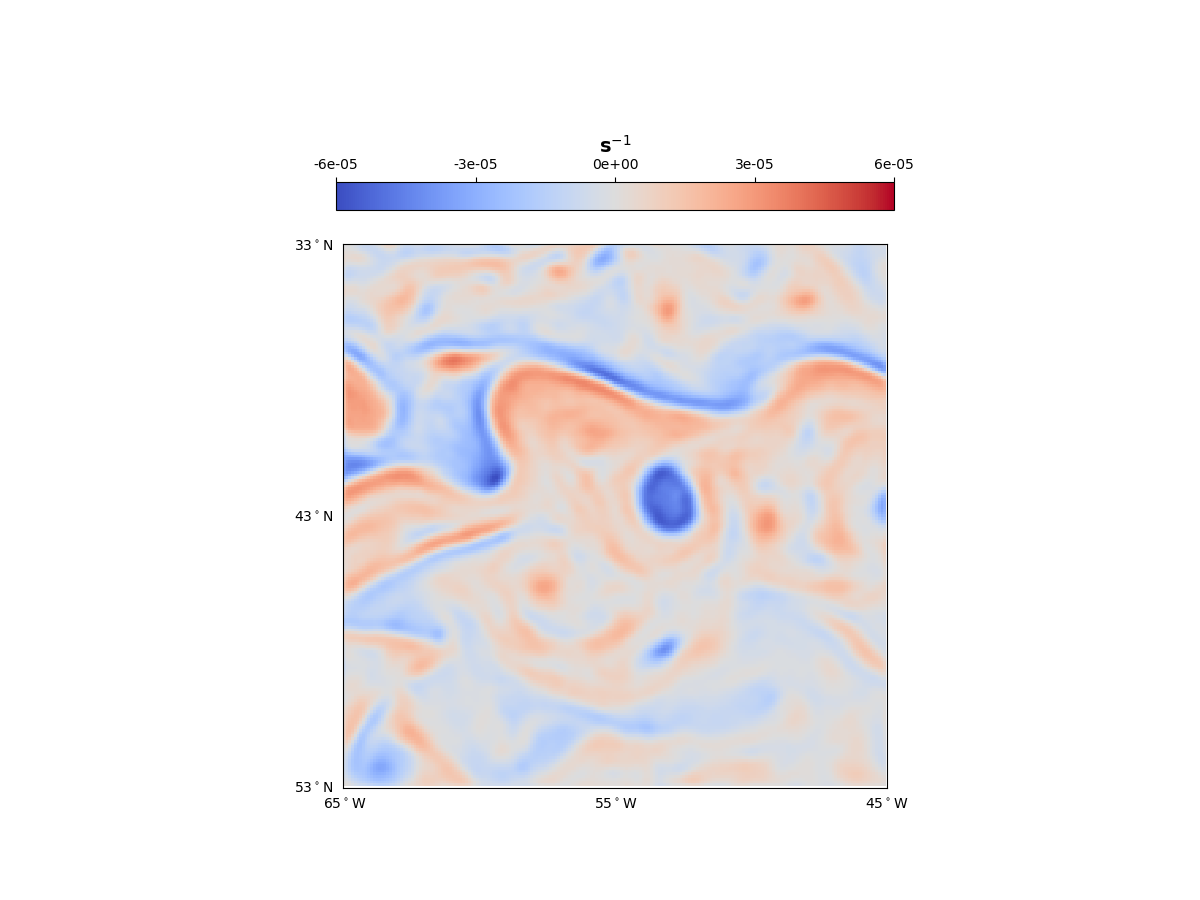}&
    \includegraphics[trim={210 65 210 155},clip,width=3.25cm]{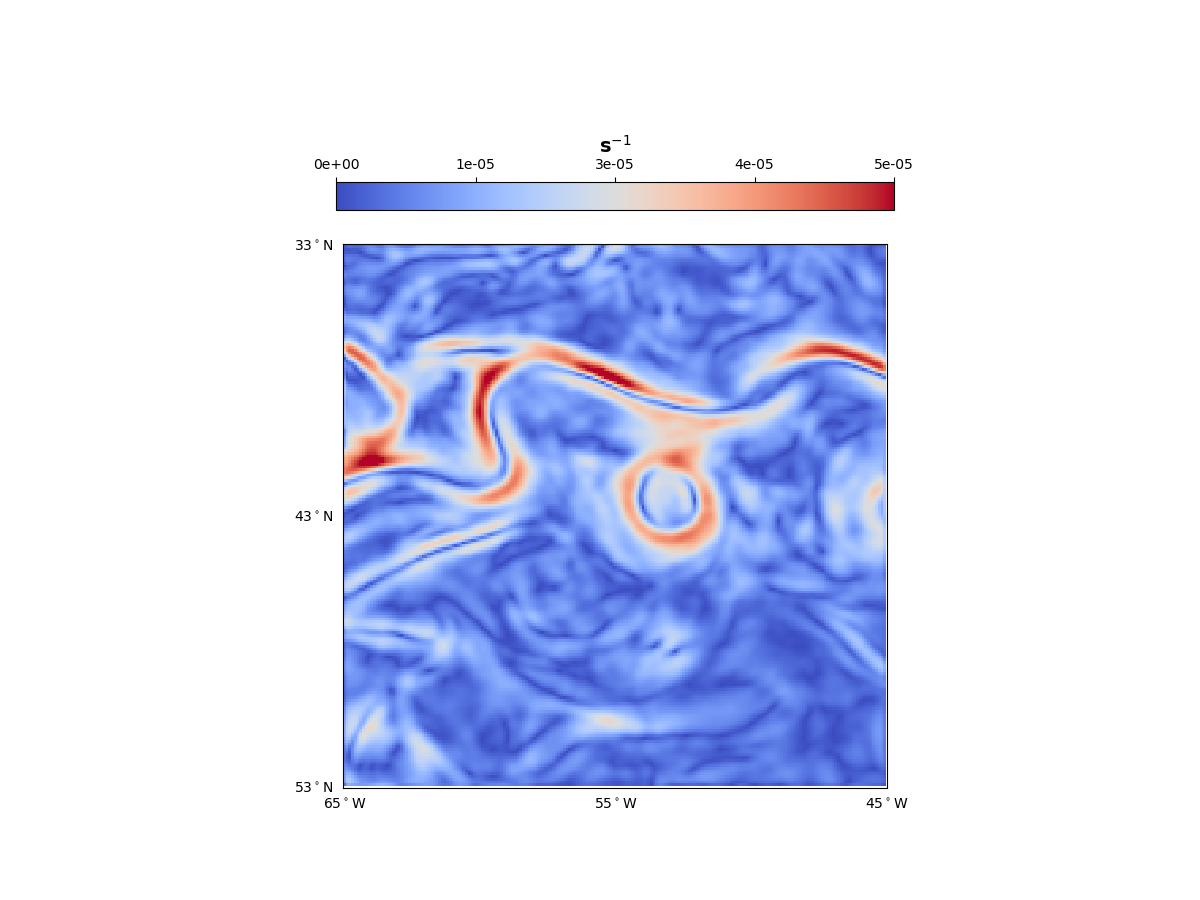}&
    \includegraphics[trim={210 65 210 155},clip,width=3.25cm]{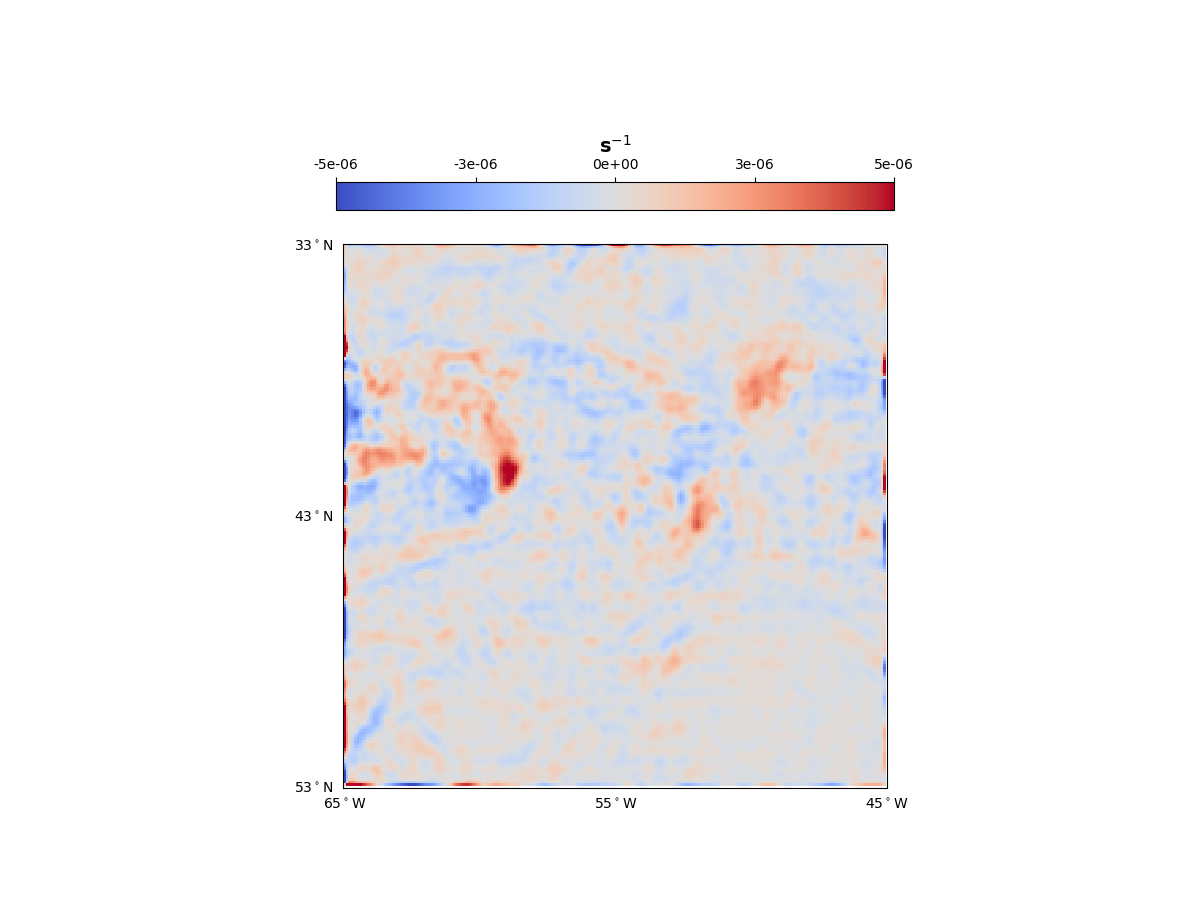}\\
    {\bf SSC norm}&{\bf Vorticity}&{\bf Strain}&{\bf Divergence}\\
    \end{tabular}}
    \caption{{\bf Reconstructed SSC fields on November 12$^\mathrm{th}$ 2012:}. From left to right, we depict the norm (i.e., velocity intensity), the vorticity, the divergence and the strain of SSC fields corresponding respectively to, from top to bottom, the true SSC field, the SSH-derived one, DUACS-derived one, and 4DVarNet-derived using SSH-SST fields and using only-SSH field.}
    \label{fig:res1}
\end{figure}

Fig.~\ref{fig:res1} further illustrates these results through the reconstructed fields on November 12$^\mathrm{th}$ 2013. This example stresses some strengthening of the sea surface current along the main meander compared with the SSH-derived geostrophic velocities. This is captured by 4DVarNet reconstruction using SSH-SST synergies. They also illustrate the improvement regarding fine-scale patterns which are smoothed by DUACS baseline. The comparison of the vorticity also indicates some correction of the SSH-derived estimation which is revealed by 4DVarNet scheme. Visually, the reconstructed divergence using 4DVarNet schemes recover important features of the true divergence, though finer-scale patterns are lost, which is in line with the performance metrics reported in Tab.~\ref{tab:res all}.

\begin{figure}[htbp]
    \centering
    \begin{tabular}{C{4cm}C{4cm}C{4cm}}
    \includegraphics[trim={210 65 200 100},clip,width=4cm]{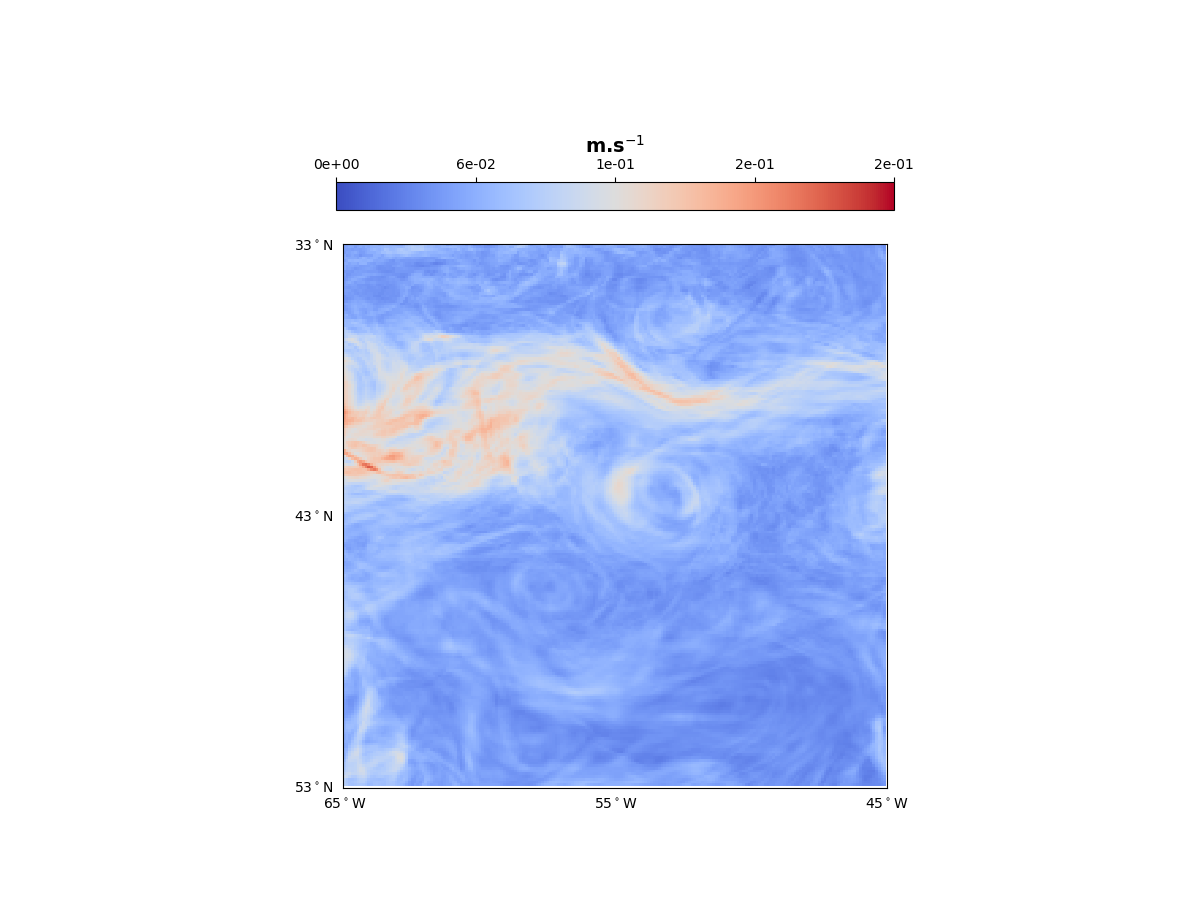}&
    \includegraphics[trim={210 65 200 100},clip,width=4cm]{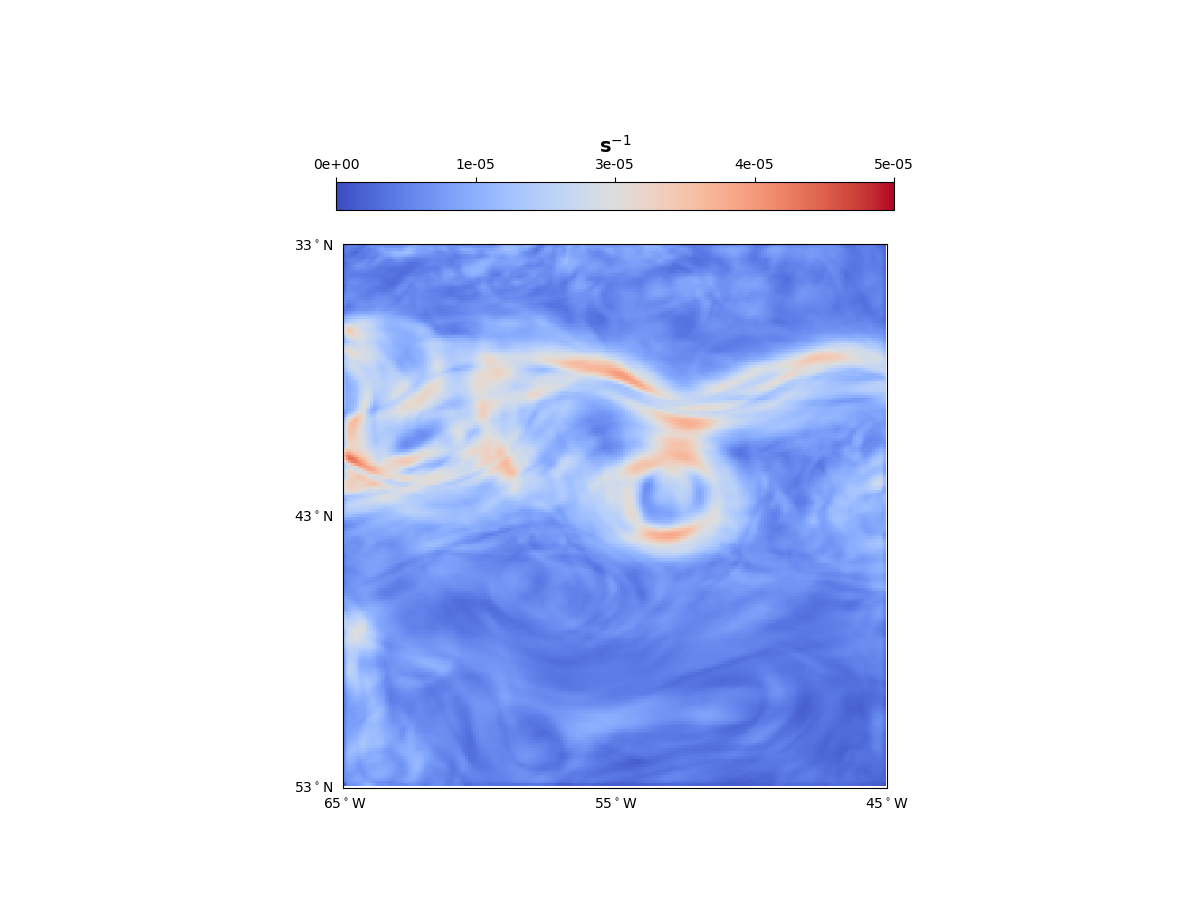}&
    \includegraphics[trim={210 65 200 100},clip,width=4cm]{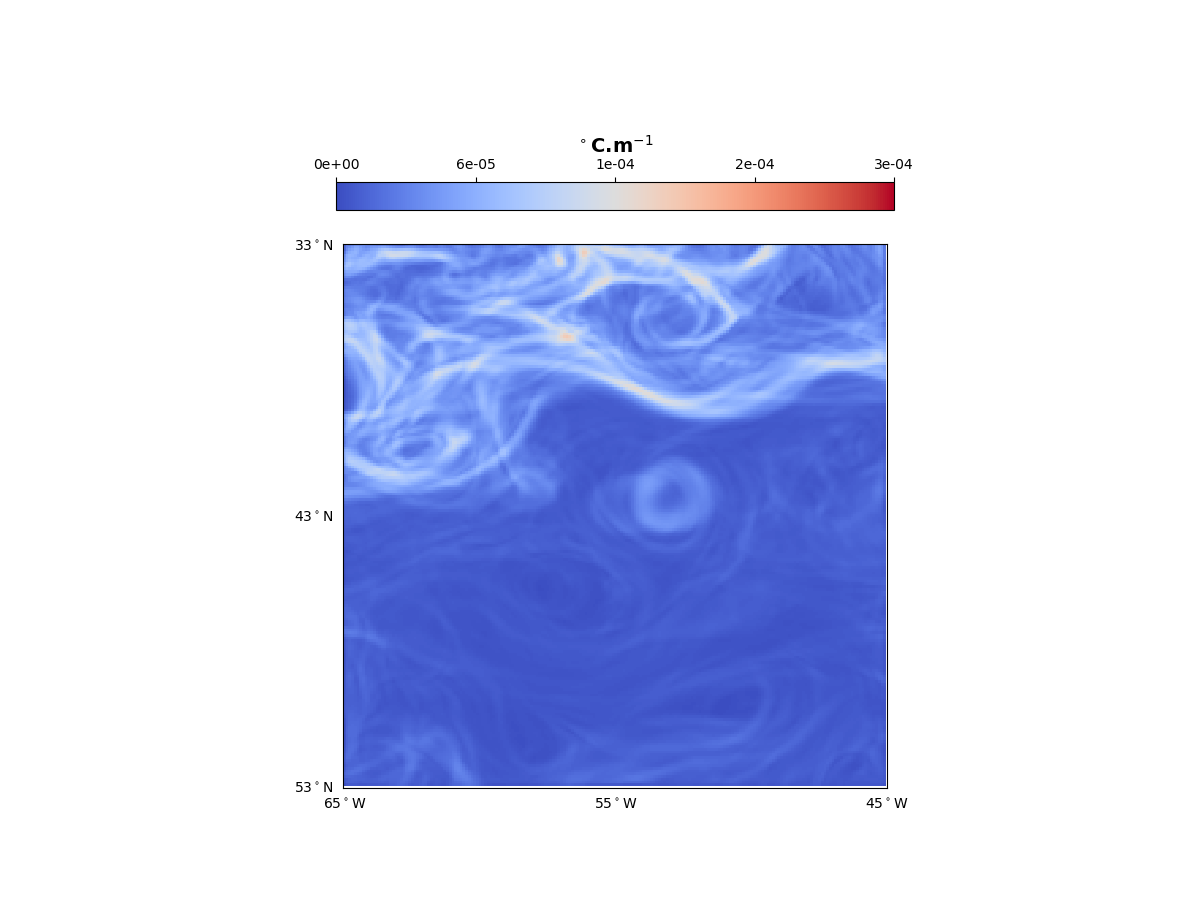}\\
    {\bf Reconstruction error}&{\bf True strain}&{\bf SST gradient}\\
    \end{tabular}
    \caption{{\bf Time-averaged reconstruction error vs. mean strain field and SST gradient over the whole test dataset:} from left to right, we display the time-averaged mean square reconstruction error field of the SSC using the 4DVarNet scheme combining altimetry and SST observations, the time-averaged true strain field and the time-averaged SST gradient. The coefficient of determination between the error field and the later fields is respectively of 59\% and 24\%. The reconstructed strain field depicted in Fig.\ref{fig:res1} leads to an explained variance of 57\% for the time-averaged reconstruction error. 
    }
    \label{fig:error2}
\end{figure}

\begin{table*}[tbh]
    \footnotesize
    \centering
    \begin{tabular}{|C{3.cm}|C{1.cm}|C{1.cm}|C{1.cm}|C{1.cm}|C{1.cm}|C{1.cm}|C{1.cm}|C{1.cm}|}
    \toprule
    \toprule
     \bf Data& \bf $\lambda_{x,u}$ ($^\circ$)&\bf $\lambda_{x,v}$ ($^\circ$)& $\lambda_{t,u}$ (d)& $\lambda_{t,v}$ (d)&$\tau_{u,v}$&$\tau_\mathrm{vort}$&$\tau_\mathrm{div}$&$\tau_\mathrm{strain}$\\
    \toprule
    \toprule
     NAlt  & 1.5 & 1.0 & 6.5 & 6.4 & 91.7\% & 79.0\% & 8.8\% & 72.7\%\\ 
     NAlt+SWOT  & 0.9 & 0.7 & 4.3 & 5.6 & 94.0\% & 86.1\% & 12.1\% & 81.3\%\\ 
     NALT+SST  & 0.81 & 0.62 & \bf 2.6 & \bf 2.5 & 97.1\% & 90.8\% & 46.3\% & 85.0\% \\ 
     NAlt+SWOT+SST  & \bf 0.76& \bf  0.61& \bf 2.7 & \bf 2.5 & \bf  97.4\% & \bf  92.1\% & \bf 46.9\% & \bf 87.2\%\\
     
    \bottomrule
    \bottomrule
    \end{tabular}
    \caption{{\bf Impact of SWOT and SST data onto the reconstruction performance w.r.t. a baseline using only nadir altimetry data}: we evaluate the performance metrics over the tests dataset for the trained 4DVarNet schemes reported in Tab.\ref{tab:res all} when
    considering nadir altimetry only (NAlt), nadir and wide-swath SWOT (NAlt+SWOT) altimetry combined or not with SST data in the proposed multimodal framework. Bold values show the highest skill.}
    \label{tab:res obs type}
\end{table*}

{\bf Impact of SWOT data:} We analyse in
Tab.~\ref{tab:res obs type} how SWOT data contribute to the reconstruction of sea surface currents. For the trained 4DVarNet models considered in Tab.~\ref{tab:res all}, we evaluate their reconstruction performance when considering only nadir altimetry data rather than nadir and SWOT altimetry data as used in the baseline configuration. These results support the relevance of wide-swath altimetry data to improve the reconstruction of sea surface currents. We report for instance a relative improvement greater than 25\% for both $\tau_{u,v}$, $\tau_\mathrm{vort}$ and $\tau_\mathrm{strain}$ metrics and improved resolved time scales of 4.3~days vs. 6.5~days for the zonal velocity. When exploiting SSH-SST synergies, we also observed some improvement though smaller. 
We observe the greatest relative improvement between 10\% and 15\% for  $\tau_{u,v}$, $\tau_\mathrm{vort}$ and $\tau_\mathrm{strain}$ metrics. This emphasizes for the considered case-study that SST fields bring most of the relevant information to retrieve the fine-scale patterns of sea surface currents. 

{\bf Impact of the spatial resolution of the SST data:} As the spatial resolution of satellite-derived SST products depends both on the satellite sensors as well as on the atmospheric conditions, we evaluate here the impact of the spatial resolution of the SST fields onto the reconstruction performance. From the trained mutimodal 4DVarNet, we fine-tune mutimodal 4DVarNet models using coarsened and subsampled versions of the original SST fields for spatial resolutions ranging from \nicefrac{1}{10}$^\circ$ to \nicefrac{1}{2}$^\circ$. We compare in Tab.\ref{tab:res sst res} the performance metrics of these models. As expected, the lower the resolution of the SST, the lower the performance. When considering SST data with a \nicefrac{1}{2}$^\circ$ resolution, we only significantly improve the reconstruction of the divergent component of the SSC (25.4\% vs. 12\% when using only altimetry data). From a \nicefrac{1}{4}$^\circ$ resolution, we report an improvement for all metrics, which is more noticeable for the resolved time scales. This is of key interest for the application to real microwave SST data \citep{ocarroll_observational_2019}. Besides, the clear gain issued from higher-resolution data also supports the potential of multispectral satellite sensors. 
Overall, Tab.\ref{tab:res sst res} suggests that the key SST features used to inform sea surface currents relate to horizontal scales between \nicefrac{1}{20}$^\circ$ and \nicefrac{1}{2}$^\circ$.  

\begin{table*}[tb]
    \footnotesize
    \centering
    \begin{tabular}{|C{3.cm}|C{1.cm}|C{1.cm}|C{1.cm}|C{1.cm}|C{1.cm}|C{1.cm}|C{1.1cm}|C{1.cm}|}
    \toprule
    \toprule
     \bf SST resolution& \bf $\lambda_{x,u}$ ($^\circ$)&\bf $\lambda_{x,v}$ ($^\circ$)& $\lambda_{t,u} (d)$& $\lambda_{t,v}$ (d)&$\tau_{u,v}$&$\tau_{vort}$&$\tau_{div}$&$\tau_{strain}$\\
    \toprule
    \toprule
     \nicefrac{1}{20}$^\circ$  & \bf 0.76&   \bf 0.61&   \bf 2.7 &  \bf 2.5 &    \bf 97.4\% &    \bf 92.1\% &  \bf 46.9\% &   \bf 87.2\%\\
      \nicefrac{1}{10}$^\circ$  & 0.81 & 0.62 & 3.0 & 2.6 & 96.9\% & 90.8\% & 41.8\% & 85.1\% \\ 
      \nicefrac{1}{5}$^\circ$  & 0.87 & 0.68 & 3.0 & 4.4 & 96.2\% & 89.0\% & 37.3\% & 83.2\% \\ 
      \nicefrac{1}{4}$^\circ$  & 0.90 & 0.70 & 3.0 & 3.5 & 95.6\% & 87.4\% & 30.0\%&81.5\% \\ 
      \nicefrac{1}{2}$^\circ$  & 0.92 & 0.82 & 4.3 & 6.3 & 94.3\% & 85.3\% & 25.4\% & 80.0\%\\ 
    \bottomrule
    \bottomrule
    \end{tabular}
    \caption{{\bf Impact of the spatial resolution of the SST observations:} we report the performance metrics of the proposed 4DVarNet models using SSH-SST synergies for SST observations with different spatial resolutions from $1/20^\circ$ to $1/2^\circ$. We refer the reader to the main text for the description of the different metrics.  Bold values show the highest skills.}
    \label{tab:res sst res}
\end{table*}

{\bf Analysis of the reconstruction error:} We further analyse the reconstruction error and illustrate in Fig.~\ref{fig:error2} the time-averaged mean square error of the total sea surface current over the considered dataset from October 20$^\mathrm{th}$ 2012 to December 4$^\mathrm{th}$ 2012 for the trained 4DVarNet scheme exploiting SSH-SST synergies. 
We also report the time-averaged strain of the true velocities along with the time-averaged amplitude of horizontal SST gradients. We observe a better match between error and strain patterns. Quantitatively, the time-averaged strain explains about 60\% of the time-averaged error over the test dataset. We reach similar correlation statistics for the reconstructed strain as expected from the visual similarity between the true and reconstructed strain depicted in Fig.~\ref{fig:res1}.

We know that the strain highlights regions of frontogenesis or gradient strengthening \citep{balwada_vertical_2021,okubo_horizontal_1970,weiss_dynamics_1991}. It could also be demonstrated that it leads to velocity error growth (Appendix). This is also consistent with kinetic energy transfer occurring due to the strain of the flow \citep{Sevellec:22}. Here we hypothesized that regions of significant error growth (i.e., strain active regions) are more difficult to reconstruct due to their chaotic nature (i.e, small disturbances will become a significant signal). However we do not expect a perfect relation between surface strain and velocity error growth (and so velocity reconstruction error) since other terms, such as viscous terms or sub-surface pressure gradients, also control the velocities, and can become sources of error growth. Interestingly, this analysis is similar for the true strain and the reconstructed one, which could provide a proxy for the quantification of the uncertainty of the reconstruction using 4DVarNet schemes.

\section{Discussion}
\label{sec:discussion}

This study has introduced a 4DVarNet deep learning scheme, backed on a variational data assimilation formulation, for the reconstruction of sea surface currents from satellite-derived SSH and SST observations. Our numerical experiments support the relevance of this learning-based approach over state-of-the-art schemes to retrieve finer-scale sea surface dynamics (typically 0.5-0.7$^\circ$ and 3~days for the resolved space-time scales), including a significant fraction of the ageostrophic component of the total current (about 47\% of the divergence of the SSC). The good skill of 4DVarNet in the reconstruction of the horizontal divergence of the flow makes it particularly relevant for tracking surface/buoyant polluant concentration (for which convergence regions act as attractors \citep{Sevellec:17b}). It is also important for the hypothetic reconstruction of vertical velocities (which vertical gradient is directly related to the horizontal divergence of the flow), with all their dynamical consequences for the export of nutrients, heat, and carbon, for instance. We have also hypothesized, through theoretical arguments, and shown the relation between the flow strain and the error of the reconstructed horizontal velocities. This suggests that strain active regions are good candidates for intense monitoring if one aims to better reconstruct horizontal velocities. We further discuss our contributions according to the following aspects: the monitoring of sea surface velocities, deep learning inversions for unobserved upper ocean dynamics and multimodal synergies for ocean monitoring and forecasting.

{\bf Reconstruction of sea surface velocities:} Satellite altimetry plays a major role in our ability to monitor sea surface dynamics on a global to local scale. Through the direct measurement of the sea level anomaly, it delivers an estimation of the geostrophic component of sea surface velocities \citep{chelton_satellite_2001}. As the scarce sampling of the sea surface by nadir altimeters limits our ability to retrieve dynamical features below $\approx 1^\circ$ and 10~days, research effort has been undertaken to observe and reconstruct finer-scale dynamics. In this context, upcoming swide-swath altimetry mission SWOT \citep{gaultier_challenge_2015} will provide the first snapshots of the sea surface height down to a \nicefrac{1}{10}$^\circ$ spatial resolution. Similarly to nadir altimeters, it will only directly inform the geostropic component of sea surface velocities. Numerous studies (e.g., \citep{baaklini_blending_2021,mahadevan_analysis_2006}) have evidenced the key role of ageostrophic dynamics in the mesoscale-to-submesoscale range. 
This has motivated a large research effort towards the exploitation of other observation sources, alone or combined with satellite altimetry, to retrieve sea surface dynamics, including among others SST \citep{fablet_improving_2018,isern-fontanet_transfer_2014,rio_improving_2016}, Ocean Colour \citep{ciani_ocean_2021}, sea surface drifters \citep{baaklini_blending_2021,sun_impacts_2022}, and SAR observations \citep{chapron_direct_2005}. From a methodological point of view, we may distinguish three main categories of approaches: optimal interpolation schemes \citep{cressie_statistics_2015,taburet_duacs_2019}, data-driven approaches \citep{fablet_data-driven_2017,fablet_multimodal_2022,manucharyan_deep_2021}, and data assimilation scheme using OGCM \citep{benkiran_assessing_2021,fujii_observing_2019} or QG dynamical priors \citep{le_guillou_mapping_2020,ubelmann_dynamic_2014}. The proposed 4DVarNet scheme benefits, on the one hand, from a variational data assimilation formulation to make explicit observation and dynamical priors, especially the expected though unknown relationship between SST and SSC features, and, on the other hand, from the computational efficiency of deep learning schemes, to learn uncalibrated terms and solvers from data in an end-to-end manner. This seems particularly promising to make the most of available multi-source observation datasets for the reconstruction of sea surface currents. Our study supports the ability to retrieve finer-scale patterns than currently achieved by operational products. While our study points out the added-value of wide-swath altimetry data, it suggests that ageostrophic components can be revealed by learning-based schemes from other sea surface tracers sampled at higher-resolution as illustrated with SST fields in our study. This likely relates to specific dynamical regimes in play in the Gulf Stream region \citep{mcwilliams_gulf_2019,reul_sea_2014}.
Key challenges for future work include the exploration of the proposed framework for the variety of dynamical regimes exhibited by upper ocean dynamics as well as the transfer from OSSE to real observation dataset. The extension to other observation datasets such as drifter trajectories \citep{baaklini_blending_2021,sun_impacts_2022}, SAR observations, as well as multispectral satellite sensors \citep{barnes_cross-calibration_2021,tilstone_performance_2021,yurovskaya_ocean_2019} also naturally arises as appealing research directions.
Whereas our OSSE provides a noise-free idealized testbed, real altimetry observations involve both observation noises and high-frequency fine-scale geophysical signals, such as internal tides and internal gravity waves \cite{arbic_concurrent_2010,xu_effects_2012}. Numerous studies support the potential robustness of deep learning schemes to noisy patterns, when the training dataset involves appropriate noise simulations \cite{shorten_survey_2019}. The availability of realistic tide-resolving submesoscale-permitting ocean simulations provides the basis to address these issues in a future work using the proposed framework. A more challenging task will be the separation of tide-related and tide-free motions at sea surface, which could benefit from an extended version of the proposed neural approach.

{\bf Deep learning inversion for unobserved upper ocean dynamics:} Our study further supports the potential for deep learning approaches to reconstruct unobserved or partially-observed variables or processes from available satellite-derived and/or {\it in situ} observation datasets. The reported case-study for horizontal sea surface currents is in line with recent studies addressing ocean processes, such as ocean eddy heat fluxes \citep{george_deep_2021}, tide-related features \citep{wang_deep_2022}, plankton dynamics \citep{martinez_neural_2020}... The underlying assumption is that deep learning can disentangle the features associated with a given process in the observations of a tracer. In this context, classic data assimilation scheme requires defining explicitly the observations and dynamical operator in play in the underlying state-space formulation (\ref{eq: state space}). Regarding upper ocean dynamics, this generally resorts to considering some discretization of the primitive equations and identity observation operators for ocean state variables and ocean-atmosphere interactions. The complexity of the inversion problem may hinder the ability of these schemes to retrieve fine-scale dynamics when dealing with a poorly-constrained observation setting as with the inversion of ocean dynamics from satellite-derived sea surface observations. By contrast, the proposed deep learning scheme focuses only on the ocean variable of interest, here observed and targeted sea surface variables. Through this learning paradigm, we reduce the complexity of the inversion problem and explore the complex relationships exhibited by sea surface tracers governed by upper ocean dynamics. We believe this approach to be generic and to show a potential to improve our understanding and monitoring of upper ocean dynamics which will remain scarcely observed.

{\bf OSSEs and Deep Learning to Bridge ocean modeling and multimodal ocean remote sensing:} We may also regard the proposed deep learning schemes as alternative means to bridge ocean modeling and ocean observation. Through the combination of simulation dataset for ocean dynamics and observing systems, OSSE settings \citep{boukabara_community_2018,fujii_observing_2019} have been widely exploited as means to assess the potential impact of new observing systems to improve the reconstruction and/or forecasting of ocean dynamics. OSSEs in the context of the upcoming SWOT mission fall into this category to evaluate the added-value of wide-swath altimetry data to inform upper ocean dynamics \citep{benkiran_assessing_2021,ubelmann_dynamic_2014}, as further supported by this study. More recently, OSSEs have provided benchmarking frameworks to intercompare the performance of different approaches for the same task such as space-time interpolation problems for sea surface dynamics \citep{fablet_multimodal_2022,le_guillou_mapping_2020,fujii_observing_2019,vient_data-driven_2021}. As illustrated here, when dealing with unobserved or scarcely observed processes, OSSEs also provide means to train a model from simulation datasets with a view to applying this model to real datasets. Previous learning-based studies dedicated to the reconstruction of ocean's interior processes, such as for instance the primary production \citep{puissant_inversion_2021} and deep geostrophic currents \citep{manucharyan_deep_2021}, provide such examples. These studies generally learn a mapping from observed gap-free sea surface tracers to the targeted ocean state dynamics. Our study supports the ability to further extend this approach to reconstruction or mapping objectives from irregularly-sampled observations. The applicability of this general framework to real observation datasets clearly depends on the quality of both the numerical simulations of ocean dynamics and of the simulations of the observing systems. As such, it advocates to pursue joint research effort into these simulation issues along with the design and evaluation of deep learning approaches. We believe these aspects to be critical for the development of multimodal observing systems combining satellite-derived observations and {\it in situ} data including among others sea surface drifters \citep{baaklini_blending_2021,sun_impacts_2022}, argo profiles \citep{cossarini_towards_2019,roemmich_future_2019}, and underwater acoustics data \citep{storto_assessing_2020,storto_neural_2021}...

\section*{\bf Appendix: Velocity error growth}
Starting from the momentum and nondivergence equations [described in (\ref{eq:momentum})] but splitting the velocity (and pressure and viscous terms) in a targeted truth and a small error ($X=\bar{X}+X'$ with $|\bar{X}|\ll|X'|$, where $X$ is any variable) we have: 
\begin{subequations}
  \begin{equation}
    D_tu'-fv'=-u'\partial_x\bar{u}-v'\partial_y\bar{u}-w'\partial_z\bar{u}-\frac{1}{\rho_0}\partial_xP'+\mathcal{F}_x',
  \end{equation}
  \begin{equation}
    D_tv'+fu'=-u'\partial_x\bar{v}-v'\partial_y\bar{v}-w'\partial_z\bar{v}-\frac{1}{\rho_0}\partial_yP'+\mathcal{F}_y'.
  \end{equation}
  \begin{equation}
    0=-\frac{1}{\rho_0}\partial_zP'-\frac{g}{\rho_0}\rho',
  \end{equation}
  \begin{equation}
    \partial_xu'+\partial_yv'+\partial_zw'=0.
  \end{equation}
\end{subequations}
So that error growth reads:
\begin{multline}  
    D_t\left(u'^2+v'^2\right)=-\left(u',v'\right)\left(\begin{array}{cc}\partial_x\bar{u}&\partial_y\bar{u}\\\partial_x\bar{v}&\partial_y\bar{v}\end{array}\right)\left(\begin{array}{c}u'\\v'\end{array}\right)\\
    -u'w'\partial_z\bar{u}-v'w'\partial_z\bar{v}-\frac{g}{\rho_0}w'\rho'\\
    -\frac{1}{\rho_0}\left[\partial_x\left(u'P'\right)+\partial_y\left(v'P'\right)+\partial_z\left(w'P'\right)\right]+\left(u'\mathcal{F}_x'+v'\mathcal{F}_y'\right), 
\end{multline}
where the left handside is the evolution of the error (or error growth), the first line of the right handside is the horizontal transfer of momentum error, the second line of the right handside is the baroclinic transfer of momentum and density errors, and the third line of the right handside is the work of the pressure and viscous force error. Here the horizontal transfer of momentum error is a scalar product of the erroneous velocities applied to the operator of horizontal targeted-truth-velocity gradient. This could be transformed as:
\begin{eqnarray}
    \left(u',v'\right)\boldSigma\left(\begin{array}{c}u'\\v'\end{array}\right)
    &=&\left(\begin{array}{cc}\partial_x\bar{u}&\partial_y\bar{u}\\\partial_x\bar{v}&\partial_y\bar{v}\end{array}\right)\left(\begin{array}{c}u'\\v'\end{array}\right),\nonumber\\
    &=&\frac{1}{2}\left(u',v'\right)\left(\begin{array}{cc}\Delta+\sigma_n&\sigma_s+\zeta\\\sigma_s-\zeta&\Delta-\sigma_n\end{array}\right)\left(\begin{array}{c}u'\\v'\end{array}\right),\nonumber\\
    &=&\frac{1}{2}\left(u',v'\right)\left(\begin{array}{cc}\Delta+\sigma_n&\sigma_s\\\sigma_s&\Delta-\sigma_n\end{array}\right)\left(\begin{array}{c}u'\\v'\end{array}\right),
\end{eqnarray}    
where $\Delta=\partial_x\bar{u}+\partial_y\bar{v}$ is the divergence, $\zeta=\partial_x\bar{v}-\partial_y\bar{u}$ is the horizontal vorticity, $\sigma_n=\partial_x\bar{u}-\partial_y\bar{v}$ is the normal strain (or stretching), and $\sigma_s=\partial_x\bar{v}+\partial_y\bar{u}$ is the shear strain (or shearing). This last expression shows the symmetric part of the gradient deformation operator, which is virtually equivalent to the one used for tracer gradient growth \citep{balwada_vertical_2021} in the Okubo-Weiss framework of frontogenesis \citep{okubo_horizontal_1970,weiss_dynamics_1991}. It is interesting to note that here the horizontal vorticity ($\zeta$) does not contribute to the error growth, though. $\boldSigma$ can be diagonalized to show the natural growth of the error, which leads to the pair of eigenvalues: $\lambda_\pm=\nicefrac{\Delta}{2}\pm\nicefrac{\sigma}{2}$, where $\sigma$ is the strain (such as $\sigma^2=\sigma_n^2+\sigma_s^2$). Since the horizontal flow is largely strain dominated, this shows that the error growth is controlled by the strain. This result is consistent with energy transfer consideration suggesting the transfer from the mean state (here the targeted-truth) to the eddy field (here the error) following the strain of the flow \citep{Sevellec:22}. However note that the total error growth will also be affected by baroclinic error transfers and pressure and viscous work.

\section*{\bf Acknowledgements}
This work was supported by LEFE program (LEFE MANU and IMAGO projects IA-OAC and ARVOR, respectively), CNES (OSTST DUACS-HR and SWOT ST DIEGO) and ANR Projects Melody (ANR-19-CE46-0011) and OceaniX (ANR-19-CHIA-0016). It benefited from HPC and GPU resources from Azure (Microsoft Azure grant) and from GENCI-IDRIS (Grant 2021-101030).  

\bibliographystyle{unsrtnat}
\bibliography{zotero,references}  

\begin{thebibliography}{69}
\providecommand{\natexlab}[1]{#1}
\providecommand{\url}[1]{\texttt{#1}}
\expandafter\ifx\csname urlstyle\endcsname\relax
  \providecommand{\doi}[1]{doi: #1}\else
  \providecommand{\doi}{doi: \begingroup \urlstyle{rm}\Url}\fi

\bibitem[Chelton et~al.(2001)Chelton, Ries, Haines, Fu, and
  Callahan]{chelton_satellite_2001}
D.B. Chelton, J.C. Ries, B.~J. Haines, L.-L. Fu, and P.~S. Callahan.
\newblock Satellite {Altimetry}.
\newblock In A.~Cazenave and L.-L. Fu, editors, \emph{International
  {Geophysics}}, volume~69 of \emph{Satellite {Altimetry} and {Earth}
  {SciencesA} {Handbook} of {Techniques} and {Applications}}. Academic Press,
  2001.

\bibitem[Ballarotta et~al.(2019)Ballarotta, Ubelmann, Pujol, Taburet, Fournier,
  Legeais, Faugère, Delepoulle, Chelton, and
  Picot]{ballarotta_resolutions_2019}
M.~Ballarotta, C.~Ubelmann, M.-I. Pujol, G.~Taburet, F.~Fournier, J.F. Legeais,
  Y.~Faugère, A.~Delepoulle, D.~Chelton, and N.~Picot.
\newblock On the resolutions of ocean altimetry maps.
\newblock \emph{Ocean Science}, 15\penalty0 (4):\penalty0 1091--1109, 2019.
\newblock \doi{https://doi.org/10.5194/os-15-1091-2019}.

\bibitem[Mahadevan and Tandon(2006)]{mahadevan_analysis_2006}
A.~Mahadevan and A.~Tandon.
\newblock An analysis of mechanisms for submesoscale vertical motion at ocean
  fronts.
\newblock \emph{Ocean Modelling}, 14\penalty0 (3):\penalty0 241--256, 2006.
\newblock ISSN 1463-5003.
\newblock \doi{10.1016/j.ocemod.2006.05.006}.

\bibitem[Baaklini et~al.(2021)Baaklini, Issa, Fakhri, Brajard, Fifani, Menna,
  Taupier-Letage, Bosse, and Mortier]{baaklini_blending_2021}
G.~Baaklini, L.~Issa, M.~Fakhri, J.~Brajard, G.~Fifani, M.~Menna,
  I.~Taupier-Letage, A.~Bosse, and L.~Mortier.
\newblock Blending drifters and altimetric data to estimate surface currents:
  {Application} in the {Levantine} {Mediterranean} and objective validation
  with different data types.
\newblock \emph{Ocean Mod.}, 166:\penalty0 101850, 2021.
\newblock \doi{10.1016/j.ocemod.2021.101850}.

\bibitem[Sun et~al.(2022)Sun, Penny, and Harrison]{sun_impacts_2022}
L.~Sun, Stephen~G. Penny, and M.~Harrison.
\newblock Impacts of the {Lagrangian} {Data} {Assimilation} of {Surface}
  {Drifters} on {Estimating} {Ocean} {Circulation} during the {Gulf} of
  {Mexico} {Grand} {Lagrangian} {Deployment}.
\newblock \emph{Mont. Weath. Rev.}, 150\penalty0 (4):\penalty0 949--965, 2022.
\newblock \doi{10.1175/MWR-D-21-0123.1}.

\bibitem[Uchida et~al.(2022)Uchida, Le~Sommer, Stern, Abernathey, Holdgraf,
  Albert, Brodeau, Chassignet, Xu, Gula, Roullet, Koldunov, Danilov, Wang,
  Menemenlis, Bricaud, Arbic, Shriver, Qiao, Xiao, Biastoch, Schubert,
  Fox-Kemper, Dewar, and Wallcraft]{uchida_cloud-based_2022}
Takaya Uchida, Julien Le~Sommer, Charles Stern, Ryan~P. Abernathey, Chris
  Holdgraf, Aurélie Albert, Laurent Brodeau, Eric~P. Chassignet, Xiaobiao Xu,
  Jonathan Gula, Guillaume Roullet, Nikolay Koldunov, Sergey Danilov, Qiang
  Wang, Dimitris Menemenlis, Clément Bricaud, Brian~K. Arbic, Jay~F. Shriver,
  Fangli Qiao, Bin Xiao, Arne Biastoch, René Schubert, Baylor Fox-Kemper,
  William~K. Dewar, and Alan Wallcraft.
\newblock Cloud-based framework for inter-comparing submesoscale-permitting
  realistic ocean models.
\newblock \emph{Geoscientific Model Development}, 15\penalty0 (14):\penalty0
  5829--5856, 2022.
\newblock \doi{10.5194/gmd-15-5829-2022}.

\bibitem[Villas~Boas et~al.(2019)Villas~Boas, Ardhuin, Ayet, Bourassa, Brandt,
  Chapron, Cornuelle, Farrar, Fewings, Fox-Kemper, Gille, Gommenginger,
  Heimbach, Hell, Li, Mazloff, Merrifield, Mouche, Rio, Rodriguez, Shutler,
  Subramanian, Terrill, Tsamados, Ubelmann, and van
  Sebille]{villas_boas_integrated_2019}
Ana~B. Villas~Boas, Fabrice Ardhuin, Alex Ayet, Mark~A. Bourassa, Peter Brandt,
  Betrand Chapron, Bruce~D. Cornuelle, J.~T. Farrar, Melanie~R. Fewings, Baylor
  Fox-Kemper, Sarah~T. Gille, Christine Gommenginger, Patrick Heimbach,
  Momme~C. Hell, Qing Li, Matthew~R. Mazloff, Sophia~T. Merrifield, Alexis
  Mouche, Marie~H. Rio, Ernesto Rodriguez, Jamie~D. Shutler, Aneesh~C.
  Subramanian, Eric~J. Terrill, Michel Tsamados, Clement Ubelmann, and Erik van
  Sebille.
\newblock Integrated {Observations} of {Global} {Surface} {Winds}, {Currents},
  and {Waves}: {Requirements} and {Challenges} for the {Next} {Decade}.
\newblock \emph{Frontiers in Marine Science}, 6, 2019.

\bibitem[Ardhuin et~al.(2019)Ardhuin, Brandt, Gaultier, Donlon, Battaglia, Boy,
  Casal, Chapron, Collard, Cravatte, Delouis, De~Witte, Dibarboure, Engen,
  Johnsen, Lique, Lopez-Dekker, Maes, Martin, Marié, Menemenlis, Nouguier,
  Peureux, Rampal, Ressler, Rio, Rommen, Shutler, Suess, Tsamados, Ubelmann,
  van Sebille, van~den Oever, and Stammer]{ardhuin_skim_2019}
Fabrice Ardhuin, Peter Brandt, Lucile Gaultier, Craig Donlon, Alessandro
  Battaglia, François Boy, Tania Casal, Bertrand Chapron, Fabrice Collard,
  Sophie Cravatte, Jean-Marc Delouis, Erik De~Witte, Gerald Dibarboure, Geir
  Engen, Harald Johnsen, Camille Lique, Paco Lopez-Dekker, Christophe Maes,
  Adrien Martin, Louis Marié, Dimitris Menemenlis, Frederic Nouguier, Charles
  Peureux, Pierre Rampal, Gerhard Ressler, Marie-Helene Rio, Bjorn Rommen,
  Jamie~D. Shutler, Martin Suess, Michel Tsamados, Clement Ubelmann, Erik van
  Sebille, Martin van~den Oever, and Detlef Stammer.
\newblock {SKIM}, a {Candidate} {Satellite} {Mission} {Exploring} {Global}
  {Ocean} {Currents} and {Waves}.
\newblock \emph{Frontiers in Marine Science}, 6, 2019.
\newblock ISSN 2296-7745.
\newblock URL
  \url{https://www.frontiersin.org/articles/10.3389/fmars.2019.00209}.

\bibitem[Moore et~al.(2019)Moore, Martin, Akella, Arango, Balmaseda, Bertino,
  Ciavatta, Cornuelle, Cummings, Frolov, Lermusiaux, Oddo, Oke, Storto,
  Teruzzi, Vidard, and Weaver]{moore_synthesis_2019}
A.M. Moore, M.J. Martin, S.~Akella, H.G. Arango, M.~Balmaseda, L.~Bertino,
  S.~Ciavatta, B.~Cornuelle, J.~Cummings, S.~Frolov, P.~Lermusiaux, P.~Oddo,
  P.R. Oke, A.~Storto, A.~Teruzzi, A.~Vidard, and A.T. Weaver.
\newblock Synthesis of {Ocean} {Observations} {Using} {Data} {Assimilation} for
  {Operational}, {Real}-{Time} and {Reanalysis} {Systems}: {A} {More}
  {Complete} {Picture} of the {State} of the {Ocean}.
\newblock \emph{Frontiers in Marine Science}, 6, 2019.

\bibitem[Storto et~al.(2019)Storto, Alvera-Azcarate, Balmaseda, Barth,
  Chevallier, Counillon, Domingues, Drevillon, Drillet, Forget, Garric, Haines,
  Hernandez, Iovino, Jackson, Lellouche, Masina, Mayer, Oke, Penny, Peterson,
  Yang, and Zuo]{storto_ocean_2019}
Andrea Storto, Aida Alvera-Azcarate, Magdalena~A. Balmaseda, Alexander Barth,
  Matthieu Chevallier, Francois Counillon, Catia~M. Domingues, Marie Drevillon,
  Yann Drillet, Gaël Forget, Gilles Garric, Keith Haines, Fabrice Hernandez,
  Doroteaciro Iovino, Laura~C. Jackson, Jean-Michel Lellouche, Simona Masina,
  Michael Mayer, Peter~R. Oke, Stephen~G. Penny, K.~Andrew Peterson, Chunxue
  Yang, and Hao Zuo.
\newblock Ocean {Reanalyses}: {Recent} {Advances} and {Unsolved} {Challenges}.
\newblock \emph{Frontiers in Marine Science}, 6, 2019.
\newblock ISSN 2296-7745.
\newblock URL
  \url{https://www.frontiersin.org/articles/10.3389/fmars.2019.00418}.

\bibitem[Taburet et~al.(2019)Taburet, Sanchez-Roman, Ballarotta, Pujol,
  Legeais, Fournier, Faugere, and Dibarboure]{taburet_duacs_2019}
G.~Taburet, A.~Sanchez-Roman, M.~Ballarotta, M.-I. Pujol, F.F. Legeais,
  F.~Fournier, Y.~Faugere, and G.~Dibarboure.
\newblock {DUACS} {DT2018}: 25 years of reprocessed sea level altimetry
  products.
\newblock \emph{Ocean Sci.}, page~18, 2019.
\newblock \doi{10.5194/os-15-1207-2019}.

\bibitem[Benkiran et~al.(2021)Benkiran, Ruggiero, Greiner, Le~Traon, Rémy,
  Lellouche, Bourdallé-Badie, Drillet, and Tchonang]{benkiran_assessing_2021}
M.~Benkiran, G.~Ruggiero, E.~Greiner, P.-Y. Le~Traon, E.~Rémy, J.M. Lellouche,
  R.~Bourdallé-Badie, Y.~Drillet, and B.~Tchonang.
\newblock Assessing the {Impact} of the {Assimilation} of {SWOT} {Observations}
  in a {Global} {High}-{Resolution} {Analysis} and {Forecasting} {System}
  {Part} 1: {Methods}.
\newblock \emph{Frontiers in Marine Science}, 8, 2021.
\newblock \doi{10.3389/fmars.2021.691955}.

\bibitem[Fujii et~al.(2019)Fujii, Rémy, Zuo, Oke, Halliwell, Gasparin,
  Benkiran, Loose, Cummings, Xie, Xue, Masuda, Smith, Balmaseda, Germineaud,
  Lea, Larnicol, Bertino, Bonaduce, Brasseur, Donlon, Heimbach, Kim,
  Kourafalou, Le~Traon, Martin, Paturi, Tranchant, and
  Usui]{fujii_observing_2019}
Y.~Fujii, E.~Rémy, H.~Zuo, P.~Oke, George Halliwell, Florent Gasparin, Mounir
  Benkiran, Nora Loose, James Cummings, Jiping Xie, Yan Xue, Shuhei Masuda,
  Gregory~C. Smith, Magdalena Balmaseda, Cyril Germineaud, Daniel~J. Lea,
  Gilles Larnicol, Laurent Bertino, Antonio Bonaduce, Pierre Brasseur, Craig
  Donlon, Patrick Heimbach, YoungHo Kim, Villy Kourafalou, Pierre-Yves
  Le~Traon, Matthew Martin, Shastri Paturi, Benoit Tranchant, and N.~Usui.
\newblock Observing {System} {Evaluation} {Based} on {Ocean} {Data}
  {Assimilation} and {Prediction} {Systems}: {On}-{Going} {Challenges} and a
  {Future} {Vision} for {Designing} and {Supporting} {Ocean} {Observational}
  {Networks}.
\newblock \emph{Front. Mar. Sc.}, 6, 2019.
\newblock \doi{10.3389/fmars.2019.00417}.

\bibitem[Ciani et~al.(2021)Ciani, Charles, Buongiorno~Nardelli, Rio, and
  Santoleri]{ciani_ocean_2021}
D.~Ciani, E.~Charles, B.~Buongiorno~Nardelli, M.-H. Rio, and R.~Santoleri.
\newblock Ocean {Currents} {Reconstruction} from a {Combination} of {Altimeter}
  and {Ocean} {Colour} {Data}: {A} {Feasibility} {Study}.
\newblock \emph{Rem. Sens.}, 13\penalty0 (12):\penalty0 2389, 2021.
\newblock \doi{10.3390/rs13122389}.

\bibitem[Isern-Fontanet et~al.(2006)Isern-Fontanet, Chapron, Lapeyre, and
  Klein]{isern-fontanet_potential_2006}
J.~Isern-Fontanet, B.~Chapron, G.~Lapeyre, and P.~Klein.
\newblock Potential use of microwave sea surface temperatures for the
  estimation of ocean currents.
\newblock \emph{Geophysical Res. Lett.}, 33\penalty0 (L24608), 2006.
\newblock URL \url{doi:10.1029/2006GL027801}.

\bibitem[Alvera-Azcárate et~al.(2007)Alvera-Azcárate, Barth, Beckers, and
  Weisberg]{alvera-azcarate_multivariate_2007}
A.~Alvera-Azcárate, A.~Barth, J.-M. Beckers, and R.~H. Weisberg.
\newblock Multivariate reconstruction of missing data in sea surface
  temperature, chlorophyll, and wind satellite fields.
\newblock \emph{Journal of Geophysical Research: Oceans}, 112\penalty0 (C3),
  2007.
\newblock \doi{10.1029/2006JC003660}.

\bibitem[Lguensat et~al.(2017)Lguensat, Tandeo, Aillot, and
  Fablet]{lguensat_analog_2017}
R.~Lguensat, P.~Tandeo, P.~Aillot, and R.~Fablet.
\newblock The {Analog} {Data} {Assimilation}.
\newblock \emph{Monthly Weather Review}, 2017.

\bibitem[Barth et~al.(2020)Barth, Alvera-Azcárate, Licer, and
  Beckers]{barth_dincae_2020}
A.~Barth, A.~Alvera-Azcárate, M.~Licer, and J.-M. Beckers.
\newblock {DINCAE} 1.0: a convolutional neural network with error estimates to
  reconstruct sea surface temperature satellite observations.
\newblock \emph{Geosci. Mod. Dev.}, 13\penalty0 (3):\penalty0 1609--1622, 2020.
\newblock \doi{10.5194/gmd-13-1609-2020}.

\bibitem[Fablet et~al.(2021{\natexlab{a}})Fablet, Amar, Febvre, Beauchamp, and
  Chapron]{fablet_end--end_2021}
R.~Fablet, M.~M. Amar, Q.~Febvre, M.~Beauchamp, and B.~Chapron.
\newblock End-to-end physics-informed representation learning for satellite
  ocean remote sensing: applications to satellite altimetry and sea surface
  currents.
\newblock In \emph{{ISPRS} {Ann}. {Photogram}. {Rem}. {Sens}. {Spat}. {Inf}.
  {Sc}.}, volume V-3-2021, pages 295--302, 2021{\natexlab{a}}.
\newblock \doi{10.5194/isprs-annals-V-3-2021-295-2021}.

\bibitem[Fablet et~al.(2022)Fablet, Febvre, and
  Chapron]{fablet_multimodal_2022}
R.~Fablet, Q.~Febvre, and B.~Chapron.
\newblock Multimodal {4DVarNets} for the reconstruction of sea surface dynamics
  from {SST}-{SSH} synergies.
\newblock \emph{arXiv:2207.01372}, 2022.
\newblock \doi{10.48550/arXiv.2207.01372}.

\bibitem[George et~al.(2021)George, Manucharyan, and
  Thompson]{george_deep_2021}
T.M. George, G.E. Manucharyan, and A.F. Thompson.
\newblock Deep learning to infer eddy heat fluxes from sea surface height
  patterns of mesoscale turbulence.
\newblock \emph{Nature Communications}, 12\penalty0 (1):\penalty0 800, 2021.
\newblock \doi{10.1038/s41467-020-20779-9}.

\bibitem[Manucharyan et~al.(2021)Manucharyan, Siegelman, and
  Klein]{manucharyan_deep_2021}
G.~E. Manucharyan, L.~Siegelman, and P.~Klein.
\newblock A {Deep} {Learning} {Approach} to {Spatiotemporal} {Sea} {Surface}
  {Height} {Interpolation} and {Estimation} of {Deep} {Currents} in
  {Geostrophic} {Ocean} {Turbulence}.
\newblock \emph{JAMES}, 13\penalty0 (1):\penalty0 e2019MS001965, 2021.
\newblock \doi{10.1029/2019MS001965}.

\bibitem[S\'{e}vellec et~al.(2022)S\'{e}vellec, Colin~de Verdi\`{e}re, and
  Kolodziejczyk]{Sevellec:22}
F.\ S\'{e}vellec, A.\ Colin~de Verdi\`{e}re, and N.~Kolodziejczyk.
\newblock Global observations of deep ocean kinetic energy transfers.
\newblock \emph{submitted in {J.\ Phys.\ Oceanogr.}}, 2022.

\bibitem[Evensen(2009)]{evensen_data_2009}
G.~Evensen.
\newblock \emph{Data {Assimilation}}.
\newblock Springer Berlin Heidelberg, 2009.

\bibitem[Carrassi et~al.(2018)Carrassi, Bocquet, Bertino, and
  Evensen]{carrassi_data_2018}
A.~Carrassi, M.~Bocquet, L.~Bertino, and G.~Evensen.
\newblock Data assimilation in the geosciences: {An} overview of methods,
  issues, and perspectives.
\newblock \emph{WIREs Climate Change}, 9\penalty0 (5):\penalty0 e535, 2018.
\newblock \doi{10.1002/wcc.535}.

\bibitem[Cressie and Wikle(2015)]{cressie_statistics_2015}
N.~Cressie and C.K. Wikle.
\newblock \emph{Statistics for {Spatio}-{Temporal} {Data}}.
\newblock John Wiley \& Sons, 2015.

\bibitem[Le~Guillou et~al.(2020)Le~Guillou, Metref, Cosme, Ubelmann,
  Ballarotta, Le~Sommer, and Verron]{le_guillou_mapping_2020}
F.~Le~Guillou, S.~Metref, E.~Cosme, C.~Ubelmann, M.~Ballarotta, J.~Le~Sommer,
  and J.~Verron.
\newblock Mapping {Altimetry} in the {Forthcoming} {SWOT} {Era} by
  {Back}-and-{Forth} {Nudging} a {One}-{Layer} {Quasigeostrophic} {Model}.
\newblock \emph{J. Atm. Ocean. Tech.}, 38:\penalty0 697 -- 710, 2020.
\newblock \doi{10.1175/jtech-d-20-0104.1}.

\bibitem[Ubelmann et~al.(2014)Ubelmann, Klein, and Fu]{ubelmann_dynamic_2014}
C.~Ubelmann, P.~Klein, and L.-L. Fu.
\newblock Dynamic {Interpolation} of {Sea} {Surface} {Height} and {Potential}
  {Applications} for {Future} {High}-{Resolution} {Altimetry} {Mapping}.
\newblock \emph{J. Atmos. Ocean. Tech.}, 32\penalty0 (1):\penalty0 177--184,
  2014.
\newblock \doi{10.1175/JTECH-D-14-00152.1}.

\bibitem[Abdalla et~al.(2021)Abdalla, Abdeh~Kolahchi, Andersen, Antich, Arbic,
  Armitage, Arnault, Artana, Aulicino, Ayoub, Badulin, Baker, Banks, Bao,
  Barbetta, Barceló-Llull, Barlier, Basu, Bauer-Gottwein, Becker, Beckley,
  Bellefond, Belonenko, Benkiran, Benkouider, Bennartz, Benveniste, Bercher,
  Berge-Nguyen, Bettencourt, Blarel, Blazquez, Blumstein, Bonnefond, Borde,
  Bouffard, Boy, Boy, Brachet, Brasseur, Braun, Brocca, Brockley, Brodeau,
  Brown, Bruinsma, Bulczak, Buzzard, Cahill, Calmant, Calzas, Camici, Cancet,
  Capdeville, Carabajal, Carrere, Cazenave, Chassignet, Chauhan, Cherchali,
  Chereskin, Cheymol, Ciani, Cipollini, Cirillo, Cosme, Coss, Cotroneo, Cotton,
  Couhert, Coutin-Faye, Crétaux, Cyr, d’Ovidio, Darrozes, David, Dayoub,
  De~Staerke, Deng, Desai, Desjonqueres, Dettmering, Di~Bella, Díaz-Barroso,
  Dibarboure, Dieng, Dinardo, Dobslaw, Dodet, Doglioli, Domeneghetti, Donahue,
  Dong, Donlon, Dorandeu, Drezen, Drinkwater, Du~Penhoat, Dushaw, Egido,
  Erofeeva, Escudier, Esselborn, Exertier, Fablet, Falco, Farrell, Faugere,
  Femenias, Fenoglio, Fernandes, Fernández, Ferrage, Ferrari, Fichen,
  Filippucci, Flampouris, Fleury, Fornari, Forsberg, Frappart, Frery, Garcia,
  Garcia-Mondejar, Gaudelli, Gaultier, Getirana, Gibert, Gil, Gilbert, Gille,
  Giulicchi, Gómez-Enri, Gómez-Navarro, Gommenginger, Gourdeau, Griffin,
  Groh, Guerin, Guerrero, Guinle, Gupta, Gutknecht, Hamon, Han, Hauser, Helm,
  Hendricks, Hernandez, Hogg, Horwath, Idzanovic, Janssen, Jeansou, Jia, Jia,
  Jiang, Johannessen, Kamachi, Karimova, Kelly, Kim, King, Kittel, Klein, Klos,
  Knudsen, Koenig, Kostianoy, Kouraev, Kumar, Labroue, Lago, Lambin, Lasson,
  Laurain, Laxenaire, Lázaro, Le~Gac, Le~Sommer, Le~Traon, Lebedev, Léger,
  Legresy, Lemoine, Lenain, Leuliette, Levy, Lillibridge, Liu, Llovel, Lyard,
  Macintosh, Makhoul~Varona, Manfredi, Marin, Mason, Massari, Mavrocordatos,
  Maximenko, McMillan, Medina, Melet, Meloni, Mertikas, Metref, Meyssignac,
  Minster, Moreau, Moreira, Morel, Morrow, Moyard, Mulet, Naeije, Nerem,
  Ngodock, Nielsen, Nilsen, Niño, Nogueira~Loddo, Noûs, Obligis, Otosaka,
  Otten, Oztunali~Ozbahceci, P.~Raj, Paiva, Paniagua, Paolo, Paris, Pascual,
  Passaro, Paul, Pavelsky, Pearson, Penduff, Peng, Perosanz, Picot, Piras,
  Poggiali, Poirier, Ponce~de León, Prants, Prigent, Provost, Pujol, Qiu,
  Quilfen, Rami, Raney, Raynal, Remy, Rémy, Restano, Richardson, Richardson,
  Ricker, Ricko, Rinne, Rose, Rosmorduc, Rudenko, Ruiz, Ryan, Salaün,
  Sanchez-Roman, Sandberg~Sørensen, Sandwell, Saraceno, Scagliola, Schaeffer,
  Scharffenberg, Scharroo, Schiller, Schneider, Schwatke, Scozzari,
  Ser-giacomi, Seyler, Shah, Sharma, Shaw, Shepherd, Shriver, Shum, Simons,
  Simonsen, Slater, Smith, Soares, Sokolovskiy, Soudarin, Spatar, Speich,
  Srinivasan, Srokosz, Stanev, Staneva, Steunou, Stroeve, Su, Sulistioadi,
  Swain, Sylvestre-baron, Taburet, Tailleux, Takayama, Tapley, Tarpanelli,
  Tavernier, Testut, Thakur, Thibaut, Thompson, Tintoré, Tison, Tourain,
  Tournadre, Townsend, Tran, Trilles, Tsamados, Tseng, Ubelmann, Uebbing,
  Vergara, Verron, Vieira, Vignudelli, Vinogradova~Shiffer, Visser, Vivier,
  Volkov, von Schuckmann, Vuglinskii, Vuilleumier, Walter, Wang, Wang, Watson,
  Wilkin, Willis, Wilson, Woodworth, Yang, Yao, Zaharia, Zakharova, Zaron,
  Zhang, Zhao, and Zinchenko]{abdalla_altimetry_2021}
S.~Abdalla, A.~Abdeh~Kolahchi, O.B. Andersen, Helena Antich, Brian Arbic,
  Thomas Armitage, Sabine Arnault, Camila Artana, Giuseppe Aulicino, Nadia
  Ayoub, Sergei Badulin, Steven Baker, Chris Banks, Lifeng Bao, Silvia
  Barbetta, Bàrbara Barceló-Llull, François Barlier, Sujit Basu, Peter
  Bauer-Gottwein, Matthias Becker, Brian Beckley, Nicole Bellefond, Tatyana
  Belonenko, Mounir Benkiran, Touati Benkouider, Ralf Bennartz, Jérôme
  Benveniste, Nicolas Bercher, Muriel Berge-Nguyen, Joao Bettencourt, Fabien
  Blarel, Alejandro Blazquez, Denis Blumstein, Pascal Bonnefond, Franck Borde,
  Jérôme Bouffard, François Boy, Jean-Paul Boy, Cédric Brachet, Pierre
  Brasseur, Alexander Braun, Luca Brocca, David Brockley, Laurent Brodeau,
  Shannon Brown, Sean Bruinsma, Anna Bulczak, Sammie Buzzard, Madeleine Cahill,
  Stéphane Calmant, Michel Calzas, Stefania Camici, Mathilde Cancet, Hugues
  Capdeville, Claudia~Cristina Carabajal, Loren Carrere, Anny Cazenave, Eric~P.
  Chassignet, Prakash Chauhan, Selma Cherchali, Teresa Chereskin, Cecile
  Cheymol, Daniele Ciani, Paolo Cipollini, Francesca Cirillo, Emmanuel Cosme,
  Steve Coss, Yuri Cotroneo, David Cotton, Alexandre Couhert, Sophie
  Coutin-Faye, Jean-François Crétaux, Frederic Cyr, Francesco d’Ovidio,
  José Darrozes, Cedric David, Nadim Dayoub, Danielle De~Staerke, Xiaoli Deng,
  Shailen Desai, Jean-Damien Desjonqueres, Denise Dettmering, Alessandro
  Di~Bella, Lara Díaz-Barroso, Gerald Dibarboure, Habib~Boubacar Dieng,
  Salvatore Dinardo, Henryk Dobslaw, Guillaume Dodet, Andrea Doglioli, Alessio
  Domeneghetti, David Donahue, Shenfu Dong, Craig Donlon, Joël Dorandeu,
  Christine Drezen, Mark Drinkwater, Yves Du~Penhoat, Brian Dushaw, Alejandro
  Egido, Svetlana Erofeeva, Philippe Escudier, Saskia Esselborn, Pierre
  Exertier, Ronan Fablet, Cédric Falco, Sinead~Louise Farrell, Yannice
  Faugere, Pierre Femenias, Luciana Fenoglio, Joana Fernandes, Juan~Gabriel
  Fernández, Pascale Ferrage, Ramiro Ferrari, Lionel Fichen, Paolo Filippucci,
  Stylianos Flampouris, Sara Fleury, Marco Fornari, Rene Forsberg, Frédéric
  Frappart, Marie-laure Frery, Pablo Garcia, Albert Garcia-Mondejar, Julia
  Gaudelli, Lucile Gaultier, Augusto Getirana, Ferran Gibert, Artur Gil, Lin
  Gilbert, Sarah Gille, Luisella Giulicchi, Jesús Gómez-Enri, Laura
  Gómez-Navarro, Christine Gommenginger, Lionel Gourdeau, David Griffin,
  Andreas Groh, Alexandre Guerin, Raul Guerrero, Thierry Guinle, Praveen Gupta,
  Benjamin~D. Gutknecht, Mathieu Hamon, Guoqi Han, Danièle Hauser, Veit Helm,
  Stefan Hendricks, Fabrice Hernandez, Anna Hogg, Martin Horwath, Martina
  Idzanovic, Peter Janssen, Eric Jeansou, Yongjun Jia, Yuanyuan Jia, Liguang
  Jiang, Johnny~A. Johannessen, Masafumi Kamachi, Svetlana Karimova, Kathryn
  Kelly, Sung~Yong Kim, Robert King, Cecile M.~M. Kittel, Patrice Klein, Anna
  Klos, Per Knudsen, Rolf Koenig, Andrey Kostianoy, Alexei Kouraev, Raj Kumar,
  Sylvie Labroue, Loreley~Selene Lago, Juliette Lambin, Léa Lasson, Olivier
  Laurain, Rémi Laxenaire, Clara Lázaro, Sophie Le~Gac, Julien Le~Sommer,
  Pierre-Yves Le~Traon, Sergey Lebedev, Fabien Léger, B.~Legresy, Frank
  Lemoine, Luc Lenain, Eric Leuliette, Marina Levy, John Lillibridge, Jianqiang
  Liu, William Llovel, Florent Lyard, Claire Macintosh, Eduard Makhoul~Varona,
  Cécile Manfredi, Frédéric Marin, Evan Mason, Christian Massari, Constantin
  Mavrocordatos, Nikolai Maximenko, Malcolm McMillan, Thierry Medina, Angelique
  Melet, Marco Meloni, Stelios Mertikas, Sammy Metref, Benoit Meyssignac,
  Jean-François Minster, Thomas Moreau, Daniel Moreira, Yves Morel, Rosemary
  Morrow, John Moyard, Sandrine Mulet, Marc Naeije, Robert~Steven Nerem, Hans
  Ngodock, Karina Nielsen, Jan Even~Øie Nilsen, Fernando Niño, Carolina
  Nogueira~Loddo, Camille Noûs, Estelle Obligis, Inès Otosaka, Michiel Otten,
  Berguzar Oztunali~Ozbahceci, Roshin P.~Raj, Rodrigo Paiva, Guillermina
  Paniagua, Fernando Paolo, Adrien Paris, Ananda Pascual, Marcello Passaro,
  Stephan Paul, Tamlin Pavelsky, Christopher Pearson, Thierry Penduff, Fukai
  Peng, Felix Perosanz, Nicolas Picot, Fanny Piras, Valerio Poggiali, Étienne
  Poirier, Sonia Ponce~de León, Sergey Prants, Catherine Prigent, Christine
  Provost, M-Isabelle Pujol, Bo~Qiu, Yves Quilfen, Ali Rami, R.~Keith Raney,
  Matthias Raynal, Elisabeth Remy, Frédérique Rémy, Marco Restano, Annie
  Richardson, Donald Richardson, Robert Ricker, Martina Ricko, Eero Rinne,
  Stine~Kildegaard Rose, Vinca Rosmorduc, Sergei Rudenko, Simón Ruiz,
  Barbara~J. Ryan, Corinne Salaün, Antonio Sanchez-Roman, Louise
  Sandberg~Sørensen, David Sandwell, Martin Saraceno, Michele Scagliola,
  Philippe Schaeffer, Martin~G. Scharffenberg, Remko Scharroo, Andreas
  Schiller, Raphael Schneider, Christian Schwatke, Andrea Scozzari, Enrico
  Ser-giacomi, Frederique Seyler, Rashmi Shah, Rashmi Sharma, Andrew Shaw,
  Andrew Shepherd, Jay Shriver, C.~K. Shum, Wim Simons, Sebatian~B. Simonsen,
  Thomas Slater, Walter Smith, Saulo Soares, Mikhail Sokolovskiy, Laurent
  Soudarin, Ciprian Spatar, Sabrina Speich, Margaret Srinivasan, Meric Srokosz,
  Emil Stanev, Joanna Staneva, Nathalie Steunou, Julienne Stroeve, Bob Su,
  Yohanes~Budi Sulistioadi, Debadatta Swain, Annick Sylvestre-baron, Nicolas
  Taburet, Rémi Tailleux, Katsumi Takayama, Byron Tapley, Angelica Tarpanelli,
  Gilles Tavernier, Laurent Testut, Praveen~K. Thakur, Pierre Thibaut, LuAnne
  Thompson, Joaquín Tintoré, Céline Tison, Cédric Tourain, Jean Tournadre,
  Bill Townsend, Ngan Tran, Sébastien Trilles, Michel Tsamados, Kuo-Hsin
  Tseng, Clément Ubelmann, Bernd Uebbing, Oscar Vergara, Jacques Verron, Telmo
  Vieira, Stefano Vignudelli, Nadya Vinogradova~Shiffer, Pieter Visser,
  Frederic Vivier, Denis Volkov, Karina von Schuckmann, Valerii Vuglinskii,
  Pierrik Vuilleumier, Blake Walter, Jida Wang, Chao Wang, Christopher Watson,
  John Wilkin, Josh Willis, Hilary Wilson, Philip Woodworth, Kehan Yang,
  Fangfang Yao, Raymond Zaharia, Elena Zakharova, Edward~D. Zaron, Yongsheng
  Zhang, Zhongxiang Zhao, and Vadim Zinchenko.
\newblock Altimetry for the future: {Building} on 25 years of progress.
\newblock \emph{Adv. Space Res.}, 68\penalty0 (2):\penalty0 319--363, 2021.
\newblock \doi{10.1016/j.asr.2021.01.022}.

\bibitem[Barthelemy et~al.(2021)Barthelemy, Brajard, Bertino, and
  Counillon]{barthelemy_super-resolution_2021}
S.~Barthelemy, J.~Brajard, L.~Bertino, and F.~Counillon.
\newblock Super-resolution data assimilation.
\newblock \emph{arXiv:2109.08017}, 2021.
\newblock \doi{10.48550/arXiv.2109.08017}.

\bibitem[Boudier et~al.(2020)Boudier, Fillion, Gratton, and
  Gürol]{boudier_dan_2020}
P.~Boudier, A.~Fillion, S.~Gratton, and S.~Gürol.
\newblock {DAN} -- {An} optimal {Data} {Assimilation} framework based on
  machine learning {Recurrent} {Networks}.
\newblock \emph{arXiv:2010.09694}, October 2020.

\bibitem[Bocquet et~al.(2020)Bocquet, Brajard, Carrassi, and
  Bertino]{bocquet_bayesian_2020}
M.~Bocquet, J.~Brajard, A.~Carrassi, and L.~Bertino.
\newblock Bayesian inference of chaotic dynamics by merging data assimilation,
  machine learning and expectation-maximization.
\newblock \emph{Foundations of Data Science}, 2\penalty0 (1):\penalty0 55--80,
  2020.
\newblock \doi{10.3934/fods.2020004}.

\bibitem[Fablet et~al.(2021{\natexlab{b}})Fablet, Chapron, Drumetz, Memin,
  Pannekoucke, and Rousseau]{fablet_learning_2021}
R.~Fablet, B.~Chapron, L.~Drumetz, E.~Memin, O.~Pannekoucke, and F.~Rousseau.
\newblock Learning {Variational} {Data} {Assimilation} {Models} and {Solvers}.
\newblock \emph{JAMES}, 13\penalty0 (e2021MS002572), 2021{\natexlab{b}}.

\bibitem[Nonnenmacher and Greenberg(2021)]{nonnenmacher_deep_2021}
M.~Nonnenmacher and D.S. Greenberg.
\newblock Deep {Emulators} for {Differentiation}, {Forecasting}, and
  {Parametrization} in {Earth} {Science} {Simulators}.
\newblock \emph{JAMES}, 13\penalty0 (7):\penalty0 e2021MS002554, 2021.
\newblock \doi{10.1029/2021MS002554}.

\bibitem[Ajayi et~al.(2020)Ajayi, Le~Sommer, Chassignet, Molines, Xu, Albert,
  and Cosme]{ajayi_spatial_2020}
A.~Ajayi, J.~Le~Sommer, E.~Chassignet, J.-M. Molines, X.~Xu, A.~Albert, and
  E.~Cosme.
\newblock Spatial and {Temporal} {Variability} of the {North} {Atlantic} {Eddy}
  {Field} {From} {Two} {Kilometric}-{Resolution} {Ocean} {Models}.
\newblock \emph{J. Geophys. Res.}, 125\penalty0 (5):\penalty0 e2019JC015827,
  2020.
\newblock \doi{10.1029/2019JC015827}.

\bibitem[Madec et~al.(2022)Madec, Bourdalle-Badie, Chanut, Clementi, Coward,
  Ethe, Iovino, Lea, Levy, Lovato, Martin, Masson, Mocavero, Rousset, Storkey,
  Mueller, Nurser, Bell, Samson, Mathiot, Mele, and Moulin]{madec_nemo_2022}
G.~Madec, R.~Bourdalle-Badie, J.~Chanut, E.~Clementi, A.~Coward, C.~Ethe,
  D.~Iovino, D.~Lea, C.~Levy, T.~Lovato, N.~Martin, S.~Masson, S.~Mocavero,
  C.~Rousset, D.~Storkey, S.~Mueller, G.~Nurser, M.~Bell, G.~Samson,
  P.~Mathiot, F.~Mele, and A.~Moulin.
\newblock {NEMO} ocean engine.
\newblock \emph{Tech. Report}, 2022.
\newblock \doi{10.5281/zenodo.6334656}.

\bibitem[Gaultier et~al.(2015)Gaultier, Ubelmann, and
  Fu]{gaultier_challenge_2015}
L.~Gaultier, C.~Ubelmann, and L.-L. Fu.
\newblock The {Challenge} of {Using} {Future} {SWOT} {Data} for {Oceanic}
  {Field} {Reconstruction}.
\newblock \emph{J. Atm. Ocean. Tech.}, 33\penalty0 (1):\penalty0 119--126,
  2015.
\newblock \doi{10.1175/JTECH-D-15-0160.1}.

\bibitem[Donlon et~al.(2012)Donlon, Martin, Stark, Roberts-Jones, Fiedler, and
  Xindong]{donlon_operational_2012}
C.~J. Donlon, M.~Martin, J.~Stark, J.~Roberts-Jones, E.~Fiedler, and
  W.~Xindong.
\newblock The {Operational} {Sea} {Surface} {Temperature} and {Sea} {Ice}
  {Analysis} ({OSTIA}) system.
\newblock \emph{Rem. Sens. Env.}, 116:\penalty0 140--158, 2012.
\newblock \doi{10.1016/j.rse.2010.10.017}.

\bibitem[O’Carroll et~al.(2019)O’Carroll, Armstrong, Beggs, Bouali, Casey,
  Corlett, Dash, Donlon, Gentemann, Høyer, Ignatov, Kabobah, Kachi, Kurihara,
  Karagali, Maturi, Merchant, Marullo, Minnett, Pennybacker, Ramakrishnan,
  Ramsankaran, Santoleri, Sunder, Saux~Picart, Vázquez-Cuervo, and
  Wimmer]{ocarroll_observational_2019}
A.G. O’Carroll, E.M. Armstrong, H.M. Beggs, M.~Bouali, K.S. Casey, Gary~K.
  Corlett, Prasanjit Dash, Craig~J. Donlon, Chelle~L. Gentemann, Jacob~L.
  Høyer, Alexander Ignatov, Kamila Kabobah, Misako Kachi, Yukio Kurihara,
  Ioanna Karagali, Eileen Maturi, Christopher~J. Merchant, Salvatore Marullo,
  Peter~J. Minnett, Matthew Pennybacker, Balaji Ramakrishnan, RAAJ Ramsankaran,
  Rosalia Santoleri, Swathy Sunder, Stéphane Saux~Picart, Jorge
  Vázquez-Cuervo, and Werenfrid Wimmer.
\newblock Observational {Needs} of {Sea} {Surface} {Temperature}.
\newblock \emph{Front. Mar. Sc.}, 6, 2019.
\newblock \doi{10.3389/fmars.2019.00420}.

\bibitem[Cicek et~al.(2016)Cicek, Abdulkadir, Lienkamp, Brox, and
  Ronneberger]{cicek_3d_2016}
O.~Cicek, A.~Abdulkadir, S.S. Lienkamp, T.~Brox, and O.~Ronneberger.
\newblock {3D} {U}-{Net}: learning dense volumetric segmentation from sparse
  annotation.
\newblock In \emph{Proc. {MICCAI}}, pages 424--432, 2016.

\bibitem[Beauchamp et~al.(2022)Beauchamp, Febvre, Georgenthum, and
  Fablet]{beauchamp_end--end_2022}
M.~Beauchamp, Q.~Febvre, H.~Georgenthum, and R.~Fablet.
\newblock End-to-end neural interpolation of satellite altimetry data using
  {4DVarNet} schemes.
\newblock \emph{Submitted to Geosc. Meth. Dev.}, 2022.
\newblock \doi{https://doi.org/10.5194/gmd-2022-241}.

\bibitem[Hospedales et~al.(2020)Hospedales, Antoniou, Micaelli, and
  Storkey]{hospedales_meta-learning_2020}
T.~Hospedales, A.~Antoniou, P.~Micaelli, and A.~Storkey.
\newblock Meta-learning in neural networks: {A} survey.
\newblock \emph{arXiv:2004.05439}, 2020.

\bibitem[Febvre et~al.(2022)Febvre, Beauchamp, Georgenthum, and
  Fablet]{febvre_pytorch_2022}
Q.~Febvre, M.~Beauchamp, H.~Georgenthum, and R.~Fablet.
\newblock Pytorch {4DVarNet} code for the reconstruction of sea surface
  currents from {SSH} and {SST} data, 2022.
\newblock doi:10.5281/zenodo.7186323.

\bibitem[Balwada et~al.(2021)Balwada, Xiao, Smith, Abernathey, and
  Gray]{balwada_vertical_2021}
D.~Balwada, Q.~Xiao, S.~Smith, R.~Abernathey, and A.R. Gray.
\newblock Vertical {Fluxes} {Conditioned} on {Vorticity} and {Strain} {Reveal}
  {Submesoscale} {Ventilation}.
\newblock \emph{J. Phys. Ocean.}, 51\penalty0 (9):\penalty0 2883--2901, 2021.
\newblock \doi{10.1175/JPO-D-21-0016.1}.

\bibitem[Okubo(1970)]{okubo_horizontal_1970}
A.~Okubo.
\newblock Horizontal dispersion of floatable particles in the vicinity of
  velocity singularities such as convergences.
\newblock \emph{Deep Sea Res.}, 17\penalty0 (3):\penalty0 445--454, 1970.

\bibitem[Weiss(1991)]{weiss_dynamics_1991}
J.~Weiss.
\newblock The dynamics of enstrophy transfer in two-dimensional hydrodynamics.
\newblock \emph{Physica D: Nonlin. Phen.}, 48\penalty0 (2):\penalty0 273--294,
  1991.

\bibitem[S\'{e}vellec et~al.(2017)S\'{e}vellec, Colin~de Verdi\`{e}re, and
  Ollitrault]{Sevellec:17b}
F.\ S\'{e}vellec, A.\ Colin~de Verdi\`{e}re, and M.~Ollitrault.
\newblock Evolution of intermediate water masses based on argo float
  displacement.
\newblock \emph{J.\ Phys.\ Oceanogr.}, 47:\penalty0 1569--1586, 2017.

\bibitem[Fablet et~al.(2018)Fablet, Verron, Mourre, Chapron, and
  Pascual]{fablet_improving_2018}
R.~Fablet, J.~Verron, B.~Mourre, B.~Chapron, and A.~Pascual.
\newblock Improving {Mesoscale} {Altimetric} {Data} {From} a {Multitracer}
  {Convolutional} {Processing} of {Standard} {Satellite}-{Derived} {Products}.
\newblock \emph{IEEE TGRS}, 56\penalty0 (5):\penalty0 2518--2525, 2018.
\newblock \doi{10.1109/TGRS.2017.2750491}.

\bibitem[Isern-Fontanet et~al.(2014)Isern-Fontanet, Shinde, and
  Andersson]{isern-fontanet_transfer_2014}
J.~Isern-Fontanet, M.~Shinde, and C.~Andersson.
\newblock On the {Transfer} {Function} between {Surface} {Fields} and the
  {Geostrophic} {Stream} {Function} in the {Mediterranean} {Sea}.
\newblock \emph{J. Phys. Ocean.}, 44\penalty0 (5):\penalty0 1406--1423, 2014.
\newblock \doi{10.1175/JPO-D-13-0186.1}.

\bibitem[Rio et~al.(2016)Rio, Santoleri, Bourdalle-Badie, Griffa, Piterbarg,
  and Taburet]{rio_improving_2016}
M.-H. Rio, R.~Santoleri, R.~Bourdalle-Badie, A.~Griffa, L.~Piterbarg, and
  G.~Taburet.
\newblock Improving the {Altimeter}-{Derived} {Surface} {Currents} {Using}
  {High}-{Resolution} {Sea} {Surface} {Temperature} {Data}: {A} {Feasability}
  {Study} {Based} on {Model} {Outputs}.
\newblock \emph{J. Atm. Ocean. Tech.}, 33\penalty0 (12):\penalty0 2769--2784,
  2016.
\newblock \doi{10.1175/JTECH-D-16-0017.1}.

\bibitem[Chapron et~al.(2005)Chapron, Collard, and
  Ardhuin]{chapron_direct_2005}
B.~Chapron, F.~Collard, and F.~Ardhuin.
\newblock Direct measurements of ocean surface velocity from space:
  {Interpretation} and validation.
\newblock \emph{J. Geophys. Res.}, 110, 2005.
\newblock \doi{10.1029/2004JC002809}.

\bibitem[Fablet et~al.(2017)Fablet, Viet, and
  Lguensat]{fablet_data-driven_2017}
R.~Fablet, P.~H. Viet, and R.~Lguensat.
\newblock Data-{Driven} {Models} for the {Spatio}-{Temporal} {Interpolation} of
  {Satellite}-{Derived} {SST} {Fields}.
\newblock \emph{IEEE Trans. on Computational Imaging}, 3\penalty0 (4):\penalty0
  647--657, 2017.
\newblock \doi{10.1109/TCI.2017.2749184}.

\bibitem[McWilliams et~al.(2019)McWilliams, Gula, and
  Molemaker]{mcwilliams_gulf_2019}
J.C. McWilliams, J.~Gula, and M.J. Molemaker.
\newblock The {Gulf} {Stream} {North} {Wall}: {Ageostrophic} {Circulation} and
  {Frontogenesis}.
\newblock \emph{Journal of Physical Oceanography}, 49\penalty0 (4):\penalty0
  893--916, 2019.
\newblock \doi{10.1175/JPO-D-18-0203.1}.

\bibitem[Reul et~al.(2014)Reul, Chapron, Lee, Donlon, Boutin, and
  Alory]{reul_sea_2014}
N.~Reul, B.~Chapron, T.~Lee, C.~Donlon, J.~Boutin, and G.~Alory.
\newblock Sea surface salinity structure of the meandering {Gulf} {Stream}
  revealed by {SMOS} sensor.
\newblock \emph{Geophysical Research Letters}, 41\penalty0 (9):\penalty0
  3141--3148, 2014.
\newblock \doi{10.1002/2014GL059215}.

\bibitem[Barnes et~al.(2021)Barnes, Hu, Bailey, Pahlevan, and
  Franz]{barnes_cross-calibration_2021}
B.B. Barnes, C.~Hu, S.W. Bailey, N.~Pahlevan, and B.A. Franz.
\newblock Cross-calibration of {MODIS} and {VIIRS} long near infrared bands for
  ocean color science and applications.
\newblock \emph{Rem. Sens. Env.}, 260:\penalty0 112439, 2021.
\newblock ISSN 0034-4257.
\newblock \doi{10.1016/j.rse.2021.112439}.

\bibitem[Tilstone et~al.(2021)Tilstone, Pardo, Dall'Olmo, Brewin, Nencioli,
  Dessailly, Kwiatkowska, Casal, and Donlon]{tilstone_performance_2021}
G.H. Tilstone, S.~Pardo, G.~Dall'Olmo, R.J.W. Brewin, F.~Nencioli,
  D.~Dessailly, E.~Kwiatkowska, T.~Casal, and C.~Donlon.
\newblock Performance of {Ocean} {Colour} {Chlorophyll} a algorithms for
  {Sentinel}-3 {OLCI}, {MODIS}-{Aqua} and {Suomi}-{VIIRS} in open-ocean waters
  of the {Atlantic}.
\newblock \emph{Rem. Sens. Env.}, 260:\penalty0 112444, 2021.
\newblock \doi{10.1016/j.rse.2021.112444}.

\bibitem[Yurovskaya et~al.(2019)Yurovskaya, Kudryavtsev, Chapron, and
  Collard]{yurovskaya_ocean_2019}
M.~Yurovskaya, V.~Kudryavtsev, B.~Chapron, and F.~Collard.
\newblock Ocean surface current retrieval from space: {The} {Sentinel}-2
  multispectral capabilities.
\newblock \emph{Rem. Sens. Env.}, 234:\penalty0 111468, 2019.
\newblock \doi{10.1016/j.rse.2019.111468}.

\bibitem[Arbic et~al.(2010)Arbic, Wallcraft, and
  Metzger]{arbic_concurrent_2010}
B.K. Arbic, A.J. Wallcraft, and E.J. Metzger.
\newblock Concurrent simulation of the eddying general circulation and tides in
  a global ocean model.
\newblock \emph{Ocean Modelling}, 32\penalty0 (3):\penalty0 175--187, 2010.
\newblock \doi{10.1016/j.ocemod.2010.01.007}.

\bibitem[Xu and Fu(2012)]{xu_effects_2012}
Yongsheng Xu and Lee-Lueng Fu.
\newblock The {Effects} of {Altimeter} {Instrument} {Noise} on the {Estimation}
  of the {Wavenumber} {Spectrum} of {Sea} {Surface} {Height}.
\newblock \emph{Journal of Physical Oceanography}, 42\penalty0 (12):\penalty0
  2229--2233, August 2012.
\newblock ISSN 0022-3670.
\newblock \doi{10.1175/JPO-D-12-0106.1}.

\bibitem[Shorten and Chin(2019)]{shorten_survey_2019}
C.~Shorten and T.M. Chin.
\newblock A survey on {Image} {Data} {Augmentation} for {Deep} {Learning}.
\newblock \emph{Journal of Big Data}, 6\penalty0 (1):\penalty0 60, July 2019.
\newblock ISSN 2196-1115.
\newblock \doi{10.1186/s40537-019-0197-0}.

\bibitem[Wang et~al.(2022)Wang, Grisouard, Salehipour, Nuz, Poon, and
  Ponte]{wang_deep_2022}
H.~Wang, N.~Grisouard, H.~Salehipour, A.~Nuz, M.~Poon, and A.L. Ponte.
\newblock A {Deep} {Learning} {Approach} to {Extract} {Internal} {Tides}
  {Scattered} by {Geostrophic} {Turbulence}.
\newblock \emph{Geophysical Res. Lett.}, 49\penalty0 (11):\penalty0
  e2022GL099400, 2022.
\newblock \doi{10.1029/2022GL099400}.

\bibitem[Martinez et~al.(2020)Martinez, Brini, Gorgues, Drumetz, Roussillon,
  Tandeo, Maze, and Fablet]{martinez_neural_2020}
E.~Martinez, A.~Brini, T.~Gorgues, L.~Drumetz, J.~Roussillon, P.~Tandeo,
  G.~Maze, and R.~Fablet.
\newblock Neural {Network} {Approaches} to {Reconstruct} {Phytoplankton}
  {Time}-{Series} in the {Global} {Ocean}.
\newblock \emph{Rem. Sens.}, 12\penalty0 (24):\penalty0 4156, 2020.
\newblock \doi{10.3390/rs12244156}.

\bibitem[Boukabara et~al.(2018)Boukabara, Ide, Zhou, Shahroudi, Hoffman,
  Garrett, Kumar, Zhu, and Atlas]{boukabara_community_2018}
S.A. Boukabara, K.~Ide, Y.~Zhou, N.~Shahroudi, R.N. Hoffman, K.~Garrett, V.K.
  Kumar, T.~Zhu, and R.~Atlas.
\newblock Community {Global} {Observing} {System} {Simulation} {Experiment}
  ({OSSE}) {Package} ({CGOP}): {Assessment} and {Validation} of the {OSSE}
  {System} {Using} an {OSSE}–{OSE} {Intercomparison} of {Summary}
  {Assessment} {Metrics}.
\newblock \emph{J. Atm. Ocean. Tech.}, 35\penalty0 (10):\penalty0 2061--2078,
  2018.
\newblock \doi{10.1175/JTECH-D-18-0061.1}.

\bibitem[Vient et~al.(2021)Vient, Jourdin, Fablet, Mengual, Lafosse, and
  Delacourt]{vient_data-driven_2021}
J.M. Vient, F.~Jourdin, R.~Fablet, B.~Mengual, L.~Lafosse, and C.~Delacourt.
\newblock Data-{Driven} {Interpolation} of {Sea} {Surface} {Suspended}
  {Concentrations} {Derived} from {Ocean} {Colour} {Remote} {Sensing} {Data}.
\newblock \emph{Rem. Sens.}, 13\penalty0 (17):\penalty0 3537, 2021.
\newblock \doi{10.3390/rs13173537}.

\bibitem[Puissant et~al.(2021)Puissant, El~Hourany, Charantonis, Bowler, and
  Thiria]{puissant_inversion_2021}
A.~Puissant, R.~El~Hourany, A.A. Charantonis, C.~Bowler, and S.~Thiria.
\newblock Inversion of {Phytoplankton} {Pigment} {Vertical} {Profiles} from
  {Satellite} {Data} {Using} {Machine} {Learning}.
\newblock \emph{Rem. Sens.}, 13\penalty0 (8):\penalty0 1445, 2021.
\newblock ISSN 2072-4292.
\newblock \doi{10.3390/rs13081445}.

\bibitem[Cossarini et~al.(2019)Cossarini, Mariotti, Feudale, Mignot, Salon,
  Taillandier, Teruzzi, and D'Ortenzio]{cossarini_towards_2019}
G.~Cossarini, L.~Mariotti, L.~Feudale, A.~Mignot, S.~Salon, V.~Taillandier,
  A.~Teruzzi, and F.~D'Ortenzio.
\newblock Towards operational {3D}-{Var} assimilation of chlorophyll
  {Biogeochemical}-{Argo} float data into a biogeochemical model of the
  {Mediterranean} {Sea}.
\newblock \emph{Ocean Mod.}, 133:\penalty0 112--128, January 2019.
\newblock \doi{10.1016/j.ocemod.2018.11.005}.

\bibitem[Roemmich et~al.(2019)Roemmich, Alford, Claustre, Johnson, King, Moum,
  Oke, Owens, Pouliquen, Purkey, Scanderbeg, Suga, Wijffels, Zilberman, Bakker,
  Baringer, Belbeoch, Bittig, Boss, Calil, Carse, Carval, Chai, Conchubhair,
  d’Ortenzio, Dall’Olmo, Desbruyeres, Fennel, Fer, Ferrari, Forget,
  Freeland, Fujiki, Gehlen, Greenan, Hallberg, Hibiya, Hosoda, Jayne, Jochum,
  Johnson, Kang, Kolodziejczyk, Körtzinger, Traon, Lenn, Maze, Mork, Morris,
  Nagai, Nash, Garabato, Olsen, Pattabhi, Prakash, Riser, Schmechtig, Schmid,
  Shroyer, Sterl, Sutton, Talley, Tanhua, Thierry, Thomalla, Toole, Troisi,
  Trull, Turton, Velez-Belchi, Walczowski, Wang, Wanninkhof, Waterhouse,
  Waterman, Watson, Wilson, Wong, Xu, and Yasuda]{roemmich_future_2019}
D.~Roemmich, M.H. Alford, H.~Claustre, Kenneth Johnson, Brian King, James Moum,
  Peter Oke, W.~Brechner Owens, Sylvie Pouliquen, Sarah Purkey, Megan
  Scanderbeg, Toshio Suga, Susan Wijffels, Nathalie Zilberman, Dorothee Bakker,
  Molly Baringer, Mathieu Belbeoch, Henry~C. Bittig, Emmanuel Boss, Paulo
  Calil, Fiona Carse, Thierry Carval, Fei Chai, Diarmuid~Ó. Conchubhair,
  Fabrizio d’Ortenzio, Giorgio Dall’Olmo, Damien Desbruyeres, Katja Fennel,
  Ilker Fer, Raffaele Ferrari, Gael Forget, Howard Freeland, Tetsuichi Fujiki,
  Marion Gehlen, Blair Greenan, Robert Hallberg, Toshiyuki Hibiya, Shigeki
  Hosoda, Steven Jayne, Markus Jochum, Gregory~C. Johnson, KiRyong Kang,
  Nicolas Kolodziejczyk, Arne Körtzinger, Pierre-Yves~Le Traon, Yueng-Djern
  Lenn, Guillaume Maze, Kjell~Arne Mork, Tamaryn Morris, Takeyoshi Nagai,
  Jonathan Nash, Alberto~Naveira Garabato, Are Olsen, Rama~Rao Pattabhi, Satya
  Prakash, Stephen Riser, Catherine Schmechtig, Claudia Schmid, Emily Shroyer,
  Andreas Sterl, Philip Sutton, Lynne Talley, Toste Tanhua, Virginie Thierry,
  Sandy Thomalla, John Toole, Ariel Troisi, Thomas~W. Trull, Jon Turton,
  Pedro~Joaquin Velez-Belchi, Waldemar Walczowski, Haili Wang, Rik Wanninkhof,
  Amy~F. Waterhouse, Stephanie Waterman, Andrew Watson, Cara Wilson, Annie
  P.~S. Wong, Jianping Xu, and I.~Yasuda.
\newblock On the {Future} of {Argo}: {A} {Global}, {Full}-{Depth},
  {Multi}-{Disciplinary} {Array}.
\newblock \emph{Front. Mar. Sc.}, 6, 2019.
\newblock ISSN 2296-7745.
\newblock \doi{10.3389/fmars.2019.00439}.

\bibitem[Storto et~al.(2020)Storto, Falchetti, Oddo, Jiang, and
  Tesei]{storto_assessing_2020}
A.~Storto, S.~Falchetti, P.~Oddo, Y.M. Jiang, and A.~Tesei.
\newblock Assessing the {Impact} of {Different} {Ocean} {Analysis} {Schemes} on
  {Oceanic} and {Underwater} {Acoustic} {Predictions}.
\newblock \emph{J. Geophys. Res.}, 125\penalty0 (7):\penalty0 e2019JC015636,
  2020.
\newblock \doi{10.1029/2019JC015636}.

\bibitem[Storto et~al.(2021)Storto, Magistris, Falchetti, and
  Oddo]{storto_neural_2021}
A.~Storto, G.D. Magistris, S.~Falchetti, and P.~Oddo.
\newblock A {Neural} {Network}–{Based} {Observation} {Operator} for {Coupled}
  {Ocean}–{Acoustic} {Variational} {Data} {Assimilation}.
\newblock \emph{Month. Weath. Rev.}, 149\penalty0 (6):\penalty0 1967--1985,
  2021.
\newblock \doi{10.1175/MWR-D-20-0320.1}.

\end{thebibliography}

\end{document}